\pdfoutput=1

\documentclass[10pt,journal,compsoc]{IEEEtran}

\usepackage[numbers]{natbib}

\usepackage{graphicx}
\usepackage{placeins}

\usepackage{amsmath}
\usepackage{amsfonts}
\usepackage{amssymb}
\usepackage{amsthm}

\usepackage{mdframed}

\usepackage{mathtools}
\usepackage[inline]{enumitem}

\usepackage{hyperref}
\hypersetup{hidelinks}
\usepackage{url}
\usepackage{cleveref}

\usepackage[detect-family, detect-mode]{siunitx}

\usepackage{multirow}

\usepackage{tabularx}
\usepackage{booktabs}

\usepackage{float}
\usepackage{pdflscape} % Allows for landscape page environments.

\usepackage{xcolor}

\Crefname{figure}{Fig.}{Fig.s}

\begin{document}
\title{%
    ARAUS: A Large-Scale Dataset and\\
    Baseline Models of Affective Responses to\\
    Augmented Urban Soundscapes%
}
\author{Kenneth~Ooi,~\IEEEmembership{Graduate~Student~Member,~IEEE,}
        Zhen-Ting~Ong,
        Karn~N.~Watcharasupat,~\IEEEmembership{Graduate~Student~Member,~IEEE,}
        Bhan~Lam,~\IEEEmembership{Member,~IEEE,}\\
        Joo~Young~Hong,
        and~Woon-Seng~Gan,~\IEEEmembership{Senior Member,~IEEE}%
\IEEEcompsocitemizethanks{%
    \IEEEcompsocthanksitem K.~Ooi, Z.-T.~Ong, B.~Lam, and W.-S. Gan are with the School of Electrical and Electronic Engineering, Nanyang Technological University (NTU), Singapore. Emails: \{wooi002, ztong, bhanlam, ewsgan\}@ntu.edu.sg.
    \IEEEcompsocthanksitem K.~N.~Watcharasupat was with the School of Electrical and Electronic Engineering, NTU, Singapore. She is currently with the Center for Music Technology, Georgia Institute of Technology, Atlanta, GA, USA. Email: kwatcharasupat@gatech.edu.
    \IEEEcompsocthanksitem J.~Y.~Hong is with the Department of Architectural Engineering, Chungnam National University, Daejeon, Republic of Korea. Email: jyhong@cnu.ac.kr.
    \IEEEcompsocthanksitem The research protocols used in this research were approved by the NTU Institutional Review Board (Ref. IRB-2020-08-035). 
    \IEEEcompsocthanksitem The ARAUS dataset (including rejected data) is publicly available in the NTU research data repository DR-NTU (Data) at \url{https://doi.org/10.21979/N9/9OTEVX}. Replication code for analysis and baseline models in this paper is available at \url{https://github.com/ntudsp/araus-dataset-baseline-models}.
    \IEEEcompsocthanksitem \textcopyright 2023 IEEE. Personal use of this material is permitted. Permission from IEEE must be obtained for all other uses, in any current or future media, including reprinting/republishing this material for advertising or promotional purposes, creating new collective works, for resale or redistribution to servers or lists, or reuse of any copyrighted component of this work in other works.
    }% <-this % stops an unwanted space
\thanks{Manuscript received Jul 3, 2022; revised Dec 22, 2022; accepted Feb 9, 2023.}
}

% The paper headers
\markboth{Submitted to IEEE Transactions on Affective Computing.}%
{Ooi \MakeLowercase{\textit{et al.}}: ARAUS: A Large-Scale Dataset and
    Baseline Models of Affective Responses to
    Augmented Urban Soundscapes}

\IEEEtitleabstractindextext{%
\begin{abstract}
Choosing optimal maskers for existing soundscapes to effect a desired perceptual change via soundscape augmentation is non-trivial due to extensive varieties of maskers and a dearth of benchmark datasets with which to compare and develop soundscape augmentation models. To address this problem, we make publicly available the ARAUS (Affective Responses to Augmented Urban Soundscapes) dataset, which comprises a five-fold cross-validation set and independent test set totaling 25,440 unique subjective perceptual responses to augmented soundscapes presented as audio-visual stimuli. Each augmented soundscape is made by digitally adding ``maskers'' (bird, water, wind, traffic, construction, or silence) to urban soundscape recordings at fixed soundscape-to-masker ratios. Responses were then collected by asking participants to rate how pleasant, annoying, eventful, uneventful, vibrant, monotonous, chaotic, calm, and appropriate each augmented soundscape was, in accordance with ISO/TS 12913-2:2018. Participants also provided relevant demographic information and completed standard psychological questionnaires. We perform exploratory and statistical analysis of the responses obtained to verify internal consistency and agreement with known results in the literature. Finally, we demonstrate the benchmarking capability of the dataset by training and comparing four baseline models for urban soundscape pleasantness: a low-parameter regression model, a high-parameter convolutional neural network, and two attention-based networks in the literature.
\end{abstract}
}

% make the title area
\maketitle

\IEEEpeerreviewmaketitle

\IEEEraisesectionheading{\section{Introduction}}
\label{sec:Introduction}
\IEEEPARstart{S}{oundscape augmentation} involves the addition of sounds, typically referred to as ``maskers'', to existing acoustic environments in an effort to change the perception that real or hypothetical listeners may have towards them. This perception-based approach to noise mitigation was developed upon findings that indicators based purely on the sound pressure level are inefficient or ineffective at influencing the perception of soundscapes \cite{Raimbault2003, Jennings2013, DePaivaVianna2015NoiseStudy, Kang2019}. For instance, based on a synthesis of studies in the literature, \cite{Miedema1998Exposure-responseNoise} found that at the same day-night level above \SI{52}{\decibel}, the percentage of people rating aircraft noise as ``highly annoying'' significantly exceeded that for road traffic noise, and the percentage for road traffic noise in turn significantly exceeded that for rail noise. This is further backed by a numerical study conducted in~\cite{DeCoensel2007ModelsPlanning}.

Changes in perception have ordinarily been measured using subjective ratings of descriptors such as ``soundscape quality'' \cite{Axelsson2014}, ``preference'' \cite{Abdalrahman2020a}, ``perceived loudness'' \cite{Hong2019b}, ``vibrancy'' \cite{Aletta2018TowardsApproach}, and other adjectives described in the ISO/TS 12913-3:2019 circumplex model of soundscape perception \cite{InternationalOrganizationforStandardization2019}. These ratings, when aggregated over multiple human participants or soundscapes, typically form a set of indicators that soundscape practitioners can use to guide interventions or analyses of a given soundscape.

However, a perennial concern lies in choosing ``optimal'' or ``appropriate'' maskers to optimize the value of a given perceptual indicator, given the large varieties of possible maskers. This is important in real-life applications of soundscape augmentation systems, such as an automatic masker system that plays back maskers in real time in time-varying urban acoustic environments \cite{VanRenterghem2020, Wong2022DeploymentAugmentation}. 

Most prior studies have primarily addressed this concern by expert-guided or post hoc analysis \cite{Leung2016, Hong2021b}, but the number of maskers and masker types under consideration at a time tends to be relatively small, on the order of tens at a time. This thus limits the choice of possible maskers for different scenarios and the generalizability of conclusions of soundscape augmentation studies to more diverse real-world scenarios. 

Hence, a dataset of subjective perceptual responses to a broad variety of augmented urban soundscapes, comprising a large set of urban soundscapes and a correspondingly diverse set of maskers, would pave the way to systematic investigations of suitable maskers and masker types for urban soundscapes in general, as well as the development of better models for soundscape augmentation. Such a dataset could also alternatively be used for soundscape studies desiring a representative sample of soundscapes from a perceptual space or with given characteristics, such as those performed by \cite{Aumond2017ModelingContext} for pleasantness, \cite{Puyana-Romero2019} for soundscape quality, and \cite{Aletta2020} for validating translations of perceptual attributes.

However, to the best of our knowledge, there is currently no common, consolidated benchmark dataset for fair cross-comparison of prediction models in the soundscape literature. According to a systematic review of prediction models by \cite{Lionello2020ASoundscapes}, the largest datasets used by prediction models for soundscape perception in the literature also appear to individually number from the thousands to about ten thousand samples. As such, this work aims to craft a large-scale, public dataset of affective responses to soundscapes which can serve as a benchmark for designing models to guide masker choice in soundscape augmentation. 

Our proposed \textbf{A}ffective \textbf{R}esponses to \textbf{A}ugmented \textbf{U}rban \textbf{S}oundscapes (ARAUS) strives to achieve this by using urban soundscapes covering a wide range of acoustic environments and maskers covering several sound types previously investigated in the soundscape literature. Additionally, the data collection protocol is compliant with the ISO/TS 12913-2:2018 \citep{InternationalOrganizationforStandardization2017} standard and is designed to be replicable with minimal specialized equipment, ensuring that this dataset can be extended by as many research groups as possible.

Due to the scope and nature of the proposed dataset, the collected data can also secondarily serve to empirically verify findings from existing soundscape literature. Additionally, the size of the dataset allows for relatively larger machine-learning models to be investigated along with traditional low-parameter models.

This article is organized as follows. \Cref{sec:Literature Review} gives an overview of related soundscape datasets in the literature. \Cref{sec:Data Collection Methodology} describes the stimuli generation and data collection method used. \Cref{sec:Analysis and Modeling Methodology} describes exploratory and confirmatory data analyses of the collected responses. \Cref{sec:Affective Response Modeling} presents benchmark models for prediction of perceptual responses to augmented soundscapes. \Cref{sec:Limitations and Future Work} discusses the limitations and future directions for this work. Finally, \Cref{sec:Conclusion} summarizes the findings of this work and makes some concluding remarks.
\section{Related Datasets}
\label{sec:Literature Review}

Publicly available, large-scale audio datasets have been made available as unlabeled data and/or labeled data for several ``objective'' acoustic tasks such as acoustic scene classification \cite{Wang2021Audio-visualSubmissions, Martin-Morato2021Low-complexitySystems}, sound event localization and detection \cite{Politis2021ADetection}, and anomaly detection \cite{Kawaguchi2021DescriptionConditions}. However, affective audio datasets tend to be smaller and rarely reach the scale of those developed for objective tasks. In particular, to the best of our knowledge, there is no publicly-available affective soundscape dataset at the scale of the proposed ARAUS dataset.

In this section, a selection of large-scale and affective audio datasets is reviewed. A brief overview of key details for the datasets discussed is shown in Table \ref{tab:audio_datasets}.

\subsection{Large-scale Audio Datasets}\label{sec:Related Work/Large-scale Audio Datasets}

The largest curated audio dataset is arguably AudioSet, which at the time of its initial release contained 1.7M 10-second segments of audio from YouTube videos organized in a systematic ontology for audio tagging \cite{Gemmeke2017AudioEvents}. Strong labels have subsequently been provided for a subset of audio \cite{Hershey2021TheClassification}, and at the time of writing, AudioSet has grown to about 2.1M segments, with 120K being strongly labeled \cite{Hershey2021AudioSet:2021}. However, the publicly available version of the data constituting AudioSet is in the form of \num{128}-dimensional VGGish features \cite{Hershey2017CNNClassification} due to potential copyright issues related to the use of raw audio from YouTube videos, which limits its use as a public dataset. 

Alternatives to AudioSet include UrbanSound8K \cite{Salamon2014AResearch}, ESC-50 \cite{Piczak2015ESC:Classification}, and FSD50K \cite{Fonseca2022FSD50K:Events}, which contain labelled Creative Commons-licensed tracks from Freesound \cite{Font2013} and serve as publicly-available benchmark datasets for weak audio classification. However, the nature of the stimuli in these datasets tends to be monophonic or same-class polyphonic. While this is useful in reducing the complexity and noise in inputs to train robust sound event classification models, the full complexities of real-life acoustic \textit{environments} necessary for soundscape research are rarely represented in these datasets.
Individual monophonic stimuli may be used to compose synthetic soundscapes like in the URBAN-SED dataset \cite{Salamon2017}, which used the tracks in UrbanSound8K, but the focus of UrbanSound8K on just 10 possible event classes may be insufficient to emulate the variety of sound sources possible in real-life urban environments.

Therefore, multiple efforts have been made to record real-life acoustic environments, which are generally polyphonic in nature, and provide them as publicly available datasets for use in sound and soundscape research. These include 
Urban Soundscapes of the World (USotW) \cite{DeCoensel2017UrbanMind};
EigenScape \cite{Ciufo2017};
SONYC Urban Sound Tagging (SONYC-UST) \cite{Cartwright2019a} and SONYC-UST-V2 \cite{Cartwright2020};
TUT Acoustic Scenes \citep{Mesaros2016TUTDetection, Mesaros2017DCASESystem} and TAU Urban Acoustic Scenes (TAU-UAS) \citep{Mesaros2018AClassification, Heittola2020AcousticSolutions, Wang2021}; Singapore Polyphonic Urban Audio (SINGA:PURA) \cite{Ooi2021AContext}; Ambisonics Recordings of Typical Environments (ARTE) \cite{Weisser2019TheDatabase}; 
and Sony-TAu Realistic Spatial Soundscapes 2022 (STARSS22)
\cite{Politis2022STARSS22:Events}. 

However, the aforementioned datasets contain no corresponding ``subjective'' labels concerning the affective perception of the recorded environments. This limits their use as datasets for soundscape augmentation, because knowing how individual soundscapes are \textit{perceived} by humans is crucial in analyzing and modeling their perception. In addition, with the exception of USotW, the recordings were not compliant with ISO/TS 12913-2:2018 \cite{InternationalOrganizationforStandardization2017} due to them preceding the publication of the standard, or being made for a different purpose. Hence, they may not be immediately suitable for use under the ISO 12913 paradigm, which the ARAUS dataset was designed under.

\subsection{Affective Sound Datasets}

Nonetheless, audio datasets with labels specific to perceptual indicators of the stimuli also exist. For example, the International Affective Digitized Sounds (IADS) dataset \cite{Bradley2007TheB-3.} has had labels for discrete emotional categories elicited by the \num{167} individual stimuli provided by \cite{Stevenson2008AffectiveCategories}, with an expanded version (IADS-E) provided by \cite{Yang2018}. The largely monophonic stimuli in IADS, however, suffer from the same drawbacks as datasets based on Freesound.
As a workaround, the Emo-Soundscapes dataset \cite{Fan2017} used individual clips from Freesound to synthetically generate \num{1213} soundscapes, each \SI{6}{\second} long, and obtained corresponding valence-arousal labels based on the Self-Assessment Manikin (SAM) \cite{Bradley1994} from participants on the CrowdFlower platform.

\begin{table*}[t]
	\centering
	\caption{Selection of datasets related to the ARAUS dataset. Datasets with multiple versions only have their latest or most complete version listed.
	}
	\label{tab:audio_datasets}
	\setlength{\tabcolsep}{3pt}
	\begin{tabularx}{\linewidth}{ l l c r r l l }
		\toprule
		Dataset && Year & \multicolumn{1}{r}{Samples} & Length & Locations/Sources & Labels \\
		\midrule
		AudioSet & \cite{Gemmeke2017AudioEvents} & 2017 & {~2.1M} & \SI{10}{\second} & From YouTube & Weak sound events \\
		AudioSet Strong & \citep{Hershey2021AudioSet:2021} & 2021 & {~120K} & \SI{10}{\second} & From AudioSet & Strong sound events \\
		\midrule
		UrbanSound8K & \cite{Salamon2014AResearch} & 2014 & 8.7K 
		& $\leq$\SI{4}{\second} & From Freesound & Weak sound events \\ 
		ESC-50 & \cite{Piczak2015ESC:Classification} & 2015 & 2.0K
		& \SI{5}{\second} & From Freesound & Weak sound events \\
		URBAN-SED & \cite{Salamon2017} & 2017 &
		10.0K 
		& \SI{10}{\second} & From UrbanSound8K & Strong sound events \\ 
		FSD50K & \cite{Fonseca2022FSD50K:Events} & 2020 & 
		51.2K
		& $\leq$\SI{30}{\second}
		& From Freesound & Weak sound events \\ 
		
		\midrule

		SONYC-UST-V2        & \cite{Cartwright2020} & 2020 &
		18.5K 
		& \SI{10}{\second} & Public locations in New York City & Weak sound events \\ 
		SINGA:PURA & \cite{Ooi2021AContext}& 2021 &
		6.5K 
		& \SI{10}{\second} & Public locations in Singapore & Strong sound events \\ 
		
		STARSS22 & \cite{Politis2022STARSS22:Events}& 2022 & 173 
		& $\leq$\SI{300}{\second}
		& Indoor environments in Tampere \& Tokyo & Strong sound events + direction of arrival \\

		\midrule

		USotW & \cite{DeCoensel2017UrbanMind} & 2017& 127 & \SI{60}{\second} & Urban public spaces in 9 cities worldwide & Acoustic scenes \\ 
		
		EigenScape & \citep{Ciufo2017} & 2017 & 64 & \SI{600}{\second} & Public locations in North England & Acoustic scenes\\
		
		ARTE & \cite{Weisser2019TheDatabase}& 2019 & 13 & \SI{120}{\second} & Indoor environments in Sydney & Acoustic scenes \\

        TAU-UAS & \citep{Heittola2020AcousticSolutions} & 2019 &
		23.0K & \SI{10}{\second} & Urban public spaces in 12 European cities & Acoustic scenes\\
		\midrule

		IADS-2 & \cite{Bradley2007TheB-3.} & 2007 & 167 & \SI{6}{\second} & From digital recordings & SAM \\ 
		IADS-E & \cite{Yang2018} & 2018 & 935 & \SI{6}{\second} & From IADS-2, Internet, or composer & SAM + basic emotion ratings \\
		Emo-Soundscapes & \cite{Fan2017} & 2017 &
		1.2K 
		& \SI{6}{\second} & From Freesound & SAM \\
		\midrule
		
		ATHUS & \cite{Giannakopoulos2019} & 2019 & 978 & $\leq$\SI{79}{\second}
		& Various locations in Athens & Subjective soundscape quality \\
		
		ISD & \cite{Mitchell2021} & 2021 & 1.2K
		& \SI{30}{\second} & Urban public spaces in 13 European cities & Affective + other labels per SSID Protocol \cite{Mitchell2020TheInformation} \\

		\midrule
		\multicolumn{2}{l}{ARAUS \hfill [Proposed]} & 2022 & 25.4K & \SI{30}{\second} & USotW, public locations in Singapore (base) & Affective responses   \\
		&&&&& Freesound, Xeno-canto (maskers)  & (+ contextual information, see \Cref{sec:Data Collection Methodology})\\
		\bottomrule
	\end{tabularx}
\end{table*}

In contrast, datasets with perceptual labels for real-life audio recordings include the Athens Urban Soundscape (ATHUS) dataset \cite{Giannakopoulos2019}, which contains \num{978} crowd-sourced recordings with corresponding labels for subjectively-rated soundscape quality on a five-point Likert scale, as well as the International Soundscape Database (ISD) \cite{Mitchell2021}, which as of v0.2.1, consists of \num{1258} \num{30}-second long recordings in \num{13} European cities and corresponding perceptual responses collected using the Soundscape Indices (SSID) Protocol \cite{Mitchell2020TheInformation}. The perceptual responses collected using the SSID Protocol were largely inspired by the Method A questionnaire in ISO/TS 12913-2:2018 \cite{InternationalOrganizationforStandardization2017}.

However, the relatively smaller sample sizes of these datasets, as compared to large-scale datasets like AudioSet, may preclude their use in developing high-parameter models, such as deep neural networks, which have shown state-of-the-art performance in various ``objective'' acoustic tasks. Hence, the ARAUS dataset was designed at a relatively large scale to be amenable to high-parameter models.
\section{Data Collection Methodology}
\label{sec:Data Collection Methodology}

Since the primary focus of this study is to create a database of affective responses to augmented soundscapes that is extensive yet extensible, the data collection methodology must necessarily be modular and repeatable to ensure valid analysis of results and enable possible future extensions to the dataset, similar to the philosophy of USotW \cite{DeCoensel2017UrbanMind} and ISD \cite{Mitchell2021}. 

Hence, the ARAUS dataset was designed as a five-fold cross-validation dataset with an independent test set, such that additional folds or data for each fold can be added following the same data collection methodology described in this section. Such a design also allows for each fold to be treated as an independent dataset for training of ensemble models or meta-learning.

\begin{figure}[t]
	\centering
\includegraphics[width=\linewidth]{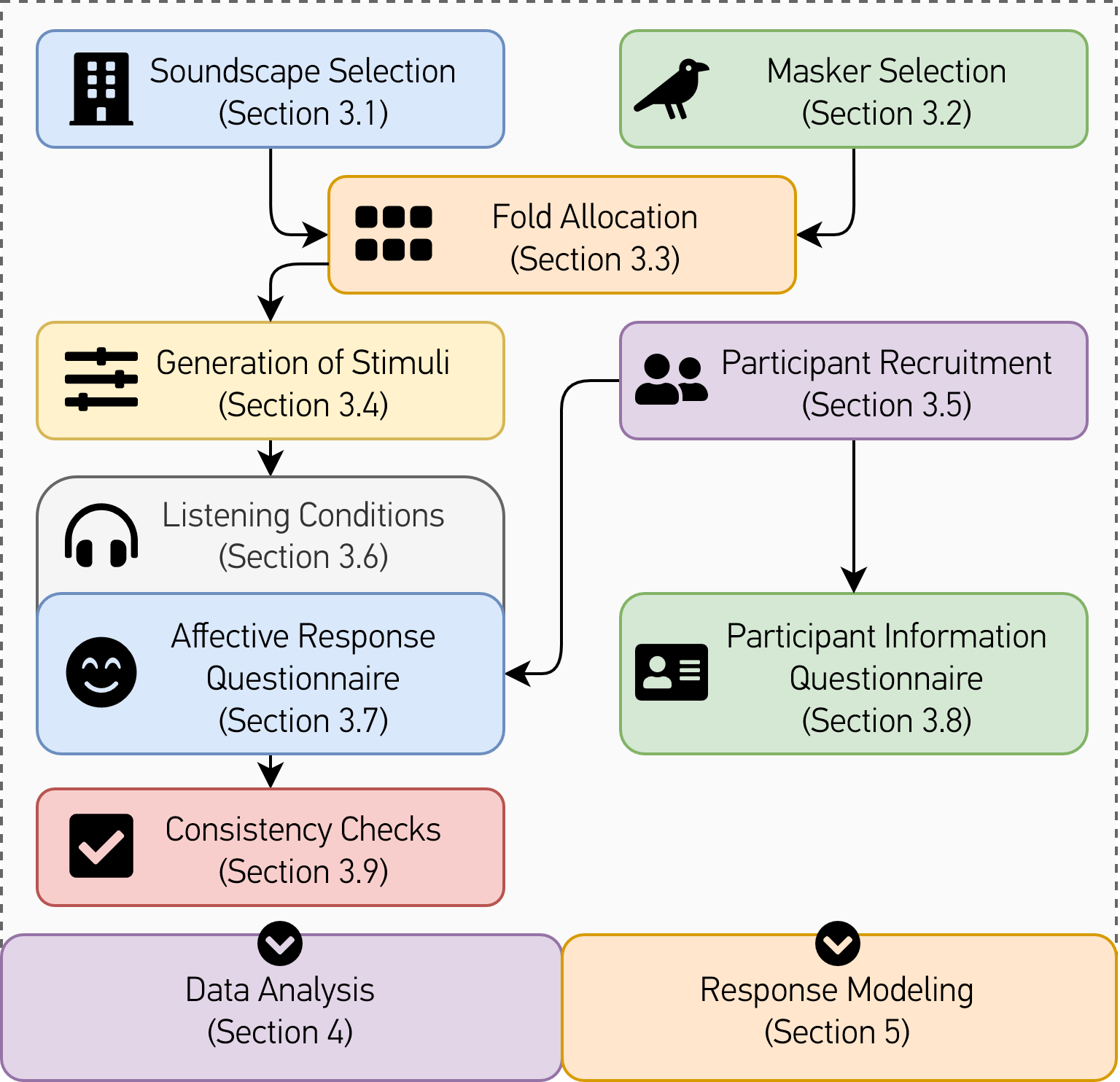}
	\caption{Framework of the study methodology.}
	\label{fig:framework}
\end{figure}

To design the cross-validation set, we prepared separate recordings of real-life ``base'' urban soundscapes and of maskers that could potentially be used to augment those ``base'' urban soundscapes. The same procedure was used to split the urban soundscape and masker recordings independently into five folds to ensure no data leakage, and combine them within their folds into augmented soundscapes that form the audio-visual stimuli in the ARAUS dataset. The audio-visual stimuli were presented to participants in laboratory conditions and their affective responses to the stimuli were collected. However, responses belonging to participants who responded in an ``overly inconsistent'' manner were dropped from the dataset (see \Cref{sec:Data Collection Methodology/Consistency Checks}) and new participants were recruited to replace the dropped responses such that each fold in the five-fold cross-validation set had an equal number of data samples. A summary of the data collection methodology is shown in \Cref{fig:framework}, and full details of each step are provided in the following subsections. 

The test set was designed in the same manner as the cross-validation set, but used no urban soundscapes, maskers, or participants already in the cross-validation set.

\subsection{Base Urban Soundscapes}\label{sec:Data Collection Methodology/Urban Soundscapes}

For the five-fold cross-validation set, all base urban soundscapes were taken from the Urban Soundscapes of the World (USotW) database \cite{DeCoensel2017UrbanMind}. The USotW database contains \num{127} publicly available recordings of urban soundscapes covering an extensive variety of urban environments from various cities around the world. The urban environments range from parks to busy streets, and the cities include those in Asia, Europe, and North America, which allows the ARAUS dataset to be broad-ranging in its coverage of real-life urban soundscapes. The urban soundscapes in the USotW dataset were also chosen by the USotW team via clustering of locations reported by local experts to be ``full of life and exciting'', ``chaotic and restless'', ``calm and tranquil'', and ``lifeless and boring'', which are adjectives spanning the perceptual space generated by the ``Pleasantness'' and ``Eventfulness'' axes of the ISO/TS 12913-3:2019 circumplex model \cite{InternationalOrganizationforStandardization2019}. This heightens the suitability of the USotW database for use in the investigation of the affective qualities of soundscapes that the ARAUS dataset aims to enable.

Each \num{60}-second binaural recording in the USotW database was split into two halves of \SI{30}{\second} for the creation of the audio-visual stimuli in the ARAUS dataset. Consequently, the audio-visual stimuli in the ARAUS dataset are all \SI{30}{\second} in length, which is in line with the stimulus length used in several soundscape studies, such as those in \cite{Hao2016, Leung2017, Aletta2020a}. Upon splitting the recordings into two halves, we discarded any half with (1)~audible electrical noise (such as those caused by a faulty microphone or loose connection), in order to reflect only accurately-captured real-life soundscapes; (2)~measured in-situ $L_\text{A,eq}$ values below \SI{52}{\decibel}, in order to ensure that reproduction levels were significantly above the noise floor of the laboratory location with the highest noise floor (about \SI{37}{\decibel}; see \Cref{fig:Rooms}) where the subjective responses were obtained; and/or (3)~measured in-situ $L_\text{A,eq}$ values above \SI{77}{\decibel} in order to ensure safe listening levels \cite{Lutman2000WhatAbove} for the participants.

For each half of the binaural recordings that was not discarded, a \SI{0}{\degree}-azimuth, \SI{0}{\degree}-elevation field of view (FoV) was cropped out of their corresponding \SI{360}{\degree}-videos from the USotW database to form the audio-visual stimulus corresponding to that urban soundscape.

For the test set, six urban soundscapes were recorded in locations within Nanyang Technological University (NTU), Singapore. The recordings were made in a similar manner as those in the USotW database and were made using equipment in accordance with the SSID Protocol \cite{Mitchell2020TheInformation}. Post-processing of the test set recordings for use as part of the stimuli for the ARAUS dataset was done in the same manner as that for the five-fold cross-validation set.

In total, this formed a base of \num{234} urban soundscape recordings for the five-fold cross-validation set of the ARAUS dataset, and an additional base of six soundscapes (not overlapping with the other \num{234}) for the independent test set. Further details on the exact recordings used can be found in Appendix A.

\subsection{Maskers}\label{sec:Data Collection Methodology/Maskers}

The masker recordings for both the five-fold cross-validation set and the independent test set were derived from source tracks found in the public databases Freesound \cite{Font2013} and Xeno-canto \cite{Planque2008Xeno-canto:Song}. Both databases host tracks with Creative Commons licenses, with Xeno-canto being a repository of bird calls and Freesound being a more general repository of sound samples and recordings.

The source tracks from both databases that we determined to be relevant to the ARAUS dataset fell into one of the following classes: bird, construction, traffic, water, and wind. Water \cite{Jeon2012, Galbrun2013AcousticalNoise, Coensel2015}, bird \cite{Hao2016, Coensel2015, Ferraroa}, and wind \cite{Hedblom2017EvaluationPreservation} sounds have previously been investigated in soundscape studies as natural-sound maskers. On the other hand, sounds from traffic and construction are ubiquitous noise sources in urban environments \cite{WorldHealthOrganizationRegionalOfficeforEurope2018}, and are commonly investigated in soundscape literature \citep{You2008, Pieretti2013a, Lu2020}. Therefore, the selection of maskers covers a variety of urban sounds investigated in the soundscape literature for the ARAUS dataset.

The source tracks for maskers corresponding to the ``bird'' class were first obtained by randomly picking a selection of high-quality tracks of birds on Xeno-canto. Each source track corresponded to bird(s) from a single species, as labeled on Xeno-canto. Additional tracks for the ``bird'' class and all other classes were obtained via the corresponding search term on Freesound, and picking a selection of ``high-quality'' tracks containing \num{30}-second sections of sound that corresponded only to that particular masker class, as determined by manual listening. However, the exact number of sources of a single masker class present in a given track was variable, so for instance, a given track of the ``bird'' class could contain vocalizations from one, two, or more birds, but all tracks of the ``bird'' class contain \textit{only} bird vocalizations. Further details on the source tracks used to create the maskers are given in Appendix A.

Each source track was then processed individually to create 30-second single-channel masker recordings. Single-channel recordings of maskers were used because \cite{Xu2019a} previously found single-channel recordings to be sufficient to replicate the perceived affective quality of soundscapes, which the ARAUS dataset aims to collect responses for. For source tracks that were originally multi-channel, only the first channel was used for consistency.
Source tracks originally longer than \SI{30}{\second} were trimmed to \SI{30}{\second}, while those originally shorter than \SI{30}{\second} were either padded with silence or looped. Finally, noise reduction via spectral gating and high-pass filtering was performed for source tracks in the ``bird'' class to reduce ambient and/or microphone noise in the track. All pre-processing was done manually using Audacity (v2.3.2).

In total, this formed a set of \num{280} masker candidates (\num{56} per fold) that were used to generate the stimuli for the five-fold cross-validation set, and a set of seven maskers for the independent test set. The breakdown of the number of masker recordings by class was \num{80} bird, \num{40} construction, \num{40} traffic, \num{80} water, and \num{40} wind for the cross-validation set; and \num{2} bird, \num{1} construction, \num{1} traffic, \num{1} water, \num{2} wind for the independent test set. 

\subsection{Fold Allocation}\label{sec:Data Collection Methodology/Fold Allocation}

After preparing the urban soundscape recordings and maskers in \Cref{sec:Data Collection Methodology/Urban Soundscapes,sec:Data Collection Methodology/Maskers}, the tracks were assigned into the five folds of the cross-validation set such that the distributions of psychoacoustic properties of the urban soundscapes and maskers were similar across the five folds. Since psychoacoustic indicators of a given soundscape have non-trivial and non-spurious correlations with corresponding perceptual indicators \cite{Engel2021}, taking psychoacoustic indicators into account was necessary to minimize distributional shifts across the folds.

The assignment procedure consisted of the following steps carried out for the urban soundscape recordings, and independently each class of masker tracks.

\subsubsection{Track Calibration} 
Each recording track was calibrated to a pre-defined A-weighted equivalent sound pressure level ($L_\text{A,eq}$). For the base urban soundscapes, the in-situ $L_\text{A,eq}$ measured at the time of recording for the urban soundscape recordings was used, while a constant value of \SI{65}{\decibel} was used for the maskers, similar to \cite{Hong2020EffectsQuality}.

\subsubsection{Acoustic and Psychoacoustic Indicator Computation}\label{sec:Data Collection Methodology/Fold Allocation/Acoustic and Psychoacoustic Indicator Computation}

For all recordings, summary statistics for acoustic and psychoacoustic indicators, as recommended by ISO/TS 12913-3:2019 \cite{InternationalOrganizationforStandardization2019}, were calculated independently for each channel using ArtemiS SUITE (HEAD Acoustics). The indicators comprised sharpness \cite{DIN456922009}, loudness \cite{ISO532-12014}, fluctuation strength \cite{Fastl2001}, roughness \cite{Fastl2001}, tonality \cite{EcmaInternational2020ECMA-418-2:2020Perception, EcmaInternational2021ECMA-74Equipment}, $L_{\text{A,eq}}$ \citep{InternationalOrganizationforStandardization2016}, and C-weighted equivalent sound pressure level ($L_{\text{C,eq}}$) \citep{InternationalOrganizationforStandardization2016}. Finally, band powers summed over third-octave bands with center frequencies from \SI{5}{\hertz} to \SI{20}{\kilo\hertz} were also calculated. \Cref{tab:sumstat} shows the summary statistics calculated for each indicator. 

With the exception of tonality in \citep{EcmaInternational2020ECMA-418-2:2020Perception}, the MATLAB Audio Toolbox\footnote{https://www.mathworks.com/products/audio.html} provides standards-compliant implementations of all other psychoacoustic indicators used. Open-source implementations of the psychoacoustic indicators used are either currently available, or have been indicated as planned, as part of the MOSQITO Toolbox \citep{SanMillan-Castillo2021MOSQITO:Education}. 

\begin{table}[t]
	\centering
	\caption{Acoustic and psychoacoustic indicators used as channel-wise summary statistics. The set of summary statistics indicated as ``common'' were the mean, maximum, exceedance levels for the 5th percentile, exceedance levels for each decile, and exceedance levels for the 95th percentile. Minimum values for sharpness, loudness, fluctuation strength, roughness, and tonality were zero for all stimuli and hence omitted from analysis.}
	\label{tab:sumstat}
	\setlength{\tabcolsep}{3pt}
	\begin{tabularx}{\linewidth}{llX}
		\toprule
		Indicator                               & Unit & Summary statistics \\
		\midrule
		Sharpness \citep{DIN456922009}          & acum            & common \\
		Loudness \citep{ISO532-12014}           & sone            & common + root mean cube \\
		Fluctuation strength \citep{Fastl2001}  & vacil           & common \\
		Roughness \citep{Fastl2001}             & asper           & common \\
		Tonality \cite{EcmaInternational2020ECMA-418-2:2020Perception, EcmaInternational2021ECMA-74Equipment}                 & tuHMS  & common \\ % 
		$L_{\text{A,eq}}$
        \citep{InternationalOrganizationforStandardization2016}                   & dB & common + minimum \\
		$L_{\text{C,eq}}$
\citep{InternationalOrganizationforStandardization2016}                   & dB & common + minimum \\
		Spectral powers                   & dB & third-octave band-wise sum\newline (center freq. \SI{5}{\hertz} to \SI{20}{\kilo\hertz})\\
		\bottomrule
	\end{tabularx}
\end{table}
	
\subsubsection{Dimensionality Reduction}
The summary statistics were then used as individual input features to a principal component analysis (PCA), and to project each recording to a principal component space with enough dimensions to achieve \SI{90}{\percent} explained variance. Further details on the PCA can be found in Appendix D.

The primary reason for using PCA in the assignment of folds was to remove correlations between multiple variables, which is desirable for noise and dimensionality reduction prior to clustering as illustrated by \cite{Flowers2021LookingClustering}, which removed correlated variables when clustering of soundscape recordings across 8 acoustic indices. 

\subsubsection{Clustering and Fold Assignment}
With the coordinates of each recording in the principal component space, the recordings were organized into clusters of five using a self-organizing map (SOM) \cite{Kohonen2001}. 

For each cluster of five recordings, each recording in the cluster was randomly assigned to a distinct fold of the cross-validation set. To prevent data leakage, the assignment was done based on psychoacoustic indicators computed from their \textit{original} \num{60}-second long binaural recordings from the USotW database, such that the \num{30}-second halves in the ARAUS dataset originating from the same original binaural recording were always assigned to the same fold.

\subsection{Generation of Stimuli}\label{sec:Data Collection Methodology/Generation of Stimuli}

\begin{figure}[t]
	\centering
	\includegraphics[width=\columnwidth]{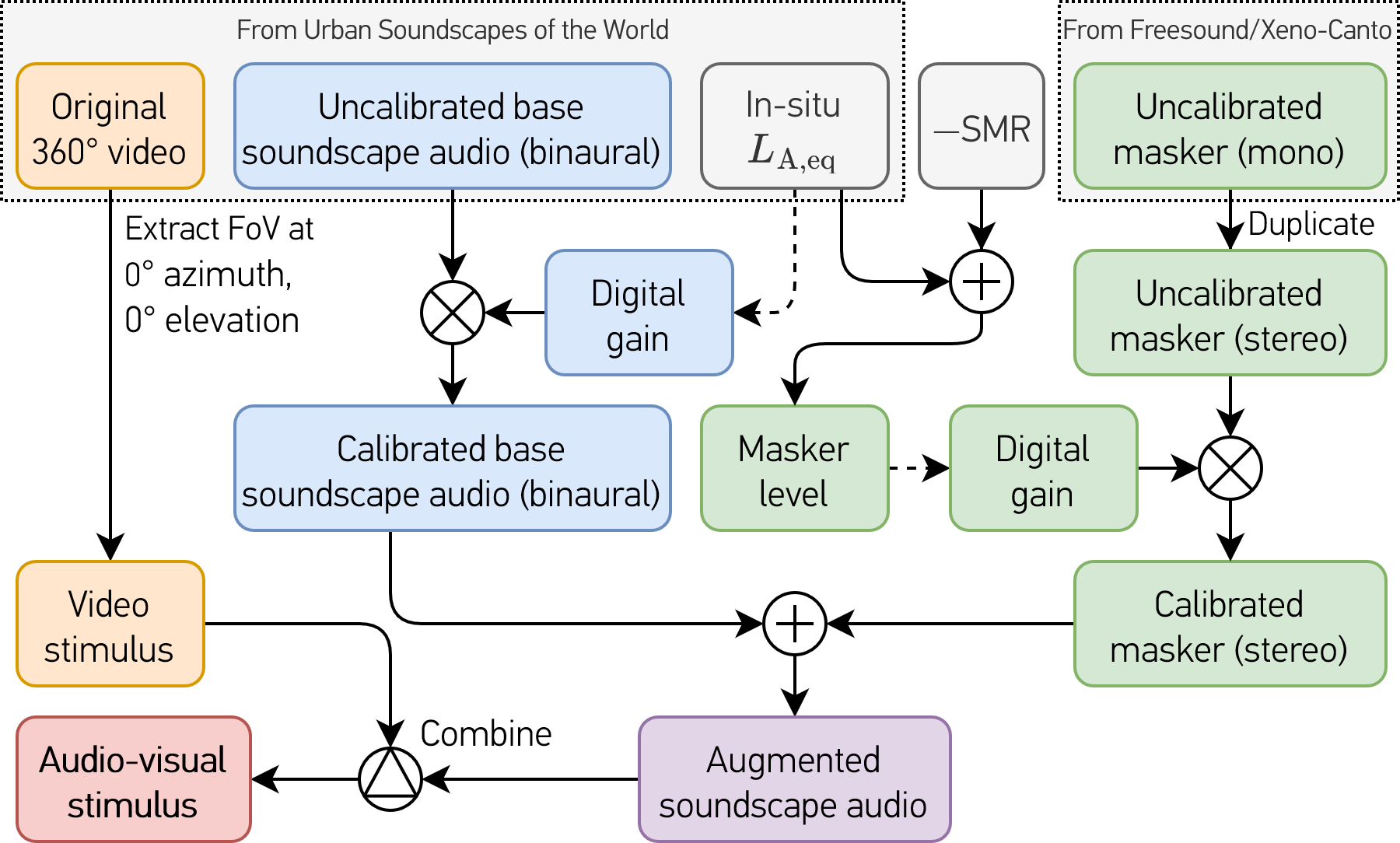}
	\caption{Illustration of stimulus generation procedure for a single stimulus. 
	}
	\label{fig:stimuli_generation}
\end{figure}

Each stimulus in the ARAUS dataset is an augmented soundscape to be presented as a \num{30}-second audio-visual stimulus to a human participant, and the procedure used to generate each stimulus is shown in \Cref{fig:stimuli_generation}.

The audio for the augmented soundscapes was made by combining the \num{30}-second binaural recordings of urban soundscapes in \Cref{sec:Data Collection Methodology/Urban Soundscapes} with the \num{30}-second single-channel recordings of maskers obtained in \Cref{sec:Data Collection Methodology/Maskers} at various gain levels, via element-wise addition of their respective gain-adjusted time-domain signals. For the purposes of stimulus generation, we also included silence in the set of possible maskers, since the addition of silence simply replicates the condition where no masker is added.

For the five-fold cross-validation set, the urban soundscapes and maskers were chosen randomly from the same fold for combination to prevent data leakage and provide a sample of all possible combinations. Since the fold allocation procedure described in \Cref{sec:Data Collection Methodology/Fold Allocation} ensured that the sets of urban soundscapes and maskers used for each fold of the cross-validation set were disjoint, the augmented soundscapes generated from them were disjoint between each fold of the cross-validation set as well. On the other hand, for the independent test set, the urban soundscapes and maskers were exhaustively combined to cover all \num{48} possible combinations. 

Before combining the urban soundscape and masker recordings for the five-fold cross-validation set, the urban soundscape recordings were calibrated to the in-situ $L_\text{A,eq}$ levels measured at the time of recording, and the maskers were calibrated to specific soundscape-to-masker ratios (SMR) with respect to the urban soundscape, chosen randomly from the set $\{$\num{-6}, \num{-3}, \num{0}, \num[retain-explicit-plus]{+3}, \num[retain-explicit-plus]{+6}$\}$ \SI{}{\decibel}. For example, if the urban soundscape recording had an in-situ $L_\text{A,eq}$ of \SI{65}{\decibel} and an SMR of \SI[retain-explicit-plus]{+3}{\decibel} was randomly chosen, then the masker would be calibrated to an $L_\text{A,eq}$ of \SI{62}{\decibel}. This gave a total of \num{65520} possible augmented soundscapes from which we sampled for the ARAUS dataset. All calibration was done via the automated method described by \cite{Ooi2021AutomationHead}. On the other hand, for the independent test set, a fixed SMR of \SI{0}{\decibel} was used for all combinations. This was to limit the number of stimuli presented to each participant in the test set and the effect of listener fatigue, since the participants assigned to the test set were required to rate all combinations exhaustively, as explained in  \Cref{sec:Data Collection Methodology/Questionnaires/Affective Response Questionnaire}.

Lastly, the single-channel recordings were added to each channel of the binaural tracks, in a manner similar to that in \cite{Abeer2021USM-SEDScenarios} with a fixed stereo panning coefficient of 0.5. This audio was then overlaid onto the \SI{0}{\degree}-azimuth, \SI{0}{\degree}-elevation field of view cropped from the \SI{360}{\degree} video recordings from the USotW database taken at the same time as the binaural urban soundscape recordings to form the audio-visual stimuli.

\subsection{Participant Recruitment}\label{sec:Data Collection Methodology/Participant Recruitment}

Prior ethical approval was obtained from the Institutional Review Board, NTU (Ref. IRB 2020-08-035) before participant recruitment and response collection. Participants were recruited via online messaging channels, posters, and emails.

In total, \num{642} unique participants were recruited to provide their responses for the ARAUS dataset, of which \num{37} (\SI{5.76}{\percent}) had their responses rejected. Hence, responses from only \num{605} participants were included in the final dataset. We rejected responses from participants who (1)~failed a hearing test (\num{19} participants, \SI{2.96}{\percent}); (2)~failed more than three out of seven consistency checks described in \Cref{sec:Data Collection Methodology/Consistency Checks} (\num{17} participants, \SI{2.65}{\percent}); or (3)~provided the same responses to any item in the Affective Response Questionnaire (ARQ) described in \Cref{sec:Data Collection Methodology/Questionnaires/Affective Response Questionnaire} for all stimuli they were presented, thereby providing no useful information to the dataset (\num{3} participants, \SI{0.47}{\percent}). Two of the rejected participants failed both the hearing test \textit{and} more than three out of seven consistency checks.

Participants aged under \num{30} were considered to have failed the hearing test if they had a mean threshold of hearing above \SI{20}{\decibel} and participants aged \num{30} and above were considered to have failed if they had a mean threshold of hearing above \SI{30}{\decibel} via pure-tone audiometry using the uHear application on a mobile phone (Apple iPhone 4S) and earbuds (Apple EarPods). The tested frequencies were \num{0.5}, \num{1}, \num{2}, \num{4}, and \SI{6}{\kilo\hertz}. These were within the standard ranges used for screening in pure tone audiometry \cite{Walker2013AudiometryInterpretation} and previous soundscape research \cite{EchevarriaSanchez2017}. The higher threshold of \SI{30}{\decibel} was applied to participants aged \num{30} and above to balance the risk of age bias (since age is highly correlated with hearing ability \cite{Wang2021ExtendedGroups}) against the need to ensure that hearing loss did not interfere with the perception of the augmented soundscapes in the ARAUS dataset.

The participants were each assigned a fold such that each fold of the five-fold cross-validation set had responses from \num{120} participants. The independent test set had responses from \num{5} participants. The age and gender distributions of the participants are shown in Table \ref{tab:age_distribution}, and further information on participant demographics can be found in Appendix C. Due to the test sites for the data collection process being located within university campuses (as shown in \Cref{fig:Rooms}), a majority of the 605 participants whose responses were included in the ARAUS dataset were students (\num{443} participants, \SI{73.2}{\percent}) who were in the process of obtaining their bachelor's degree (\num{380} participants, \SI{62.8}{\percent}). Participants were also relatively young (mean age \num{26.7} years, standard deviation \num{10.0} years) compared to those in the ISD (mean age \num{33.8} years, standard deviation \num{14.57} years) \cite{Mitchell2021b}, but slightly older compared to those in the IADS-2 (college students) \cite{Bradley2007TheB-3.} and IADS-E (mean age 21.32 years, standard deviation 2.38 years) \cite{Yang2018} datasets.

\begin{table}[t]
	\centering
	\caption{
		Age and gender\protect\footnotemark{} distribution of participants\\in ARAUS dataset by fold and in entirety.
		}
	\label{tab:age_distribution}
	\def\arraystretch{1.0}
	\begin{tabularx}{\linewidth}{Xccccccc}
		\toprule
		& \multicolumn{6}{c}{Fold} & \\
		\cmidrule{2-7}
		Statistic        & Test &    1 &    2 &    3 &    4 &    5 &  All \\
		\midrule
		Sample size      &    5 &  120 &  120 &  120 &  120 &  120 &  605 \\
		\# female        &    4 &   62 &   56 &   69 &   68 &   69 &  328 \\   
		\# male          &    1 &   58 &   64 &   51 &   52 &   51 &  277 \\
		\midrule
		Mean age         & 22.4 & 27.4 & 26.7 & 26.3 & 27.3 & 26.3 & 26.7 \\
		Std. dev. of age &  5.5 & 10.2 &  9.8 & 10.4 & 11.0 &  8.7 & 10.0 \\
		Minimum age      &   18 &   19 &   18 &   18 &   18 &   18 &   18 \\
		Median age       &   21 &   24 &   24 &   23 &   23 &   24 &   24 \\
		Maximum age      &   32 &   63 &   71 &   68 &   65 &   60 &   71 \\
		\bottomrule
	\end{tabularx}
\end{table}

\footnotetext{To safeguard their identities, the small number (2 participants, \SI{0.33}{\percent})
% -- Kenneth, please check 
of participants identifying as neither male nor female were randomly assigned as ``male'' or ``female'' in the public release of the ARAUS dataset, each with a probability of \SI{50}{\percent}.}

\subsection{Listening Conditions}\label{sec:Data Collection Methodology/Listening Conditions}

Participants were presented with all audio-visual stimuli via closed-back headphones (Beyerdynamic Custom One Pro) powered by an external sound card (SoundBlaster E5). The video was presented via a \num{23}-inch monitor (Philips 236E SoftBlue) and measured \SI{21.5}{\centi\meter} by \SI{12}{\centi\meter} on the screen. Participants sat facing about \SI{1}{\meter} from the monitor.

\begin{figure}[t]
	\centering
	\includegraphics[width=0.49\columnwidth]{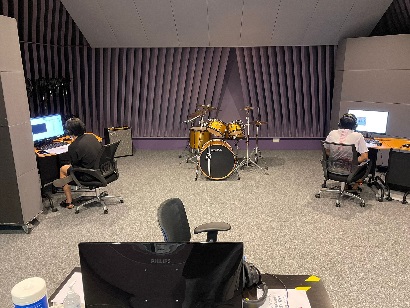} 
	\includegraphics[width=0.49\columnwidth]{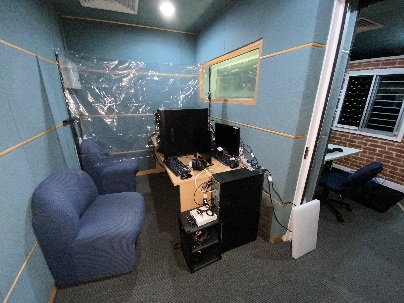} 
	\includegraphics[width=0.49\columnwidth]{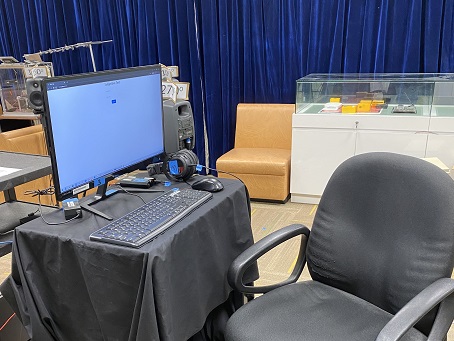} 
	\includegraphics[width=0.49\columnwidth]{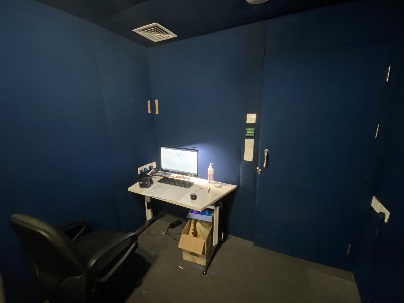} 
	\caption{Test sites at (top left) Academic Media Studio, SUTD, (top right) Media Technology Laboratory, NTU, (bottom left) Demo Room, NTU, (bottom right) Interactive Soundscape Room, NTU. Their noise floors, measured as $L_{\text{A,eq,3-min}}$ values with a B\&K Sound Level Meter Type 2240, were \SI{20.6}{\decibel}, \SI{26.0}{\decibel}, \SI{36.9}{\decibel}, and \SI{30.2}{\decibel}, respectively.}
	\label{fig:Rooms}
\end{figure}

Due to the large number of participants involved in the experiment, participants listened to the stimuli in one of four quiet rooms, three located in NTU and one in the Singapore University of Technology and Design (SUTD). Photos of each of the four quiet rooms are shown in \Cref{fig:Rooms}.

\subsection{Affective Response Questionnaire}\label{sec:Data Collection Methodology/Questionnaires/Affective Response Questionnaire}

For each audio-visual stimulus, participants were instructed to perform their evaluations given the following prompt:
\begin{quote}
	Imagine that you are standing at the location shown in the video, listening to the sound environment playing through the headphones.
\end{quote}
After completely experiencing the stimulus at least once, participants were presented with a set of \num{9} questions from the Method A questionnaire in ISO/TS 12913-2:2018 \cite{InternationalOrganizationforStandardization2017}, which we refer to as the ``Affective Response Questionnaire'' (ARQ). The first \num{8} questions were related to the perceived affective quality of the sound they heard over the headphones:
\begin{quote}
	To what extent do you agree or disagree that the present surrounding sound environment is \{pleasant, eventful, chaotic, vibrant, uneventful, calm, annoying, monotonous\}?
\end{quote}
The last question was related to the appropriateness of the location depicted in the video they saw on the monitor with respect to the sound they heard over the headphones:
\begin{quote}
	To what extent is the present surrounding sound environment appropriate to the pleasant place?
\end{quote}

\begin{figure}[t]
	\centering
	\includegraphics[width=\columnwidth]{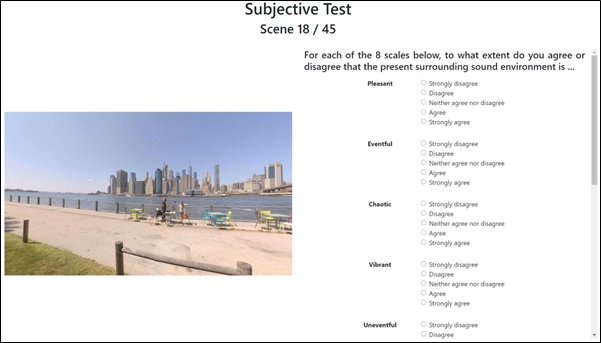}
	\caption{GUI used to administer the ARQ for the ARAUS dataset.}
	\label{fig:gui}
\end{figure}

Participants responded to all questions on a five-point Likert scale via a computerized graphical user interface (GUI) depicted in \Cref{fig:gui}, and we coded their responses to values in $\{1,2,3,4,5\}$ to match the scale.

For the cross-validation set, participants responded to the ARQ for \num{42} unique, randomly-selected stimuli from the fold that they were assigned to. For the independent test set, all participants responded to the ARQ for the same 48 exhaustively-generated stimuli. 

In addition to the ``main'' \num{42} (\num{48}) stimuli shown to each participant in the cross-validation (test) set, we presented each participant with three auxiliary stimuli. Participants were also required to provide responses to the ARQ for these stimuli, but the responses are not part of the ARAUS dataset, because these stimuli were the same for every participant, and did not serve the same purpose as the main stimuli in the ARAUS dataset. These stimuli were namely:
\begin{enumerate}[label=(\arabic*)]
	\item A ``pre-experiment'' stimulus, which was shown as the first stimulus \textit{before} presenting the main stimuli for the cross-validation/test set.
	\item An ``attention'' stimulus, which was identical to the ``pre-experiment'' stimulus and shown \textit{in between} two randomly selected ``main'' stimuli for the cross-validation/test set. For this stimulus, the GUI had special instructions for the participant to choose specific options for the ARQ before they could proceed.
	\item A ``post-experiment'' stimulus, which was identical to the ``pre-experiment'' stimulus and shown as the last stimulus \textit{after} presenting the main stimuli for the cross-validation/test set. ARQ responses to this stimulus were used for internal consistency checks, as described in \Cref{sec:Data Collection Methodology/Consistency Checks}.
\end{enumerate}

\subsection{Participant Information Questionnaire}\label{sec:Data Collection Methodology/Questionnaires/Participant Information Questionnaire} 

Since the ISO 12913-1:2014 definition of ``soundscape'' mandates that acoustic environments must be perceived \textit{in context} \cite{InternationalOrganizationforStandardization2014}, information specific to the listeners who participated in the study was also necessary to understand the context behind listener perceptions of the augmented soundscapes presented as stimuli. Hence, we administered another questionnaire, which we dub the ``Participant Information Questionnaire'' (PIQ), to each participant after they had completed the ARQs for the stimuli shown to them. The PIQ consisted of items related to basic demographic information and standard psychological questionnaires. 

Items related to basic demographic information consisted of age, gender, spoken languages, citizenship status, education status, occupational status, dwelling type, ethnicity, and length of residence of the participant.

The psychological questionnaires administered were
(1)~a shortened, \num{10}-item version of the Weinstein Noise Sensitivity Scale (WNSS-10) \cite{Weinstein1978};
(2)~a shortened, \num{10}-item version of the Perceived Stress Scale (PSS-10) \cite {Cohen1983AStress};
(3)~the WHO-5 Well-Being Index \cite{WorldHealthOrganization1998WHO-5Index}; and
(4)~the Positive and Negative Affect Schedule (PANAS) \cite{Watson1988}.

The relevance of the questionnaires is evident in the fact that noise sensitivity \cite{Aletta2018} and stress level \cite{Ratcliffe2021} play a significant role in the affective perception of soundscapes, the WHO-5 Well-Being Index is used in the Soundscape Indices Protocol (SSID) \cite{Mitchell2020TheInformation}, and \cite{Masullo2021a} previously used PANAS as a measure of participants' mood in a study of the emotional salience of sounds in soundscapes. The PSS-10 has been validated by the same authors of the original 14-item questionnaire \cite{Cohen1988PerceivedStates}, but the particular version (WNSS-10) of the Weinstein Noise Sensitivity Scale used for the ARAUS dataset has not previously been validated in the literature. Nonetheless, the internal reliability of WNSS-10 is affirmed based on the results of the analysis obtained in \Cref{sec:Analysis and Modeling Methodology/Questionnaires/Participant Information Questionnaire}. 

Full details of the PIQ with individual questions, options, and response coding for each section of the questionnaire can be found in Appendix B.

\subsection{Consistency Checks}\label{sec:Data Collection Methodology/Consistency Checks}

In order to ensure a baseline level of data quality in the responses to the ARQ, seven consistency checks were performed on each participant's responses. The consistency checks were designed as single-value metrics, and a participant was considered to have failed a consistency check if the corresponding value of the metric was at least \num{1}. If a participant failed more than three out of seven consistency checks, their responses were dropped from the dataset.

\newif\ifconsistency
\consistencytrue

To describe the single-value metrics, we first define $r_{\text{pl}}, r_{\text{ev}}, r_{\text{ch}}, r_{\text{vi}}, r_{\text{ue}}, r_{\text{ca}}, r_{\text{an}}, r_{\text{mo}}, r_{\text{ap}} \in \{1,2,3,4,5\}$ as the responses to the ARQ in \Cref{sec:Data Collection Methodology/Questionnaires/Affective Response Questionnaire} regarding the extent to which the sound environment in a given stimulus was respectively pleasant, eventful, chaotic, vibrant, uneventful, calm, annoying, monotonous, and appropriate. 

The seven consistency checks consist of three types of checks. The first check measures the mean absolute difference (MAD) between ARQ responses on the ``pre-experiment'' and ``post-experiment'' stimuli. Since the ``pre-experiment'' and ``post-experiment'' stimuli were identical, a perfectly consistent participant would have provided the same responses to both presentations.

Since the affective descriptors ``pleasant'' and ``annoying'' are on opposite axes of the ISO/TS 12913-3:2019 circumplex model of soundscape perception \cite{InternationalOrganizationforStandardization2019}, a perfectly consistent participant would have provided the same response for ``pleasant''  and ``annoying'' if the Likert coding of either were to be reversed. The same considerations apply to the other three pairs of opposite attributes on the circumplex model. Therefore, the next four checks consider the MADs between the four pairs of $r_{p}$ and $(6-r_{q})$, where $(p,q)$ is a pair of opposite descriptors.

Lastly, the mean squared error (MSE) between $r_{\text{pl}}$ and $(3+2P)$, and that between $r_{\text{ev}}$ and $(3+2E)$ across all stimuli presented was computed, where 
\begin{align}
	P &= k^{-1}\left(\sqrt{2}r_{\text{pl}}-\sqrt{2}r_{\text{an}}+r_{\text{ca}}-r_{\text{ch}}+r_{\text{vi}}-r_{\text{mo}}\right), \label{eq:ISOPl}\\
    E &= k^{-1}\left(\sqrt{2}r_{\text{ev}}-\sqrt{2}r_{\text{ue}}-r_{\text{ca}}+r_{\text{ch}}+r_{\text{vi}}-r_{\text{mo}}\right), \label{eq:ISOEv}
\end{align}
respectively are the normalized values of ``ISO Pleasantness'' and ``ISO Eventfulness'' as suggested in \cite{Mitchell2022}, and $k=8+\sqrt{32}$ is a normalization constant such that $P,E\in[-1,1]$. 
Since the affective descriptor ``pleasant'' theoretically parallels the principal axis of ``Pleasantness'' in the ISO/TS 12913-3:2019 circumplex model of soundscape perception, a perfectly consistent participant would have provided responses matching the magnitude in both directions. The rescaling of $P$ as $(3+2P)$ was necessary to match the range of values of $r_{\text{pl}}$ for a valid comparison. Similar considerations apply for the affective descriptor ``eventful'' to the principal axis ``Eventfulness''.
\section{Data Analysis}
\label{sec:Analysis and Modeling Methodology}

After the data collection described in \Cref{sec:Data Collection Methodology} had been completed, we performed analyses to verify data quality and empirical consistency with known literature.

Firstly, the ARQ responses were analyzed to compare the effect of different maskers on the normalized ISO Pleasantness $P$ of the augmented soundscapes, and validate that the stimuli in the ARAUS dataset spanned the perceptual space generated by the ISO Pleasantness and ISO Eventfulness axes. Statistical and internal reliability tests were then performed on the PIQ responses and consistency check metrics to ensure that the distribution of data and responses in each fold of the cross-validation set did not significantly differ. This allowed us to assess the degree to which the methodological efforts to minimize domain shift between folds were successful.

\subsection{Affective Response Questionnaire}\label{sec:Analysis and Modeling Methodology/Questionnaires/Affective Response Questionnaire}

To investigate the effect each masker had on the ISO Pleasantness of the urban soundscapes it was augmented to, we first used the ARQ responses to compute the ISO Pleasantness values for all the stimuli in the ARAUS dataset using Equation \eqref{eq:ISOPl}. The difference in ISO Pleasantness for each base urban soundscape (i.e., augmented with silence) and the same urban soundscape augmented with each masker was computed. These differences were averaged across all soundscapes for which the same masker, regardless of the SMR, was presented to obtain the mean change in ISO Pleasantness effected by each masker across different soundscapes. This allowed us to determine which maskers were optimal for augmentation, at least on average in a naive sense across the urban soundscapes in the ARAUS dataset.

\begin{figure*}[!ht]
	\centering
	\includegraphics[width=\linewidth]{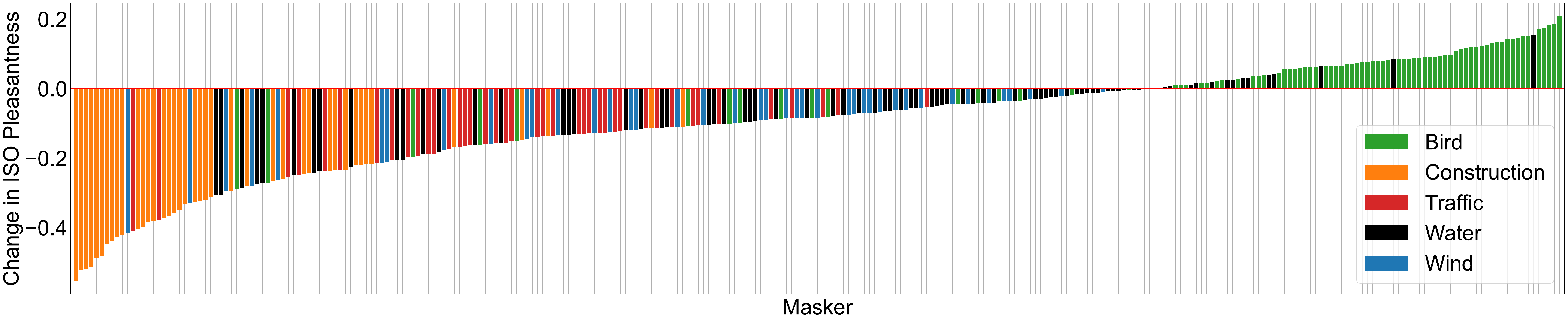}
	\caption{Mean change in ISO Pleasantness value as a function of each of the 287 (280 cross-validation, 7 test set) maskers used to augment soundscapes in the ARAUS dataset, aggregated over soundscapes and SMRs used.}
	\label{fig:isopl_delta}
\end{figure*}

Figure \ref{fig:isopl_delta} shows the mean change in ISO Pleasantness value as a function of masker used to augment soundscape, aggregated over soundscapes and SMRs used. Out of the \num{287} maskers in the ARAUS dataset, mean positive changes in ISO Pleasantness were only observed in maskers belonging to the bird and water classes. However, \textit{not all} of the bird and water maskers showed mean positive changes, corroborating the findings by \cite{Ratcliffe2021} that not all natural sounds are necessarily perceived as pleasant. Nevertheless, augmentation with \num{64} (\SI{78.0}{\percent}) of the bird maskers and \num{16} (\SI{19.5}{\percent}) of the water maskers, each out of \num{82}, resulted in effective mean positive changes in ISO Pleasantness, which supports findings in \cite{Leung2016,Hao2016,Hong2020EffectsQuality} where specific bird and water sounds were found to have improved the perceived pleasantness of urban soundscapes.

In contrast, mean negative changes in ISO Pleasantness were observed for \textit{all} wind, traffic, and construction maskers in the ARAUS dataset. While the decrease in ISO Pleasantness due to the addition of traffic \cite{Guski2017} and construction \cite{Hong2020NoiseAnnoyance} maskers is expected, the decrease in ISO Pleasantness for all wind sounds used as maskers was contrary to the results in narrative interviews reported by \cite{Ratcliffe2021} that people tended to perceive wind rustling through trees as pleasant. This could be due to the range of SMRs used for the ARAUS dataset, which ranged only from \num{-6} to \SI[retain-explicit-plus]{+6}{\decibel}. The mean in-situ $L_{A,eq}$ of the tracks in the USotW database used for the ARAUS dataset was about \SI{65}{\decibel}, 
but natural wind sounds in real-life environments tend to have a mean $L_{A,eq}$ of about \SI{55}{\decibel} or less \cite{VanRenterghem2002EffectWind}, which means that an SMR of \SI[retain-explicit-plus]{+10}{\decibel} or higher may have been more appropriate to achieve an increase in perceived pleasantness for wind maskers instead. At the SMRs used for the ARAUS dataset, the wind maskers may have been added at overly high levels, rendering them to be perceived similarly to the traffic maskers due to their similar spectral characteristics and potentially contributing to the decrease in ISO Pleasantness. Additionally, the laboratory-based nature of the data collection process caused the wind maskers to be heard without perceiving natural movements of the air that would be present in an in-situ study, which could have resulted in an artificial or unpleasant situation for the participants.

\begin{figure}[h]
	\centering
	\includegraphics[width=0.75\columnwidth]{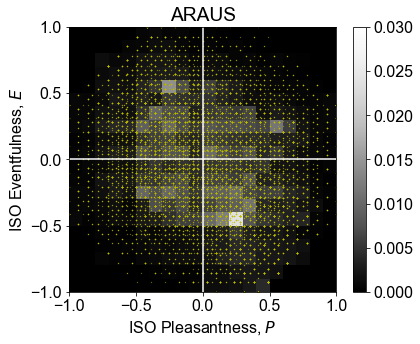}
	\includegraphics[width=0.75\columnwidth]{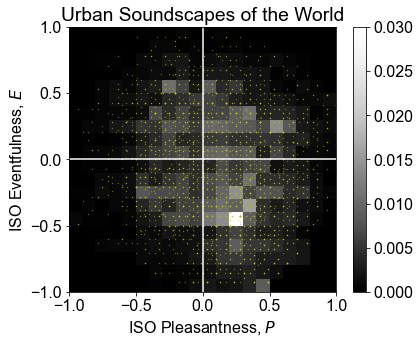}
	\caption{Heat map (grayscale boxes) and scatter plot (yellow points) of ARQ responses in the ``ISO Pleasantness'' and ``ISO Eventfulness'' axes to (top) ARAUS dataset stimuli, (bottom) USotW soundscapes in ARAUS dataset. The brightness of grayscale boxes denotes the proportion of responses belonging to that region.}
	\label{fig:circumplex_distribution}
\end{figure}

To investigate how well the ARAUS dataset stimuli and the base urban soundscapes from the USotW database spanned the perceptual space generated by the ``Pleasantness--Eventfulness'' axes, we first computed the normalized values of $P$ and $E$ according to Equations \eqref{eq:ISOPl} and \eqref{eq:ISOEv} for each individual ARQ response in the ARAUS dataset. Then, we plotted a heat map and scatter plot of the responses on the $P$--$E$ axes, as shown in \Cref{fig:circumplex_distribution}. We can see that both the ARAUS dataset stimuli and the USotW soundscapes covered the positive, neutral, and negative regions of both the ISO Pleasantness and the ISO Eventfulness axes, thereby validating their use in analysis related to the ISO 12913 standard.

\subsection{Participant Information Questionnaire}\label{sec:Analysis and Modeling Methodology/Questionnaires/Participant Information Questionnaire}

Since items in the PIQ such as the participant's age \cite{Yang2005AcousticSpaces} and education status \cite{Fang2021SoundscapeParks} can affect the affective perception of soundscapes, the participants in each fold of the cross-validation set of the ARAUS dataset should ideally be drawn from distributions with similar demographics. This was not enforced during the participant recruitment and fold allocation process, thus a post hoc analysis of the PIQ responses was performed to assess the demographic distributions. The post hoc analysis was performed via standard statistical tests for equality of distributions.

For items in the PIQ coded as categorical variables, $\chi^2$-tests were conducted between the responses obtained in each fold of the cross-validation set (treated as observed frequencies) and those obtained in the entire cross-validation set (treated as expected frequencies). No significant differences were observed, at \SI{5}{\percent} significance levels, in the distribution of all categorical variables between folds, with the lowest $p$-value being \num{0.1070} for the participants' gender.

For items in the PIQ coded as continuous variables, Kruskal-Wallis tests were conducted by treating each fold as an independent group. The Kruskal-Wallis tests showed no significant differences at \SI{5}{\percent} significance levels in the distribution of all continuous variables between folds, with the lowest $p$-value being \num{0.3090} for the extent to which participants were annoyed by noise over the past \num{12} months.

Together, these results indicate that the distribution of participants in a given fold did not significantly differ from the participants in any other fold, which reinforces the validity of using the five-fold cross-validation set of the ARAUS dataset to generate models for populations sharing similar characteristics to that of the entirety of the cross-validation set. Full results of the statistical tests and explicit distributions of responses by fold and PIQ item are detailed in Appendix C.

For the psychological questionnaires used in the PIQ (WNSS-10, PSS-10, WHO-5, and PANAS), Cronbach's $\alpha$ \cite{Cronbach1951CoefficientTests} and McDonald's $\omega$ \cite{McDonald1999TestTreatment} were computed as standard internal reliability coefficients to independently verify if their items in aggregate were indeed measuring the same construct. This served as a validation study for WNSS-10 and a replication study for the other questionnaires.

The internal reliability study results are shown in Table \ref{tab:internal consistency}. All values of the internal reliability coefficients are above 0.8, which can be considered ``high'' given the number of items in each questionnaire \cite{Taber2018TheEducation}. Therefore, all items in each questionnaire indeed measured the same construct. In particular, the Cronbach's $\alpha$ value for of \num{0.835} for WNSS-10 parallels similar validation studies for different iterations of the questionnaires of the original \num{21}-item version \cite{Worthington2018WeinsteinWNSS}, thereby confirming the reliability of the previously unvalidated version used for the ARAUS dataset.

\begin{table}[t]
	\centering
	\caption{Reliability metrics for psychological questionnaires in the PIQ. Larger values are desirable, as denoted by the up arrow $(\uparrow)$.}
	\label{tab:internal consistency}
	\def\arraystretch{1.0}
	\begin{tabularx}{\linewidth}{Xrr}
		\toprule
		Questionnaire & Cronbach's $\alpha \hspace{1mm} (\uparrow)$ & McDonald's $\omega  \hspace{1mm} (\uparrow)$ \\
		\midrule
		WNSS-10          & 0.835 & 0.837 \\
		PSS-10           & 0.875 & 0.874 \\
		WHO-5            & 0.854 & 0.857 \\
		PANAS (Positive) & 0.886 & 0.891 \\
		PANAS (Negative) & 0.891 & 0.891 \\
		\bottomrule
	\end{tabularx}
\end{table}

\subsection{Consistency Checks}\label{sec:Analysis and Modeling Methodology/Consistency Checks}

In a similar manner to the PIQ, Kruskal-Wallis tests for each of the seven single-value consistency metrics were performed by treating the ARQ responses in each fold as independent groups. Full results are presented in Appendix C.

No significant differences were observed, at \SI{5}{\percent} significance levels, in the distribution of all continuous variables between folds, with the lowest $p$-value being \num{0.2545} for the MAD between ``vibrant'' and reversed ``monotonous'' ratings. This indicates that the distribution of response consistency in a given fold did not significantly differ from the responses in any other fold, reinforcing the validity of using the five-fold cross-validation set of the ARAUS dataset to generate generalizable and unbiased models, at least from the perspective of consistency with the ISO 12913-3:2019 circumplex model of soundscape perception.
\section{Affective Response Modeling}\label{sec:Affective Response Modeling}

To illustrate how the ARAUS dataset can be used for systematic and fair benchmarking of models for affective perception, four models were trained to predict the ISO Pleasantness, as defined in Equation \eqref{eq:ISOPl}, of a given augmented soundscape using the ARAUS dataset. The models were a relatively low-parameter regression model, a
convolutional neural network (CNN), and two different probabilistic perceptual attribute predictor (PPAP) models \citep{Ooi2022ProbablyAugmentation, Watcharasupat2022AutonomousGain}.
A dummy model that always predicts the mean of the training data labels was additionally used as a naive benchmark.

Using a five-fold cross-validation scheme, each model was trained five times, each with a different fold of the cross-validation set used as the validation set (\num{5040} samples) and the remaining four folds used as the training set (\num{20160} samples). For models sensitive to random initialization, results from \num{10} different seeds per validation fold were recorded, totalling \num{50} runs. After training, each model was evaluated on the independent test set (\num{48} samples).
\subsection{Models}

\subsubsection{Regularized Linear Regression}

The linear regression models used acoustic and psychoacoustic indicators of an augmented soundscape as input features to predict its ISO Pleasantness. Firstly, the statistics detailed in Section \ref{sec:Data Collection Methodology/Fold Allocation/Acoustic and Psychoacoustic Indicator Computation} were computed for each binaural \textit{augmented} soundscape in the ARAUS dataset, which gave \num{264} possible input features to be used for regression. The features, without prior dimensionality reduction, were used to train elastic net models \cite{Zou2005RegularizationNet} with $L_1$ and $L_2$ regularization weights of \num{0.5} and \num{0.25}, respectively.

Elastic net models are designed for parameter sparsity, since the elastic net loss function will cause most weights to be set to zeroes after training concludes. This indirectly allows the method to automatically choose suitable parameters for regression from an initial larger selection, and makes it more computationally efficient than stepwise regression.

\subsubsection{Convolutional Neural Network}\label{sec:Affective Response Modeling/Models/Convolutional Neural Network}

The CNN models were trained to predict the ISO Pleasantness of an augmented soundscape using its log-mel spectrogram. Firstly, the channel-wise log-mel spectrograms were computed for each binaural augmented soundscape in the ARAUS dataset, with a Hann window length of \num{4096} samples, \SI{50}{\percent} overlap between windows, and \num{64} mel frequency bands from \SI{20}{\hertz} to \SI{20}{\kilo\hertz}.

\begin{figure}[t]
	\centering
	\includegraphics[width=0.82\linewidth]{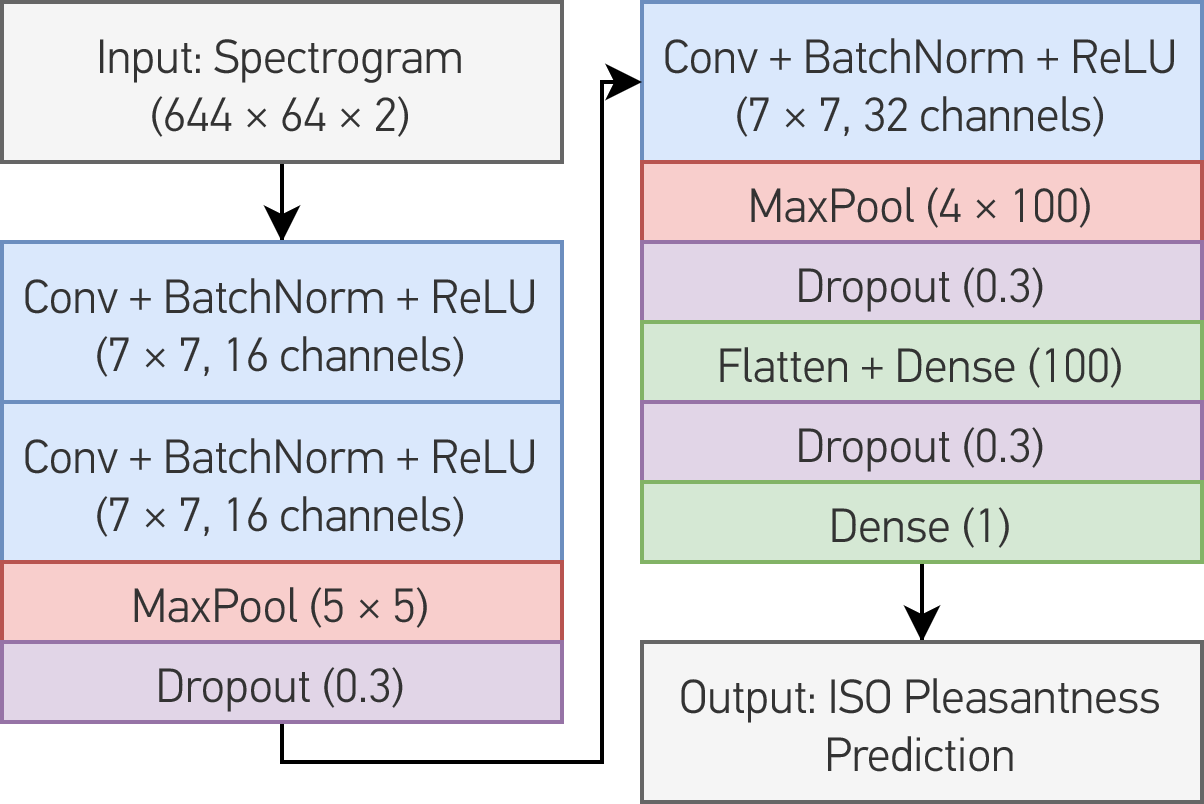}
	\caption{Baseline CNN model architecture, adapted from \cite{Heittola2020AcousticSolutions}}
	\label{fig:CNN Architecture}
\end{figure}

The log-mel spectrograms (as \num{644}-by-\num{64}-by-\num{2} tensors) were then used as input to the CNN models with the model architecture shown in \Cref{fig:CNN Architecture}. The architecture is identical to the baseline model architecture used for the acoustic scene classification task (Task 1B) of the DCASE 2020 Challenge \cite{Heittola2020AcousticSolutions}. However, due to the slightly larger input dimensions of the ARAUS data, the modified models shown in \Cref{fig:CNN Architecture} contained about 142K parameters, compared to the original 116K in \cite{Heittola2020AcousticSolutions}, with the only difference being the input dimensions of the final dense layer.

The CNN models were trained over \num{100} epochs with an Adam optimizer \cite{Kingma2015Adam:Optimization} with learning rate \num{1e-4} and batch size \num{32}. The training was stopped early if the validation set MSE did not improve for \num{10} consecutive epochs. This process was repeated for \num{10} runs per validation fold to obtain the mean performance of the models.

\subsubsection{Probabilistic Perceptual Attribute Predictor}

The PPAP models were used to model the subjectivity in ARQ responses by outputting predictions for ISO Pleasantness based on probability distributions rather than deterministic values. As described in \cite{Ooi2022ProbablyAugmentation,Watcharasupat2022AutonomousGain}, we trained the PPAP models using the normal distribution $N(\mu,\sigma^2)$, with the loss function being the log-probability of the ground-truth response being observed, given the output distribution parameters $\mu$ and $\sigma$.

Using the ARAUS dataset responses, we trained two different variations of the PPAP models: one performing augmentation in the time domain (i.e., taking in log-mel spectrograms of the augmented soundscapes as initial input) \cite{Ooi2022ProbablyAugmentation}, and one performing augmentation in the feature domain (i.e., taking in log-mel spectrograms of base urban soundscapes and maskers separately as initial inputs, and subsequently performing augmentation on features extracted from them) \citep{Watcharasupat2022AutonomousGain}, both with multi-head attention in the feature mapping blocks. The parameters for the computation of the log-mel spectrograms and the training procedure for the PPAP models were similar to that of the CNN models.

\subsection{Results and Discussion}\label{sec:Models/Results and Discussion}

\begin{table*}[t]
	\centering
	\caption{Mean squared error values ($\pm$ standard deviation over 10 runs where applicable) for models described in Section \ref{sec:Affective Response Modeling}. Smaller values are desirable, as denoted by the down arrow $(\downarrow)$.}
	\label{tab:baseline model metrics}
	\begin{tabularx}{\textwidth}{Xr*{3}{l@{\hspace{3pt}}l@{\hspace{3pt}}l}}
		\toprule
		&& \multicolumn{9}{c}{Mean squared error $(\downarrow)$} \\
		\cmidrule(lr){3-11}
		Model & Params &
		\multicolumn{3}{c}{Train} & 
		\multicolumn{3}{c}{Validation} & 
		\multicolumn{3}{c}{Test} \\
		\midrule
		Dummy (mean of labels) & --- & 0.1551 &&& 0.1553 &&& 0.1192 \\
		Elastic net & 4 & 0.1351 &&& 0.1357 &&& 0.0930 \\ 
		CNN based on DCASE 2020 Task 1B baseline \cite{Heittola2020AcousticSolutions} & 142K &
		0.1152 &$\pm$& 0.0020& 
		\textbf{0.1212} &$\pm$& 0.0010& 
		0.0865 &$\pm$& 0.0063 \\
		Multi-head attention PPAP, additive time-domain augmentation \cite{Ooi2022ProbablyAugmentation} & 123K &
		0.1098&$\pm$&0.0047 &
		0.1216&$\pm$&0.0016 &
		0.0889&$\pm$&0.0064\\
		Multi-head attention PPAP, additive feature-domain augmentation \citep{Watcharasupat2022AutonomousGain} & 515K &
		\textbf{0.1086}&$\pm$& 0.0036 &
		0.1238&$\pm$& 0.0022 &
		\textbf{0.0838}&$\pm$& 0.0103\\
		\bottomrule
	\end{tabularx}
\end{table*}

\begin{figure}[t]
	\centering
	\includegraphics[width=\linewidth]{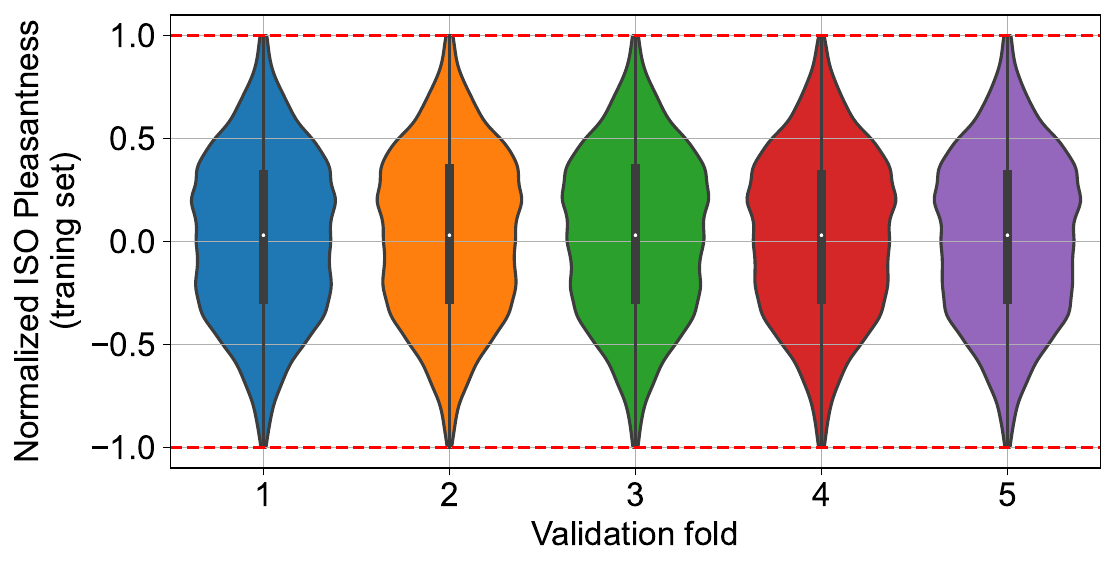}
	\caption{Violin plots of distributions of normalized ISO Pleasantness (ground-truth labels) in training set, by fold left out as validation fold.}
	\label{fig:label_means}
\end{figure}

The label means for the normalized ISO Pleasantness $P$ were \num{0.0208}, \num{0.0317}, \num{0.0307}, \num{0.0281}, and \num{0.0225} when folds \num{1}, \num{2}, \num{3}, \num{4}, and \num{5} were respectively left out as the validation fold for the derivation of the dummy models. Hence, these were also the exact predictions given by all dummy models regardless of the soundscape presented. Considering that $P\in[-1,1]$, the difference in label means by fold is also insubstantial, since they are within \SI{0.5}{\percent} of the full range. In fact, the distributions of the training set labels themselves have no significant difference by fold, as shown in \Cref{fig:label_means}.

After training the elastic net models, the weights for all but three input features were set to zero regardless of the fold used as the validation set. Taking the mean of weights for the regression models across the 5-fold cross-validation set, this gave the general regression model 
\begin{equation}
    \widehat{P} =
    \begin{bmatrix*}[l]
        +0.3035 \\
		-7.8731\cdot 10^{-3}\\
        -0.4512\cdot 10^{-3}\\
        -1.7326\cdot 10^{-3}
        \end{bmatrix*}^\mathsf{T}\begin{bmatrix*}[l]
        1\\
        N_{\text{max}}\\
        M_{\text{\SI{10}{\kilo\hertz}}}\\
        M_{\text{\SI{12.5}{\kilo\hertz}}}
    \end{bmatrix*},\label{eq:regression}
\end{equation}
where $N_{\text{max}}$ is the maximum loudness (in sone), $M_{f}$ is the power (in \SI{}{\decibel}) at the one-third octave band with center frequency $f$, the prediction for the normalized ISO Pleasantness value $P$ is denoted as $\widehat{P}$, and the regression coefficients are not normalized. 

This implies that of the \num{264} acoustic and psychoacoustic parameters used as input features to train the regression models, the features that most significantly impacted the ISO Pleasantness were $N_{\text{max}}$, $M_{\text{\SI{10}{\kilo\hertz}}}$ and $M_{\text{\SI{12.5}{\kilo\hertz}}}$, at least based on the responses in the ARAUS dataset. Incidentally, for the stimuli in the ARAUS dataset, we observed that the ranges of these features were $N_{\text{max}}\in[8.36,94.1]$, $M_{\text{\SI{10}{\kilo\hertz}}}\in[29.62,85.15]$ and $M_{\text{\SI{12.5}{\kilo\hertz}}}\in[32.51,78.67]$, which meant that the predictions had a range of $\widehat{P}\in[-0.6121,0.1680]$. The negative coefficients of the features in Equation \eqref{eq:regression} also suggest that increasing the peak loudness and high-frequency components of an augmented soundscape correlate with a decrease in its perceived pleasantness, corroborating the respective findings by \cite{Ma2021} and \cite{Kidd2003TheSounds} regarding general acoustic environments.

Lastly, Table \ref{tab:baseline model metrics} compares the performance of the four baseline models and the dummy model with respect to the MSEs achieved across their training, validation, and test sets in the \num{5}-fold cross-validation scheme used to train them. All baseline models performed better than the dummy model in the training, validation, and test sets, which suggests that the features trained in all the baseline models to make ISO Pleasantness predictions were indeed meaningful.

In addition, the CNN and PPAP models performed significantly better than the elastic net models, with the lowest mean train, validation, and test set MSE of \num{0.1086}, \num{0.1212}, and \num{0.0838} observed with the PPAP performing feature-domain augmentation \cite{Watcharasupat2022AutonomousGain}, the CNN \cite{Heittola2020AcousticSolutions}, and the PPAP performing feature-domain augmentation \cite{Watcharasupat2022AutonomousGain}, respectively. This is expected due to the vast difference in the numbers of parameters used by the CNN and PPAP models (142--515K) for prediction, in comparison to the regression models (\num{4} parameters) and dummy model. Nevertheless, none of the models were overfitted to the dataset, as can be seen from the relatively similar values for training and validation set MSEs across all models, which demonstrates the versatility of the ARAUS dataset for use in training generalizable high-parameter models. 

Notably, the test set MSE values are also smaller than that of the validation set, which could be due to the relative size of the test set (\num{48} samples) as compared to the validation set (\num{5040} samples) making the test set ``easier'' from the perspective of the prediction models than their respective validation sets.
\section{Limitations and Future Work}
\label{sec:Limitations and Future Work}

While the ARAUS dataset utilized recordings of in-situ urban soundscapes as part of its stimuli, the laboratory-based data collection process means that models trained using the ARAUS dataset may require confirmation of their ecological validity with follow-up in-situ experiments or soundwalks. The range of SMRs used for the generation of the augmented soundscapes in the ARAUS dataset (from \num{-6} to \SI[retain-explicit-plus]{+6}{\decibel}) is also a potential limitation of the dataset, since real-life or virtual sound sources used for augmentation in general could have relative differences outside of the chosen range, so future data collection related to the ARAUS dataset could consider increasing the range of SMRs used for the generation of stimuli, which could be done by changing input parameter settings in the provided replication code. Not all combinations of urban soundscapes, maskers, and SMRs were exhaustively generated for the current iteration of the ARAUS dataset, so responses could also be collected exhaustively in the future to make the dataset fully comprehensive. Responses could alternatively be collected continuously over time, instead of just once after the presentation of each stimulus, in order to study the consistency of participants' affective responses over time. The visual content of the stimuli could also be varied, such as by using video recordings captured at completely different locations, for the same audio recording, in order to investigate changes in affective responses due purely to changes in the visual content.

Moreover, since the participants in the ARAUS dataset were mostly young university students, the results and analysis obtained may not necessarily translate equivalently to a more general population of people exposed to urban soundscapes. The inclusion of only five participants in the test set, as compared to \num{120} in each fold of the cross-validation set, is also a primary limitation of the dataset. At the scale of the test set, person-to-person differences in perception could dominate any other factor contributing to perception, thereby causing the test set to serve more as a small focus group of participants rather than a general sample representative of the same population as the cross-validation set. Hence, future extensions to the ARAUS dataset could concentrate on enlarging the test set and obtaining responses from participants from an older demographic, to allow for improved benchmarking of models trained using the dataset.

The models described in this article were also trained using the individual augmented soundscapes in the ARAUS dataset in isolation from each other, so they did not account for the effects of temporal successions of different maskers, and hence different augmented soundscapes. Na\"{i}vely applying them to choose a time series of optimal maskers for an extended duration of time could lead to a dissonant succession of maskers that may inadvertently result in an unpleasant augmented soundscape overall, despite the individual maskers being predicted as optimal for individual time windows. Further work on these models could thus look into generalizing them to extended durations of time. The models could also be fine-tuned via transfer learning methods on the affective soundscape datasets in \Cref{tab:audio_datasets} (which possess labels of a different nature), and be compared with other benchmarks to assess the amenability of the ARAUS dataset to transfer learning.

Lastly, other perceptual indicators other than those defined in ISO/TS 12913-2:2018, such as the perceived loudness or tranquility, could also be used as part of the ARQ to expand its scope and allow the ARAUS dataset to possess greater generalizability.
\section{Conclusion}
\label{sec:Conclusion}

In conclusion, we presented the Affective Responses to Augmented Urban Soundscapes (ARAUS) dataset, which functions as a benchmark dataset for comparisons of prediction models for the affective perception of urban and augmented soundscapes. We first presented a systematic methodology for the collection of data, which can be replicated or extended upon. Subsequently, we analyzed the responses obtained for the ARAUS dataset, and provided benchmark models for predicting the perceptual attribute of ISO Pleasantness. To the best of our knowledge, the ARAUS dataset is currently the largest soundscape dataset with perceptual labels, but is not without its inherent limitations as described in \Cref{sec:Limitations and Future Work}.

Nonetheless, we hope that the ARAUS dataset becomes a beneficial and enduring resource for the soundscape community, by assisting soundscape researchers in developing more accurate, robust models for soundscape perception.

% TO ACCESS SUPPLEMENTARY: 
% click on supplementary.tex to make it the file in the editor, then compile.

% \FloatBarrier

\section*{Acknowledgments}

We would like to thank Prof. Chen Jer-Ming, Mr. Tan Yi Xian, and Ms. Cindy Lin for assisting with the administrative arrangements and setup of the test site at the Singapore University of Technology and Design.

This research is supported by the Singapore Ministry of National Development and the National Research Foundation, Prime Minister's Office under the Cities of Tomorrow Research Programme (Award No. COT-V4-2020-1). Any opinions, findings and conclusions or recommendations expressed in this material are those of the authors and do not reflect the view of National Research Foundation, Singapore, and Ministry of National Development, Singapore.

% Can use something like this to put references on a page
% by themselves when using endfloat and the captionsoff option.
\ifCLASSOPTIONcaptionsoff
  \newpage
\fi

% trigger a \newpage just before the given reference
% number - used to balance the columns on the last page
% adjust value as needed - may need to be readjusted if
% the document is modified later
%\IEEEtriggeratref{8}
% The "triggered" command can be changed if desired:
%\IEEEtriggercmd{\enlargethispage{-5in}}

% references section

% can use a bibliography generated by BibTeX as a .bbl file
% BibTeX documentation can be easily obtained at:
% http://mirror.ctan.org/biblio/bibtex/contrib/doc/
% The IEEEtran BibTeX style support page is at:
% http://www.michaelshell.org/tex/ieeetran/bibtex/
\bibliographystyle{IEEEtran}

\begin{thebibliography}{100}
	\providecommand{\url}[1]{#1}
	\def\UrlFont{\rmfamily}
	\providecommand{\newblock}{\relax}
	\providecommand{\bibinfo}[2]{#2}
	\providecommand\BIBentrySTDinterwordspacing{\spaceskip=0pt\relax}
	\providecommand\BIBentryALTinterwordstretchfactor{4}
	\providecommand\BIBentryALTinterwordspacing{\spaceskip=\fontdimen2\font plus
		\BIBentryALTinterwordstretchfactor\fontdimen3\font minus
		\fontdimen4\font\relax}
	\providecommand\BIBforeignlanguage[2]{{%
			\expandafter\ifx\csname l@#1\endcsname\relax
			\typeout{** WARNING: IEEEtran.bst: No hyphenation pattern has been}%
			\typeout{** loaded for the language `#1'. Using the pattern for}%
			\typeout{** the default language instead.}%
			\else
			\language=\csname l@#1\endcsname
			\fi
			#2}}
	
	\bibitem{Raimbault2003}
	M.~Raimbault, C.~Lavandier, and M.~B{\'{e}}rengier, ``{Ambient sound assessment
		of urban environments: Field studies in two French cities},''
	\emph{{Applied Acoustics}}, vol.~64, no.~12, pp.
	1241--1256, 2003.
	
	\bibitem{Jennings2013}
	P.~Jennings and R.~Cain, ``{A framework for improving urban soundscapes},''
	\emph{{Applied Acoustics}}, vol.~74, no.~2, pp.
	293--299, 2013.
	
	\bibitem{DePaivaVianna2015NoiseStudy}
	K.~M. De~Paiva~Vianna, M.~R. Alves~Cardoso, and R.~M.~C. Rodrigues, ``{Noise
		pollution and annoyance: An urban soundscapes study},''
	\emph{{Noise and Health}}, vol.~17, no.~76, pp.
	125--133, 2015.
	
	\bibitem{Kang2019}
	J.~Kang, \emph{{et~al.}}, ``{Towards soundscape
		indices},'' in \emph{{23rd International Congress on
			Acoustics}}, 2019, pp. 2488--2495.
	
	\bibitem{Miedema1998Exposure-responseNoise}
	H.~M.~E. Miedema and H.~Vos, ``{Exposure-response relationships for
		transportation noise},'' \emph{{The Journal of the
			Acoustical Society of America}}, vol. 104, no.~6, pp. 3432--3445, 1998.
	
	\bibitem{DeCoensel2007ModelsPlanning}
	B.~De~Coensel and D.~Botteldooren, ``{Models for soundscape perception and
		their use in planning},'' in \emph{{Proceedings of
			Inter-Noise 2007}}, 2007.
	
	\bibitem{Axelsson2014}
	O.~Axelsson, M.~E. Nilsson, B.~Hellstr{\"{o}}m, and P.~Lund{\'{e}}n, ``{A field
		experiment on the impact of sounds from a jet-and-basin fountain on
		soundscape quality in an urban park},'' \emph{{Landscape
			and Urban Planning}}, vol. 123, pp. 49--60, 2014.
	
	\bibitem{Abdalrahman2020a}
	Z.~Abdalrahman and L.~Galbrun, ``{Audio-visual preferences, perception, and use
		of water features in open-plan offices},''
	\emph{{Journal of the Acoustical Society of America}},
	vol. 147, no.~3, pp. 1661--1672, 2020.
	
	\bibitem{Hong2019b}
	J.~Y. Hong, \emph{{et~al.}}, ``{The effects of spatial
		separations on water sound and traffic noise sources on soundscape
		assessment},'' \emph{{Building and Environment}}, vol.
	167, no. 106423, 2020.
	
	\bibitem{Aletta2018TowardsApproach}
	F.~Aletta and J.~Kang, ``{Towards an urban vibrancy model: A soundscape
		approach},'' \emph{{International Journal of
			Environmental Research and Public Health}}, vol.~15, no.~8, 2018.
	
	\bibitem{InternationalOrganizationforStandardization2019}
	{International Organization for Standardization}, \emph{{ISO 12913-3:2019 -
			Acoustics - Soundscape - Part 3: Data analysis}}.\hskip 1em plus 0.5em minus
	0.4em\relax Geneva, Switzerland: International Organization for
	Standardization, 2019.
	
	\bibitem{VanRenterghem2020}
	T.~Van~Renterghem, \emph{{et~al.}}, ``{Interactive
		soundscape augmentation by natural sounds in a noise polluted urban park},''
	\emph{{Landscape and Urban Planning}}, vol. 194, no.
	October 2019, p. 103705, 2020.
	
	\bibitem{Wong2022DeploymentAugmentation}
	T.~Wong, \emph{{et~al.}}, ``{Deployment of an IoT System
		for Adaptive In-Situ Soundscape Augmentation},'' in
	\emph{{Proceedings of Inter-Noise 2022}}, 2022.
	
	\bibitem{Leung2016}
	T.~M. Leung, C.~K. Chau, and S.~K. Tang, ``{On the study of effects on
		different types of natural sounds on the perception of combined sound
		environment with road traffic noise},'' in
	\emph{{Proceedings of Inter-Noise 2016}}, 2016, pp.
	1764--1770.
	
	\bibitem{Hong2021b}
	J.~Y. Hong, \emph{{et~al.}}, ``{A mixed-reality approach
		to soundscape assessment of outdoor urban environments augmented with natural
		sounds},'' \emph{{Building and Environment}}, vol. 194,
	no. July 2020, p. 107688, 2021.
	
	\bibitem{Aumond2017ModelingContext}
	P.~Aumond, A.~Can, B.~De~Coensel, D.~Botteldooren, C.~Ribeiro, and
	C.~Lavandier, ``{Modeling soundscape pleasantness using perceptual
		assessments and acoustic measurements along paths in urban context},''
	\emph{{Acta Acustica united with Acustica}}, vol. 103,
	no.~3, pp. 430--443, 2017.
	
	\bibitem{Puyana-Romero2019}
	V.~Puyana-Romero, G.~Ciaburro, G.~Brambilla, C.~Garz{\'{o}}n, and L.~Maffei,
	``{Representation of the soundscape quality in urban areas through
		colours},'' \emph{{Noise Mapping}}, vol.~6, no.~1, pp.
	8--21, 2019.
	
	\bibitem{Aletta2020}
	F.~Aletta, \emph{{et~al.}}, ``{Soundscape assessment:
		Towards a validated translation of perceptual attributes in different
		languages},'' in \emph{{Proceedings of Inter-Noise
			2020}}, 2020.
	
	\bibitem{Lionello2020ASoundscapes}
	\BIBentryALTinterwordspacing
	M.~Lionello, F.~Aletta, and J.~Kang, ``{A systematic review of prediction
		models for the experience of urban soundscapes},''
	\emph{{Applied Acoustics}}, vol. 170, p. 107479, 2020.
	\BIBentrySTDinterwordspacing
	
	\bibitem{InternationalOrganizationforStandardization2017}
	{International Organization for Standardization}, \emph{{ISO 12913-2 Acoustics
			- Soundscape - Part 2: Data collection and reporting requirements}}.\hskip
	1em plus 0.5em minus 0.4em\relax Geneva, Switzerland: International
	Organization for Standardization, 2018.
	
	\bibitem{Wang2021Audio-visualSubmissions}
	\BIBentryALTinterwordspacing
	S.~Wang, T.~Heittola, A.~Mesaros, and T.~Virtanen, ``{Audio-visual scene
		classification: analysis of DCASE 2021 Challenge submissions},'' in
	\emph{{Proceedings of DCASE 2021 Workshop}}, 2021, pp.
	45--49.
	
	\bibitem{Martin-Morato2021Low-complexitySystems}
	\BIBentryALTinterwordspacing
	I.~Mart{\'{i}}n-Morat{\'{o}}, T.~Heittola, A.~Mesaros, and T.~Virtanen,
	``{Low-complexity acoustic scene classification for multi-device audio:
		Analysis of DCASE 2021 challenge systems},'' in
	\emph{{Proceedings of DCASE 2021 Workshop}}, 2021, pp.
	85--89.
	
	\bibitem{Politis2021ADetection}
	\BIBentryALTinterwordspacing
	A.~Politis, S.~Adavanne, D.~Krause, A.~Deleforge, P.~Srivastava, and
	T.~Virtanen, ``{A Dataset of Dynamic Reverberant Sound Scenes with
		Directional Interferers for Sound Event Localization and Detection},'' in
	\emph{{Proceedings of DCASE 2021 Workshop}}, 2021, pp.
	125--129.
	
	\bibitem{Kawaguchi2021DescriptionConditions}
	\BIBentryALTinterwordspacing
	Y.~Kawaguchi, \emph{{et~al.}}, ``{Description and
		Discussion on DCASE 2021 Challenge Task 2: Unsupervised Anomalous Sound
		Detection for Machine Condition Monitoring under Domain Shifted
		Conditions},'' in \emph{{Proceedings of DCASE 2021
			Workshop}}, 2021, pp. 186--190.
	
	\bibitem{Gemmeke2017AudioEvents}
	J.~F. Gemmeke, \emph{{et~al.}}, ``{Audio Set: An ontology
		and human-labeled dataset for audio events},'' in
	\emph{{Proceedings of IEEE ICASSP 2017}}, 2017, pp.
	776--780.
	
	\bibitem{Hershey2021TheClassification}
	S.~Hershey, \emph{{et~al.}}, ``{The benefit of
		temporally-strong labels in audio event classification},'' in
	\emph{{Proceedings of IEEE ICASSP 2021}}, 2021, pp.
	366--370.
	
	\bibitem{Hershey2021AudioSet:2021}
	\BIBentryALTinterwordspacing
	------, ``{AudioSet: Temporally-Strong Labels Download (May 2021)},'' 2021.
	[Online]. Available:
	\url{https://research.google.com/audioset/download_strong.html}
	\BIBentrySTDinterwordspacing
	
	\bibitem{Hershey2017CNNClassification}
	------, ``{CNN architectures for large-scale audio classification},'' in
	\emph{{Proceedings of IEEE ICASSP 2017}}, 2017, pp.
	131--135.
	
	\bibitem{Salamon2014AResearch}
	J.~Salamon, C.~Jacoby, and J.~P. Bello, ``{A dataset and taxonomy for urban
		sound research},'' in \emph{{Proceedings of the 2014 ACM
			Multimedia Conference}}, 2014, pp. 1041--1044.
	
	\bibitem{Piczak2015ESC:Classification}
	K.~J. Piczak, ``{ESC: Dataset for environmental sound classification},'' in
	\emph{{Proceedings of the 2015 ACM Multimedia
			Conference}}, 2015, pp. 1015--1018.
	
	\bibitem{Fonseca2022FSD50K:Events}
	E.~Fonseca, X.~Favory, J.~Pons, F.~Font, and X.~Serra, ``{FSD50K: An Open
		Dataset of Human-Labeled Sound Events},''
	\emph{{IEEE/ACM Transactions on Audio Speech and
			Language Processing}}, vol.~30, pp. 829--852, 2022.
	
	\bibitem{Font2013}
	F.~Font, G.~Roma, and X.~Serra, ``{Freesound Technical Demo},'' in
	\emph{{Proceedings of the 2013 ACM Multimedia
			Conference}}, 2013, pp. 411--412.
	
	\bibitem{Salamon2017}
	\BIBentryALTinterwordspacing
	J.~Salamon, D.~MacConnell, M.~Cartwright, P.~Li, and J.~P. Bello, ``{Scaper: a
		library for soundscape synthesis and augmentation},'' 2017, pp. 344--348.
	
	\bibitem{DeCoensel2017UrbanMind}
	B.~De~Coensel, K.~Sun, and D.~Botteldooren, ``{Urban Soundscapes of the World:
		Selection and reproduction of urban acoustic environments with soundscape in
		mind},'' in \emph{{Proceedings of Inter-Noise 2017}},
	2017.
	
	\bibitem{Ciufo2017}
	M.~Ciufo and D.~Thomas, ``{EigenScape: A Database of Spatial Acoustic Scene
		Recordings},'' \emph{{Applied Sciences}}, vol.~7, 2017.
	
	\bibitem{Cartwright2019a}
	M.~Cartwright, \emph{{et~al.}}, ``{SONYC Urban Sound
		Tagging (SONYC-UST): A Multilabel Dataset from an Urban Acoustic Sensor
		Network},'' in \emph{{Proceedings of DCASE 2019
			Workshop}}, 2019.
	
	\bibitem{Cartwright2020}
	\BIBentryALTinterwordspacing
	------, ``{SONYC-UST-V2: An Urban Sound Tagging Dataset with Spatiotemporal
		Context},'' in \emph{{Proceedings of DCASE 2020
			Workshop}}, 2020, pp. 16--20.
	
	\bibitem{Mesaros2016TUTDetection}
	A.~Mesaros, T.~Heittola, and T.~Virtanen, ``{TUT database for acoustic scene
		classification and sound event detection},'' in
	\emph{{Proceedings of EUSIPCO 2016}}, 2016, pp.
	1128--1132.
	
	\bibitem{Mesaros2017DCASESystem}
	A.~Mesaros, \emph{{et~al.}}, ``{DCASE 2017 challenge
		setup: tasks, datasets and baseline system},'' in
	\emph{{Proceedings of DCASE 2017 Workshop}}, 2017.
	
	\bibitem{Mesaros2018AClassification}
	\BIBentryALTinterwordspacing
	A.~Mesaros, T.~Heittola, and T.~Virtanen, ``{A multi-device dataset for urban
		acoustic scene classification},'' in \emph{{Proceedings
			of DCASE 2018 Workshop}}, 2018.
	
	\bibitem{Heittola2020AcousticSolutions}
	\BIBentryALTinterwordspacing
	T.~Heittola, A.~Mesaros, and T.~Virtanen, ``{Acoustic scene classification in
		DCASE 2020 Challenge: generalization across devices and low complexity
		solutions},'' in \emph{{Proceedings of DCASE 2020
			Workshop}}, 2020.
	
	\bibitem{Wang2021}
	S.~Wang, A.~Mesaros, T.~Heittola, and T.~Virtanen, ``{A curated dataset of
		urban scenes for audio-visual scene analysis},'' in
	\emph{{Proceedings of IEEE ICASSP 2021}}, 2021, pp.
	626--630.
	
	\bibitem{Ooi2021AContext}
	K.~Ooi, \emph{{et~al.}}, ``{A Strongly-Labelled Polyphonic
		Dataset of Urban Sounds with Spatiotemporal Context},'' in
	\emph{{Proceedings of APSIPA ASC 2021}}, 2021.
	
	\bibitem{Weisser2019TheDatabase}
	A.~Weisser, \emph{{et~al.}}, ``{The Ambisonic Recordings
		of Typical Environments (ARTE) database},'' \emph{{Acta
			Acustica united with Acustica}}, vol. 105, no.~4, pp. 695--713, 2019.
	
	\bibitem{Politis2022STARSS22:Events}
	\BIBentryALTinterwordspacing
	A.~Politis, \emph{{et~al.}}, ``{STARSS22: A dataset of
		spatial recordings of real scenes with spatiotemporal annotations of sound
		events},'' 2022. [Online]. Available:
	\url{https://doi.org/10.5281/zenodo.6387880}
	\BIBentrySTDinterwordspacing
	
	\bibitem{Bradley2007TheB-3.}
	M.~M. Bradley and P.~J. Lang, ``{The International Affective Digitized Sounds
		(2nd Edition; IADS-2): Affective ratings of sounds and instruction manual.
		Technical report B-3.}'' University of Florida, Gainesville, Fl., Tech. Rep.,
	2007.
	
	\bibitem{Stevenson2008AffectiveCategories}
	R.~A. Stevenson and T.~W. James, ``{Affective auditory stimuli:
		Characterization of the International Affective Digitized Sounds (IADS) by
		discrete emotional categories},'' \emph{{Behavior
			Research Methods}}, vol.~40, no.~1, pp. 315--321, 2008.
	
	\bibitem{Yang2018}
	W.~Yang, \emph{{et~al.}}, ``{Affective auditory stimulus
		database: An expanded version of the International Affective Digitized Sounds
		(IADS-E)},'' \emph{{Behavior Research Methods}},
	vol.~50, no.~4, pp. 1415--1429, 2018.
	
	\bibitem{Fan2017}
	J.~Fan, M.~Thorogood, and P.~Pasquier, ``{Emo-soundscapes: A dataset for
		soundscape emotion recognition},'' in \emph{{Proceedings
			of ACII 2017}}, 2017, pp. 196--201.
	
	\bibitem{Bradley1994}
	M.~M. Bradley and P.~J. Lang, ``{Measuring emotion: The self-assessment manikin
		and the semantic differential},'' \emph{{Journal of
			Behavior Therapy and Experimental Psychiatry}}, vol.~25, no.~1, pp. 49--59,
	1994.
	
	\bibitem{Giannakopoulos2019}
	T.~Giannakopoulos, M.~Orfanidi, and S.~Perantonis, ``{Athens Urban Soundscape
		(ATHUS): A Dataset for Urban Soundscape Quality Recognition},'' in
	\emph{{Proceedings of 25th International Conference on
			Multimedia Modeling}}, 2019, pp. 338--348.
	
	\bibitem{Mitchell2021}
	\BIBentryALTinterwordspacing
	A.~Mitchell, \emph{{et~al.}}, ``{The International
		Soundscape Database: An integrated multimedia database of urban soundscape
		surveys -- questionnaires with acoustical and contextual information},''
	2021. [Online]. Available: \url{https://doi.org/10.5281/Zenodo.5914762}
	\BIBentrySTDinterwordspacing
	
	\bibitem{Mitchell2020TheInformation}
	------, ``{The Soundscape Indices (SSID) Protocol : A Method for Urban
		Soundscape Surveys — Questionnaires with Acoustical and Contextual
		Information},'' \emph{{Applied Sciences}}, vol.~10, no.
	2397, pp. 1--27, 2020.
	
	\bibitem{Hao2016}
	\BIBentryALTinterwordspacing
	Y.~Hao, J.~Kang, and H.~Wortche, ``{Assessment of the masking effects of
		birdsong on the road traffic noise environment},''
	\emph{{Journal of the Acoustical Society of America}},
	vol. 140, no.~2, pp. 978--987, 2016.
	\BIBentrySTDinterwordspacing
	
	\bibitem{Leung2017}
	\BIBentryALTinterwordspacing
	T.~M. Leung, C.~K. Chau, S.~K. Tang, and J.~M. Xu, ``{Developing a multivariate
		model for predicting the noise annoyance responses due to combined water
		sound and road traffic noise exposure},'' \emph{{Applied
			Acoustics}}, vol. 127, pp. 284--291, 2017.
	\BIBentrySTDinterwordspacing
	
	\bibitem{Aletta2020a}
	F.~Aletta, T.~Oberman, A.~Mitchell, H.~Tong, and J.~Kang, ``{Assessing the
		changing urban sound environment during the COVID-19 lockdown period using
		short-term acoustic measurements},'' \emph{{Noise
			Mapping}}, vol.~7, no.~1, pp. 123--134, 2020.
	
	\bibitem{Lutman2000WhatAbove}
	M.~E. Lutman, ``{What is the risk of noise-induced hearing loss at 80, 85, 90
		dB(A) and above?}'' \emph{{Occupational Medicine}},
	vol.~50, no.~4, pp. 274--275, 2000.
	
	\bibitem{Planque2008Xeno-canto:Song}
	B.~Planqu{\'{e}} and W.-P. Vellinga, ``{Xeno-canto: a 21st century way to
		appreciate Neotropical bird song},'' \emph{{Neotropical
			Birding}}, vol.~3, no. January, pp. 17--23, 2008.
	
	\bibitem{Jeon2012}
	\BIBentryALTinterwordspacing
	J.~Y. Jeon, P.~J. Lee, J.~You, and J.~Kang, ``{Acoustical characteristics of
		water sounds for soundscape enhancement in urban open spaces},''
	\emph{{Journal of the Acoustical Society of America}},
	vol. 131, no.~3, pp. 2101--2109, 2012.
	\BIBentrySTDinterwordspacing
	
	\bibitem{Galbrun2013AcousticalNoise}
	L.~Galbrun and T.~T. Ali, ``{Acoustical and perceptual assessment of water
		sounds and their use over road traffic noise},''
	\emph{{Journal of the Acoustical Society of America}},
	vol. 133, no.~1, pp. 227--237, 2013.
	
	\bibitem{Coensel2015}
	B.~De~Coensel, S.~Vanwetswinkel, and D.~Botteldooren, ``{Effects of natural
		sounds on the perception of road traffic noise},''
	\emph{{JASA Express Letters}}, vol. 129, no.~4, pp.
	148--153, 2011.
	
	\bibitem{Ferraroa}
	\BIBentryALTinterwordspacing
	D.~M. Ferraro, \emph{{et~al.}}, ``{The phantom chorus:
		birdsong boosts human well-being in protected areas},''
	\emph{{Proceedings of the Royal Society B}}, vol. 287,
	no. 1941, 2020.
	\BIBentrySTDinterwordspacing
	
	\bibitem{Hedblom2017EvaluationPreservation}
	M.~Hedblom, I.~Knez, {Ode Sang}, and B.~Gunnarsson, ``{Evaluation of natural
		sounds in urban greenery: Potential impact for urban nature preservation},''
	\emph{{Royal Society Open Science}}, vol.~4, no.~2,
	2017.
	
	\bibitem{WorldHealthOrganizationRegionalOfficeforEurope2018}
	{World Health Organization Regional Office for Europe}, \emph{{Environmental
			Noise Guidelines for the European Region}}.\hskip 1em plus 0.5em minus
	0.4em\relax Copenhagen: The Regional Office for Europe of the World Health
	Organization, 2018.
	
	\bibitem{You2008}
	J.~You and J.~Y. Jeon, ``{Sound-masking technique for combined noise exposure
		in open public spaces},'' in \emph{{Proceedings of ICBEN
			2008}}, 2008.
	
	\bibitem{Pieretti2013a}
	\BIBentryALTinterwordspacing
	N.~Pieretti and A.~Farina, ``{Application of a recently introduced index for
		acoustic complexity to an avian soundscape with traffic noise},''
	\emph{{Journal of the Acoustical Society of America}},
	vol. 134, no.~1, pp. 891--900, 2013.
	\BIBentrySTDinterwordspacing
	
	\bibitem{Lu2020}
	\BIBentryALTinterwordspacing
	X.~Lu, J.~Tang, P.~Zhu, F.~Guo, J.~Cai, and H.~Zhang, ``{Spatial variations in
		pedestrian soundscape evaluation of traffic noise},''
	\emph{{Environmental Impact Assessment Review}},
	vol.~83, 2020.
	\BIBentrySTDinterwordspacing
	
	\bibitem{Xu2019a}
	C.~Xu and J.~Kang, ``{Soundscape evaluation: Binaural or monaural?}''
	\emph{{Journal of the Acoustical Society of America}},
	vol. 145, no.~5, pp. 3208--3217, 2019.
	
	\bibitem{Engel2021}
	M.~S. Engel, A.~Fiebig, C.~Pfaffenbach, and J.~Fels, ``{A Review of the Use of
		Psychoacoustic Indicators on Soundscape Studies},''
	\emph{{Current Pollution Reports}}, vol.~7, no.~3, pp.
	359--378, 2021.
	
	\bibitem{Hong2020EffectsQuality}
	J.~Y. Hong, \emph{{et~al.}}, ``{Effects of adding natural
		sounds to urban noises on the perceived loudness of noise and soundscape
		quality},'' \emph{{Science of the Total Environment}},
	vol. 711, 2020.
	
	\bibitem{DIN456922009}
	{German Institute for Standardization}, \emph{{DIN 45692: Measurement technique
			for the simulation of the auditory sensation of sharpness}}.\hskip 1em plus
	0.5em minus 0.4em\relax Beuth Verlag GmbH, 2009.
	
	\bibitem{ISO532-12014}
	{International Organization for Standardization}, \emph{{ISO 532-1: Acoustics -
			Methods for calculating loudness - Part 1: Zwicker method}}, Geneva, 2014.
	
	\bibitem{Fastl2001}
	H.~Fastl and E.~Zwicker, \emph{{Psychoacoustics - Facts and Models}}, T.~S.
	Huang, M.~R. Schroeder, and T.~Kohonen, Eds.\hskip 1em plus 0.5em minus
	0.4em\relax Springer, 2001.
	
	\bibitem{EcmaInternational2020ECMA-418-2:2020Perception}
	{Ecma International}, \emph{{ECMA-418-2:2020 - Psychoacoustic metrics for ITT
			equipment - Part 2 (models based on human perception)}}, 1st~ed., Geneva,
	Switzerland, 2020.
	
	\bibitem{EcmaInternational2021ECMA-74Equipment}
	------, \emph{{ECMA-74 - Acoustics - Measurement of airborne noise emitted by
			information technology and telecommunications equipment}}, 19th~ed., Geneva,
	Switzerland, 2021.
	
	\bibitem{InternationalOrganizationforStandardization2016}
	{International Organization for Standardization}, \emph{{ISO 1996-1:2016
			Acoustics — Description , measurement and assessment of environmental noise
			— Part 1: Basic quantities and assessment procedures}}.\hskip 1em plus
	0.5em minus 0.4em\relax Geneva: International Organization for
	Standardization, 2016.
	
	\bibitem{SanMillan-Castillo2021MOSQITO:Education}
	R.~San Mill{\'{a}}n-Castillo, E.~Latorre-Iglesias, D.~Jim{\'{e}}nez-Caminero,
	J.~M. {\'{A}}lvarez-Jimeno, M.~Glesser, and S.~Wanty, ``{MOSQITO: An
		open-source and free toolbox for sound quality metrics in the industry and
		education},'' in \emph{{Proceedings of Inter-Noise
			2021}}, 2021.
	
	\bibitem{Flowers2021LookingClustering}
	C.~Flowers, F.~M. Le~Tourneau, N.~Merchant, B.~Heidorn, R.~Ferriere, and
	J.~Harwood, ``{Looking for the -scape in the sound: Discriminating
		soundscapes categories in the Sonoran Desert using indices and clustering},''
	\emph{{Ecological Indicators}}, vol. 127, 2021.
	
	\bibitem{Kohonen2001}
	T.~Kohonen, \emph{{Self-organizing maps}}.\hskip 1em plus 0.5em minus
	0.4em\relax Springer-Verlag Berlin Heidelberg, 2001.
	
	\bibitem{Ooi2021AutomationHead}
	\BIBentryALTinterwordspacing
	K.~Ooi, Y.~Xie, B.~Lam, and W.~S. Gan, ``{Automation of binaural headphone
		audio calibration on an artificial head},''
	\emph{{MethodsX}}, vol.~8, no. February, pp. 1--12,
	2021.
	\BIBentrySTDinterwordspacing
	
	\bibitem{Abeer2021USM-SEDScenarios}
	\BIBentryALTinterwordspacing
	J.~Abe{\ss}er, ``{USM-SED - A Dataset for Polyphonic Sound Event Detection in
		Urban Sound Monitoring Scenarios},'' 2021. [Online]. Available:
	\url{http://arxiv.org/abs/2105.02592}
	\BIBentrySTDinterwordspacing
	
	\bibitem{Walker2013AudiometryInterpretation}
	J.~J. Walker, L.~M. Cleveland, J.~L. Davis, and J.~S. Seales, ``{Audiometry
		screening and interpretation},'' \emph{{American Family
			Physician}}, vol.~87, no.~1, pp. 41--47, 2013.
	
	\bibitem{EchevarriaSanchez2017}
	\BIBentryALTinterwordspacing
	G.~M. Echevarria~Sanchez, T.~Van~Renterghem, K.~Sun, B.~De~Coensel, and
	D.~Botteldooren, ``{Using Virtual Reality for assessing the role of noise in
		the audio-visual design of an urban public space},''
	\emph{{Landscape and Urban Planning}}, vol. 167, pp.
	98--107, 2017.
	\BIBentrySTDinterwordspacing
	
	\bibitem{Wang2021ExtendedGroups}
	M.~Wang, Y.~Ai, Y.~Han, Z.~Fan, P.~Shi, and H.~Wang, ``{Extended high-frequency
		audiometry in healthy adults with different age groups},''
	\emph{{Journal of Otolaryngology - Head and Neck
			Surgery}}, vol.~50, no.~1, pp. 1--6, 2021.

	\bibitem{Mitchell2021b}
	M.~Andrew, T.~Oberman, F.~Aletta, M.~Kachlicka, M.~Lionello, M.~Erfanian, and
	J. Kang, ``{Investigating urban soundscapes of the COVID-19 lockdown: A 
		predictive soundscape modeling approach},''
	\emph{{The Journal of the Acoustical Society of America}}, vol.~150, no.~6, pp. 4474--4488, 2021.

	\bibitem{InternationalOrganizationforStandardization2014}
	{International Organization for Standardization}, \emph{{ISO 12913-1:2014 -
			Acoustics - Soundscape - Part 1: Definition and conceptual framework}}.\hskip
	1em plus 0.5em minus 0.4em\relax Geneva, Switzerland: International
	Organization for Standardization, 2014.
	
	\bibitem{Weinstein1978}
	N.~D. Weinstein, ``{Individual differences in reactions to noise: A
		longitudinal study in a college dormitory},''
	\emph{{Journal of Applied Psychology}}, vol.~63, no.~4,
	pp. 458--466, 1978.
	
	\bibitem{Cohen1983AStress}
	S.~Cohen, T.~Kamarck, and R.~Mermelstein, ``{A Global Measure of Perceived
		Stress},'' \emph{{Journal of Health and Social
			Behavior}}, vol.~24, no.~4, pp. 385--396, 1983.
	
	\bibitem{WorldHealthOrganization1998WHO-5Index}
	{WHO Regional Office for Europe}, \emph{{Wellbeing measures in primary health
			care}}, Cophenhagen, Denmark, 1998.
	
	\bibitem{Watson1988}
	D.~Watson, L.~A. Clark, and A.~Tellegan, ``{Development and
		Validation of Brief Measures of Positive and Negative Affect: The PANAS
		Scales},'' \emph{{Journal of Personality and Social
			Psychology}}, vol.~54, no.~6, pp. 1063--1070, 1988.
	
	\bibitem{Aletta2018}
	\BIBentryALTinterwordspacing
	F.~Aletta, \emph{{et~al.}}, ``{The relationship between
		noise sensitivity and soundscape appraisal of care professionals in their
		work environment: a case study in Nursing Homes in Flanders, Belgium},'' in
	\emph{{Proceedings of Euro-Noise 2018}}, 2018.
	
	\bibitem{Ratcliffe2021}
	E.~Ratcliffe, ``{Sound and Soundscape in Restorative Natural Environments: A
		Narrative Literature Review.}'' \emph{{Frontiers in
			Psychology}}, vol.~12, p. 570563, 2021.
	
	\bibitem{Masullo2021a}
	\BIBentryALTinterwordspacing
	M.~Masullo, \emph{{et~al.}}, ``{A questionnaire
		investigating the emotional salience of sounds},''
	\emph{{Applied Acoustics}}, vol. 182, p. 108281, 2021.
	\BIBentrySTDinterwordspacing
	
	\bibitem{Cohen1988PerceivedStates}
	\BIBentryALTinterwordspacing
	S.~Cohen and G.~Williamson, ``{Perceived stress in a probability sample of the
		United States},'' \emph{{The Social Psychology of
			Health}}, vol.~13, pp. 31--67, 1988.
	\BIBentrySTDinterwordspacing

	\bibitem{Mitchell2022}
	\BIBentryALTinterwordspacing
	A.~Mitchell, F.~Aletta, and J.~Kang, ``{How to analyse and represent
		quantitative soundscape data},'' \emph{{JASA Express
			 Letters}}, vol.~2, p. 037201, 2022.
	\BIBentrySTDinterwordspacing

	\bibitem{Guski2017}
	R.~Guski, D.~Schreckenberg, and R.~Schuemer, ``{A systematic review on
		environmental noise and annoyance},''
	\emph{{International Journal of Environmental Research
			and Public Health}}, vol.~14, no.~12, pp. 1--39, 2017.
	
	\bibitem{Hong2020NoiseAnnoyance}
	A.~Hong, B.~Kim, and M.~Widener, ``{Noise and the city: Leveraging crowdsourced
		big data to examine the spatio-temporal relationship between urban
		development and noise annoyance},'' \emph{{Environment
			and Planning B: Urban Analytics and City Science}}, vol.~47, no.~7, pp.
	1201--1218, 2020.
	
	\bibitem{VanRenterghem2002EffectWind}
	T.~Van~Renterghem and D.~Botteldooren, ``{Effect of a row of trees behind noise
		barriers in wind},'' \emph{{Acta Acustica united with
			Acustica}}, vol.~88, no.~6, pp. 869--878, 2002.
	
	\bibitem{Yang2005AcousticSpaces}
	W.~Yang and J.~Kang, ``{Acoustic comfort evaluation in urban open public
		spaces},'' \emph{{Applied Acoustics}}, vol.~66, no.~2,
	pp. 211--229, 2005.
	
	\bibitem{Fang2021SoundscapeParks}
	X.~Fang, \emph{{et~al.}}, ``{Soundscape Perceptions and
		Preferences for Different Groups of Users in Urban Recreational Forest
		Parks},'' \emph{{Forests}}, vol.~12, no.~4, p. 468,
	2021.
	
	\bibitem{Cronbach1951CoefficientTests}
	L.~J. Cronbach, ``{Coefficient alpha and the internal structure of tests},''
	\emph{{Psychometrika}}, vol.~16, no.~3, pp. 297--334,
	1951.
	
	\bibitem{McDonald1999TestTreatment}
	R.~P. McDonald, \emph{{Test Theory: A Unified Treatment}}.\hskip 1em plus 0.5em
	minus 0.4em\relax Mahwah, New Jersey: Lawrence Erlbaum Associates, Inc.,
	1999.
	
	\bibitem{Taber2018TheEducation}
	K.~S. Taber, ``{The Use of Cronbach’s Alpha When Developing and Reporting
		Research Instruments in Science Education},''
	\emph{{Research in Science Education}}, vol.~48, no.~6,
	pp. 1273--1296, 2018.
	
	\bibitem{Worthington2018WeinsteinWNSS}
	D.~Worthington, ``{Weinstein Noise Sensitivity Scale (WNSS)},'' in \emph{The
		Sourcebook of Listening Research: Methodology and Measures}, 1st~ed.,
	D.~Worthington and G.~Bodie, Eds.\hskip 1em plus 0.5em minus 0.4em\relax John
	Wiley and Sons Ltd, 2018, pp. 475--481.
	
	\bibitem{Ooi2022ProbablyAugmentation}
	K.~Ooi, K.~N. Watcharasupat, B.~Lam, Z.-T. Ong, and W.-S. Gan, ``{Probably
		Pleasant? A Neural-Probabilistic Approach to Automatic Masker Selection for
		Urban Soundscape Augmentation},'' in \emph{{Proceedings
			of IEEE ICASSP 2022}}, 2022, p.~5.
	
	\bibitem{Watcharasupat2022AutonomousGain}
	\BIBentryALTinterwordspacing
	K.~N. Watcharasupat, K.~Ooi, B.~Lam, T.~Wong, Z.-T. Ong, and W.-S. Gan,
	``{Autonomous In-Situ Soundscape Augmentation via Joint Selection of Masker
		and Gain},'' pp. 1--5, 2022. [Online]. Available:
	\url{http://arxiv.org/abs/2204.13883}
	\BIBentrySTDinterwordspacing
	
	\bibitem{Zou2005RegularizationNet}
	H.~Zou and T.~Hastie, ``{Regularization and variable selection via the elastic
		net},'' \emph{{Journal of the Royal Statistical Society.
			Series B: Statistical Methodology}}, vol.~67, no.~5, p. 768, 2005.
	
	\bibitem{Kingma2015Adam:Optimization}
	D.~P. Kingma and J.~L. Ba, ``{Adam: A method for stochastic optimization},'' in
	\emph{{Proceedings of ICLR 2015}}, 2015, pp. 1--15.
	
	\bibitem{Ma2021}
	K.~Ma, C.~Mak, and H.~Wong, ``{Effects of environmental sound quality on
		soundscape preference in a public urban space},''
	\emph{{Applied Acoustics}}, vol. 171, 2021.
	
	\bibitem{Kidd2003TheSounds}
	G.~R. Kidd and C.~S. Watson, ``{The perceptual dimensionality of environmental
		sounds},'' \emph{{Noise Control Engineering Journal}},
	vol.~51, no.~4, pp. 216--231, 2003.
	
\end{thebibliography}
% argument is your BibTeX string definitions and bibliography database(s)

\newpage
\newpage

\begin{IEEEbiography}[{\includegraphics[width=1in,height=1.25in,clip,keepaspectratio]{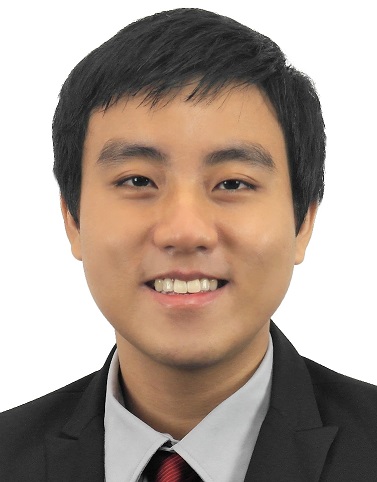}}]{Kenneth Ooi}
(S'21) received the B.Sc. (Hons.) degree in mathematical sciences from Nanyang Technological University (NTU), Singapore, in 2019, and received the Lee Kuan Yew Gold Medal as the top graduand of his cohort. He is currently pursuing the Ph.D. degree in electrical engineering in NTU under the supervision of Prof. Woon-Seng Gan. His research interests include deep learning for acoustic scene and event classification, as well as corresponding applications to the field of soundscape research.
\end{IEEEbiography}

\begin{IEEEbiography}[{\includegraphics[width=1in,height=1.25in,clip,keepaspectratio]{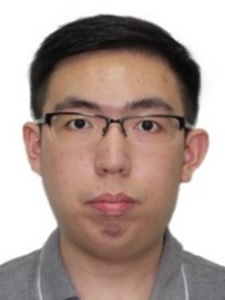}}]{Zhen-Ting Ong} received the B.Eng. degree in electrical and electronic engineering from Nanyang Technological University (NTU), Singapore, in 2016. He is currently a Research Engineer at the School of Electrical and Electronic Engineering, Nanyang Technological University (NTU), Singapore. Since 2016, he has published more than 30 works on soundscape engineering in international conferences and journals.
\end{IEEEbiography}

\begin{IEEEbiography}[{\includegraphics[width=1in,height=1.25in,clip,keepaspectratio]{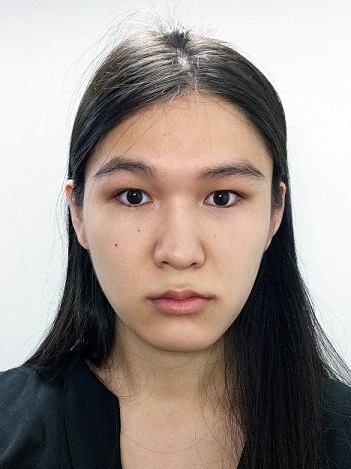}}]{Karn N. Watcharasupat} (S'19-GS'22) received the B.Eng. (Hons.) degree in electrical and electronic engineering, from Nanyang Technological University (NTU), Singapore, in 2022. She was a recipient of the Nanyang Scholarship (CN Yang Scholars Programme) and the Lee Kuan Yew Gold Medal for the Class of 2022.

She is currently pursuing a Ph.D. in music technology at the Music Informatics Group, Center for Music Technology (GTCMT), Georgia Institute of Technology, Atlanta, GA, USA, where she was also a visiting student (2020; 2021-2022 remote). At NTU, she was with the Media Technology Laboratory (2018-2020), and later the Digital Signal Processing Laboratory (2020-2022).  Her research interests are in signal processing, machine learning, and artificial intelligence for music and audio applications. She is a co-inventor of two patent applications and has published more than 20 papers in international conferences and journals on music information retrieval, soundscapes, spatial audio, speech enhancement, and blind source separation. 

She currently serves as the Treasurer for Women in Music Technology at Georgia Tech, and was a tech volunteer for the 22nd International Society for Music Information Retrieval Conference (ISMIR). She also serves as a reviewer for the IEEE International Conference on Acoustics, Speech and Signal Processing (ICASSP; 2022); the EURASIP Journal on Audio, Speech, and Music Processing (2022-); Digital Signal Processing (2022-); and Applied Acoustics (2022-). 
\end{IEEEbiography}

\begin{IEEEbiography}[{\includegraphics[width=1in,height=1.25in,clip,keepaspectratio]{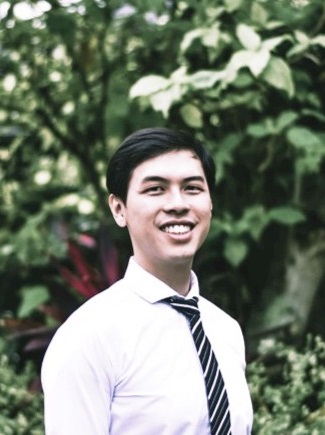}}]{Bhan Lam} (S'18-M'19) received the B.Eng. (Hons.) and Ph.D. degrees both from the School of Electrical and Electronic Engineering, Nanyang Technological University, Singapore, in 2013 and 2019, respectively. He was awarded the NTU Research Scholarship and EEE Graduate Award to undertake his PhD under the supervision of Prof. Woon-Seng Gan. In 2015, he was a visiting postgrad in the signal processing and control group at the Institute of Sound and Vibration Research, University of Southampton, UK. He was an invited representative at the 2020 Global Young Scientist Summit and an invited tutorial speaker at APSIPA ASC 2020.

He is currently a Research Assistant Professor at the School of Electrical and Electronic Engineering, NTU, Singapore. He has authored more than 60 refereed journal articles and conference papers in the areas of acoustics, soundscape, and signal processing for active control. His work on anti-noise windows has been patented (WO2022055432) and recognised as a top-100 paper in Nature Scientific Reports in 2020. He is currently a guest editor with MDPI Sustainability and served as a special session co-chair in the 2022 IEEE International Conference on Acoustics, Speech and Signal Processing (ICASSP). He was appointed by the Singapore Standards Council as the National Mirror Committee Chair of ISO TC43/SC1/WG54 on ``Perceptual assessment of soundscape quality''. His current research interests include 
active noise control, soundscape, and signal processing for active control. 
\end{IEEEbiography}

\begin{IEEEbiography}[{\includegraphics[width=1in,height=1.25in,clip,keepaspectratio]{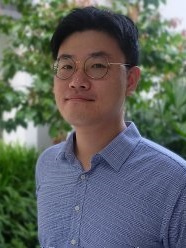}}]{Joo Young Hong}
received the B.Sc. degree in Architectural Engineering and Ph.D. degree in Architectural Acoustics from Hanyang University, Seoul, Korea in 2010 and 2015, respectively. He was invited as one of the Korean participants to the Global Young Scientist Summit @one-north 2015, Singapore. He also won the Best Paper prize at the INTER-NOISE conference in San Francisco, USA, 2015, sponsored by the International Institute of Noise Control Engineering. He was also invited as a plenary speaker at the Urban Noise Symposium, 2016, Shanghai to talk about the soundscape design approach. In 2017, he was a recipient of the prestigious Lee Kuan Yew Post-Doctoral Fellowship at the School of Electrical and Electronic Engineering, Nanyang Technological University, Singapore. In 2020, he was an Assistant Professor in Architecture and Sustainable Design at the Singapore University of Technology and Design. Since 2021, he has been an Assistant Professor at the Department of Architectural Engineering, Chungnam National University, Korea.

His research interests fall under the umbrella of soundscape research, which is a new paradigm that emphasizes a holistic perspective of perceived acoustic environments in a given context. He has investigated relationships between physical acoustic phenomena and human auditory perception in indoor and outdoor environments through multidisciplinary approaches from acoustics, psychology, architecture, and urban planning. He has been involved as an assistant acoustic consultant in several practical auditorium design projects. He has published 28 peer-reviewed articles in international journals and 2 invited book chapters. He has also served as a technical committee member of Korean Industrial Standards (KS) on building and environmental noises.
\end{IEEEbiography}

\begin{IEEEbiography}[{\includegraphics[width=1in,height=1.25in,clip,keepaspectratio]{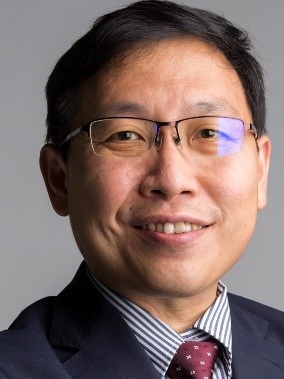}}]{Woon-Seng Gan}
(S’90–M’93-SM’00)
received his B.Eng. (1st Class Hons.) and Ph.D. degrees, both in electrical and electronic engineering from the University of Strathclyde, United Kingdom, in 1989 and 1993 respectively. He is currently a Professor of Audio Engineering and the Director of the Smart Nation Lab in the School of Electrical and Electronic Engineering in Nanyang Technological University (NTU). He also served as the Head of the Information Engineering Division in the School of Electrical and Electronic Engineering in NTU (2011-2014), and the Director of the Centre for Infocomm Technology (2016-2019). His research concerns the connections between the physical world, signal processing and sound control, which has resulted in the practical demonstration and licensing of spatial audio algorithms, directional sound beam, and active noise control for headphones and open windows.

Prof.\ Gan has published more than 400 international refereed journals and conferences, and has translated his research into 6 granted patents.
He has co-authored three books on Subband Adaptive Filtering: Theory and Implementation (John Wiley, 2009); Embedded Signal Processing with the Micro Signal Architecture, (Wiley-IEEE, 2007); and Digital Signal Processors: Architectures, Implementations, and Applications (Prentice Hall, 2005).
In 2017, he won the APSIPA Sadaoki Furui Prize Paper Award.
He is a Fellow of the Audio Engineering Society (AES), a Fellow of the Institute of Engineering and Technology (IET), and a Senior Member of the IEEE. He served as an Associate Editor of the IEEE/ACM Transactions on Audio, Speech, and Language Processing (TASLP; 2012-15) and was presented with an Outstanding TASLP Editorial Board Service Award in 2016. He also served as the Associate Editor for the IEEE Signal Processing Letters (2015-19). He is currently serving as a Senior Area Editor of the IEEE Signal Processing Letters (2019-); Associate Technical Editor of the Journal of Audio Engineering Society (JAES; 2013-); Editorial member of the Asia Pacific Signal and Information Processing Association (APSIPA; 2011-) Transaction on Signal and Information Processing; Associate Editor of the EURASIP Journal on Audio, Speech and Music Processing (2007-). He served as the Technical Program Chair of the 2022 IEEE International Conference on Acoustics, Speech and Signal Processing (ICASSP), held in Singapore.
\end{IEEEbiography}

% You can push biographies down or up by placing
% a \vfill before or after them. The appropriate
% use of \vfill depends on what kind of text is
% on the last page and whether or not the columns
% are being equalized.

\vfill % Comment if not needed

% Can be used to pull up biographies so that the bottom of the last one
% is flush with the other column.
\enlargethispage{-4in}

% insert where needed to balance the two columns on the last page with
% biographies
%\newpage

% Uncomment this section to also output appendices (but without ``Appendixes'' heading)
\onecolumn

\newpage

\appendices

\renewcommand{\thefigure}{\thesection.\arabic{figure}}
\renewcommand{\thetable}{\thesection.\arabic{table}}
\setcounter{figure}{0}
\setcounter{table}{0}

\section{Source Files for ARAUS Dataset}
\label{sec:Details of source files used for augmented soundscapes in ARAUS dataset}

\begin{figure*}[!ht]
	\centering
	\includegraphics[width=0.60\linewidth]{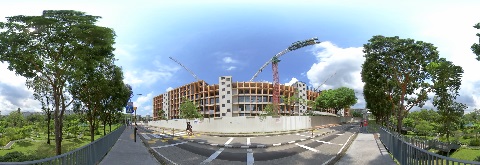}
	\includegraphics[width=0.60\linewidth]{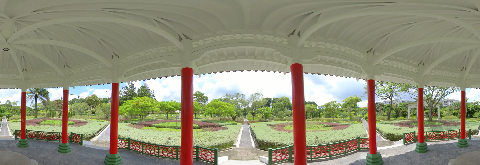}
	\includegraphics[width=0.60\linewidth]{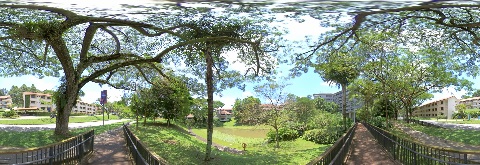}
	\includegraphics[width=0.60\linewidth]{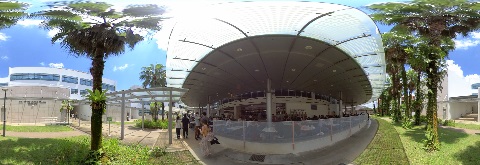}
	\includegraphics[width=0.60\linewidth]{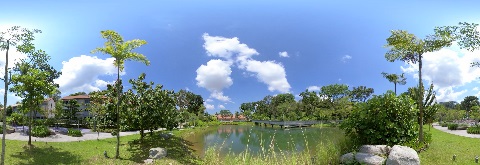}
	\includegraphics[width=0.60\linewidth]{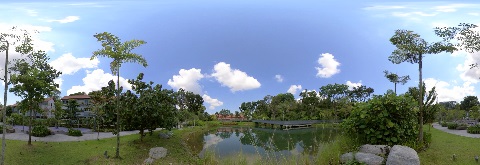}
	\caption{Panoramic photographs of locations where recordings for test set of ARAUS dataset were made, with coordinates specified as (latitude, longitude) pairs. From top to bottom: (a) A road facing a construction site (1.342123, 103.683790), (b) A gazebo in a park (1.342780, 103.684824), (c) A walkway facing a lake (1.346492, 103.687056), (d) A walkway facing a crowded canteen (1.342259, 103.682418), (e) A path facing a lake (1.344954, 103.684667), (f) A path facing a lake with an aircraft flying overhead (1.344954, 103.684667).}
	\label{fig:test-set-locations}
\end{figure*}

This appendix contains details of the exact source files used to generate the augmented soundscapes in the ARAUS dataset. Table \ref{tab:soundscapes} shows the breakdown (by fold) of the soundscape recordings from the Urban Sounds of the World (USotW) database used as base soundscapes for the five-fold cross-validation set of the ARAUS dataset. The first half of R0091 was used as the ``pre-experiment'', ``attention'', and ``post-experiment'' stimulus, so R0091 is not considered to be part of any fold in the ARAUS dataset. Table \ref{tab:maskers} shows the breakdown (by fold) of the source recordings from Freesound and Xeno-Canto used as maskers for both the five-fold cross-validation set and independent test set. Panoramic photos and GPS coordinates of the locations recorded for the independent test set of the ARAUS dataset are shown in \Cref{fig:test-set-locations}.

\begin{table}[h] 
	\centering
	\caption{List of Urban Soundscape of the World (USotW) recordings used for cross-validation set of ARAUS dataset. Recordings are denoted using their index numbers in the USotW database. Single asterisks ($^{*}$) and daggers ($^{\dagger}$) respectively denote recordings whose first and second halves were omitted from the ARAUS dataset. As of publication time, R0021, R0077, R0086, R0093, R0100, R0102 are not listed in the USotW database.}
	\label{tab:soundscapes}
		\begin{tabularx}{\textwidth}{ X *{13}{X} }
		\toprule
		Fold & \multicolumn{4}{l}{USotW identifiers} \\
		\midrule
		Fold 1	&	R0005	&	R0011	&	R0017	&	R0018	&	R0027	&	R0046	&	R0048	&	R0055	&	R0056	&	R0071	&	R0079	&	R0080	\\	
		&	R0081	&	R0084$^{*}$	&	R0089	&	R0092	&	R0095	&	R0111$^{\dagger}$	&	R0115	&	R0116	&	R0122	&	R0124$^{*}$	&	R0126	&	R0129	\\	\midrule
		Fold 2	&	R0001	&	R0006	&	R0016	&	R0020	&	R0025	&	R0026	&	R0033	&	R0037	&	R0040	&	R0044	&	R0052	&	R0053	\\	
		&	R0067	&	R0076	&	R0078	&	R0082	&	R0088	&	R0103	&	R0106	&	R0107	&	R0110	&	R0117	&	R0123	&	R0133	\\	
		\midrule
		Fold 3	&	R0009	&	R0012	&	R0024$^{*}$	&	R0028	&	R0029	&	R0032	&	R0034	&	R0041	&	R0042	&	R0045	&	R0051	&	R0057	\\	
		&	R0059	&	R0063	&	R0065	&	R0070	&	R0085$^{\dagger}$	&	R0094	&	R0104	&	R0109	&	R0114	&	R0118	&	R0120	&	R0132	\\	
		\midrule
		Fold 4	&	R0002	&	R0003	&	R0004	&	R0013	&	R0015	&	R0019	&	R0022	&	R0031	&	R0039	&	R0043	&	R0047$^{\dagger}$	&	R0050	\\	
		&	R0054	&	R0061	&	R0072	&	R0074	&	R0087	&	R0097	&	R0098	&	R0099	&	R0112	&	R0113	&	R0127	&	R0131	\\	
		\midrule
		Fold 5	&	R0007	&	R0008	&	R0010	&	R0023	&	R0030	&	R0035	&	R0036	&	R0038	&	R0049	&	R0058	&	R0060	&	R0062	\\	
		&	R0064	&	R0069	&	R0073	&	R0075	&	R0090	&	R0096	&	R0105	&	R0108	&	R0119	&	R0121	&	R0128	&	R0130	\\	
		\bottomrule
		\end{tabularx}
	
\end{table}

\begin{table*}[t]
    \centering
    \caption{List of tracks from Freesound (denoted as ``FS'') and Xeno-canto (denoted as ``XC'') used as maskers in ARAUS dataset. Numbers after ``FS'' and ``XC'' denote the index numbers of the tracks in the Freesound and Xeno-canto databases, respectively.}
    \label{tab:maskers}
    \newcommand{\thistablescale}{0.9}
    \def\arraystretch{\thistablescale}
    \begin{tabularx}{\thistablescale\textwidth}{>{\centering\arraybackslash}X *{7}{>{\centering\arraybackslash}X}}
        \toprule
        & \multicolumn{7}{c}{Masker track identifier}\\
        \cmidrule(lr){2-8}
        Fold & \multicolumn{2}{c}{Bird} & Construction & Traffic & \multicolumn{2}{c}{Water} & Wind \\
        \midrule
        Test & XC568124 & XC640568 & FS586168 & FS587219 & FS587000 & FS587759 & FS587205 \\ 
        \midrule
        1 & XC109203 & XC482053 & FS218748 & FS84646 & FS202915 & FS463265 & FS181255 \\ 
        & XC134886 & XC503057 & FS246171 & FS235531 & FS345649 & FS516934 & FS244942 \\ 
        & XC184374 & XC518767 & FS400991 & FS243720 & FS376801 & FS541717 & FS403051 \\ 
        & XC185560 & XC518843 & FS421050 & FS330288 & FS410927 & FS547136 & FS444921 \\ 
        & XC311306 & XC556166 & FS455683 & FS337095 & FS412308 & FS547892 & FS454373 \\ 
        & XC370500 & XC571000 & FS553476 & FS426886 & FS415151 & FS548476 & FS483076 \\ 
        & XC419391 & XC612865 & FS555037 & FS454864 & FS433589 & FS550930 & FS546527 \\ 
        & XC470089 & XC613796 & FS555039 & FS504138 & FS450755 & FS553051 & FS548173 \\ 
        \midrule
        2 & FS257445 & XC477488 & FS74505 & FS67704 & FS260056 & FS441152 & FS144083 \\ 
        & FS317450 & XC481933 & FS134896 & FS77016 & FS336848 & FS457565 & FS185070 \\ 
        & XC85417 & XC509271 & FS193351 & FS160015 & FS346641 & FS459983 & FS242064 \\ 
        & XC133059 & XC537855 & FS194866 & FS191350 & FS365915 & FS533932 & FS422579 \\ 
        & XC301810 & XC553169 & FS522584 & FS322231 & FS400402 & FS534124 & FS423800 \\ 
        & XC332810 & XC562095 & FS553475 & FS370009 & FS401277 & FS544183 & FS444852 \\ 
        & XC370485 & XC600722 & FS555025 & FS433869 & FS411509 & FS550756 & FS475448 \\ 
        & XC420908 & XC608496 & FS555040 & FS448092 & FS423622 & FS552457 & FS540192 \\ 
        \midrule
        3 & FS478637 & XC566219 & FS171400 & FS93329 & FS56771 & FS388706 & FS62056 \\ 
        & XC122469 & XC575300 & FS289478 & FS118046 & FS169181 & FS414762 & FS73714 \\ 
        & XC137556 & XC591803 & FS335963 & FS160002 & FS185062 & FS415027 & FS346106 \\ 
        & XC140239 & XC598469 & FS361777 & FS173154 & FS243629 & FS436813 & FS397947 \\ 
        & XC350943 & XC601752 & FS376405 & FS336649 & FS249485 & FS537986 & FS405561 \\ 
        & XC485841 & XC602571 & FS383442 & FS394689 & FS256009 & FS543579 & FS423914 \\ 
        & XC552316 & XC602677 & FS545183 & FS439218 & FS329680 & FS546236 & FS438857 \\ 
        & XC554222 & XC603552 & FS555038 & FS546474 & FS384276 & FS551989 & FS491296 \\ 
        \midrule
        4 & XC50850 & XC480035 & FS62394 & FS65807 & FS260980 & FS478439 & FS104875 \\ 
        & XC112734 & XC570338 & FS149777 & FS84645 & FS264598 & FS511098 & FS131032 \\ 
        & XC368749 & XC579905 & FS169098 & FS149814 & FS396056 & FS542260 & FS180025 \\ 
        & XC375911 & XC580986 & FS170873 & FS252216 & FS438925 & FS544838 & FS182837 \\ 
        & XC445899 & XC600707 & FS173106 & FS274830 & FS448766 & FS546107 & FS211820 \\ 
        & XC449522 & XC601772 & FS192176 & FS434302 & FS454283 & FS546804 & FS345681 \\ 
        & XC457222 & XC604251 & FS483556 & FS463636 & FS459850 & FS548381 & FS457318 \\ 
        & XC467315 & XC611766 & FS555035 & FS512142 & FS460441 & FS553159 & FS546526 \\ 
        \midrule
        5 & XC203258 & XC489739 & FS112637 & FS23273 & FS62395 & FS489073 & FS84111 \\ 
        & XC242969 & XC505573 & FS118042 & FS110309 & FS167034 & FS536223 & FS104952 \\ 
        & XC255139 & XC545465 & FS328141 & FS180156 & FS169250 & FS547012 & FS441866 \\ 
        & XC348370 & XC555184 & FS362068 & FS259629 & FS261445 & FS547232 & FS469280 \\ 
        & XC376468 & XC578478 & FS384837 & FS261344 & FS352902 & FS547720 & FS533930 \\ 
        & XC450847 & XC600292 & FS488828 & FS262830 & FS365919 & FS549929 & FS546759 \\ 
        & XC475279 & XC604437 & FS555026 & FS503456 & FS376709 & FS552691 & FS548235 \\ 
        & XC482274 & XC614063 & FS555034 & FS551451 & FS469009 & FS553135 & FS551318 \\ 
        \midrule
    \end{tabularx}
\end{table*}

\setcounter{figure}{0}
\setcounter{table}{0}

\FloatBarrier

\vspace*{-9.5mm}

\section{Participant Information Questionnaire}
\label{sec:Appendix/Participant Information Questionnaire}

This appendix lists the all the items for the participant information questionnaire (PIQ), which contains demographic and auxiliary information that we collected from the participants who contributed their responses to the ARAUS dataset. 

The questionnaire was administered to all participants via a digital form. All questions were multiple-choice questions except for Question 2(a), which asked for the age of the participant and accepted positive integers. Possible options for the participants and numerical values used to code the responses for the dataset are shown in itemized square brackets. Yes/no questions are indicated by [Y/N] at the end of the questions. ``No'' is coded as 0 and ``Yes'' is coded as 1.

\vspace*{-5.5mm}

\renewcommand{\dotfill}{%
  \leavevmode\cleaders\hbox to 1.00em{\hss $\cdot$\hss }\hfill\kern0pt }

\vspace*{6mm}

\begin{mdframed}
\begin{enumerate}
	\item Spoken languages
	\begin{enumerate}
		\item Do you speak fluently in any languages/dialects other than English? \hfill[Y/N]\\
		\textit{If your response is ``No'', please go to Question 2.}
		\item Is English your first or native language? \hfill[Y/N]
		\item Among the languages/dialects you speak, would you consider yourself to be most fluent in English? \hfill[Y/N]
	\end{enumerate}
\end{enumerate}
\end{mdframed}
\begin{mdframed}
\begin{enumerate}[start=2]
	\item Demographic information
	\begin{enumerate}
		\item What is your age?
		\item What is your gender?\\
		\begin{enumerate*}[label={[\arabic*]}, start=0, itemjoin=; \qquad]
			\item Male
			\item Female
			\item[{[0 or 1]}] Other/prefer not to say\footnote{Due to the risk of identification considering a very small number of participants chose this option, where participants responded ``Other/prefer not to say'', we randomly coded the response as 0 or 1, each with a probability of \SI{50}{\percent}.}
		\end{enumerate*}
		\item What is your ethnic group?\\
		\begin{enumerate*}[label={[\arabic*]}, start=0, itemjoin=; \qquad]
			\item Other
			\item Chinese
			\item Malay
			\item Indian
		\end{enumerate*}
		\item What is the highest level of education you have \textit{completed}?\\
		\begin{enumerate*}[label={[\arabic*]}, start=0, itemjoin=;\newline]
		    \item Other
		    \item No qualification
		    \item Primary (PSLE), elementary school or equivalent
		    \item Secondary (GCE `N' \& `O' level), middle school or equivalent
		    \item Institute of Technical Education or equivalent
		    \item Junior College (`A' level), high school or equivalent
		    \item Polytechnic and Arts Institution (Diploma level) or equivalent
		    \item University (Bachelor's Degree) or equivalent
		    \item University (Master's Degree) or equivalent
		    \item University (PhD)
		\end{enumerate*}
		\item What is your occupational status?\\
		\begin{enumerate*}[label={[\arabic*]}, start=0, itemjoin=; \qquad]
			\item Other 
			\item Student
			\item Employed 
			\item Retired
			\item Unemployed
		\end{enumerate*}\\
		\textit{If your response is ``Student'', please go to Question 2(g)}
		\item What is the highest level of education you are \textit{currently} undergoing?\\
		\begin{enumerate*}[label={[\arabic*]}, start = 0, itemjoin=;\newline]
		    \item Other
		\end{enumerate*}\\
		\begin{enumerate*}[label={[\arabic*]}, start = 2, itemjoin=;\newline]
		    \item Primary (PSLE), elementary school or equivalent
		    \item Secondary (GCE `N' \& `O' level), middle school or equivalent
		    \item Institute of Technical Education or equivalent
		    \item Junior College (`A' level), high school or equivalent
		    \item Polytechnic and Arts Institution (Diploma level) or equivalent
		    \item University (Bachelor's Degree) or equivalent
		    \item University (Master's Degree) or equivalent
		    \item University (PhD)
		\end{enumerate*}
		\item What dwelling type is your current \textit{main} residence in Singapore?\\
		\begin{enumerate*}[label={[\arabic*]}, start=0, itemjoin=;\newline]
		    \item Other
		    \item Housing Development Board (HDB) flat or other public apartment
		    \item Hall of Residence or other student dormitory
		    \item Landed property
		    \item Condominium or other private apartment
		\end{enumerate*}
		\item Are you a Singapore citizen? \hfill [Y/N]
		\item Have you resided in Singapore for more than 10 years? \hfill [Y/N]
	\end{enumerate}
\end{enumerate}
\end{mdframed}
\begin{mdframed}
\begin{enumerate}[start=3]
	\item How much has indoor/outdoor noise bothered, disturbed, or annoyed you over the past 12 months?\\\\
	\parbox{6.5em}{Not at all [0]} 
	\dotfill 11-point Likert \dotfill 
	\parbox{6.5em}{\hfill [10] Extremely}\\
\end{enumerate}
\end{mdframed}
\begin{mdframed}
\begin{enumerate}[start=4]
	\item How would you describe your satisfaction of the overall quality of the acoustic environment in Singapore?\\\\
	\parbox{12em}{Extremely dissatisfied [0]}
	\dotfill 11-point Likert \dotfill 
	\parbox{12em}{\hfill [10] Extremely satisfied}\\
\end{enumerate}
\end{mdframed}
\begin{mdframed}
\begin{enumerate}[start=5]
	\item \textbf{Shortened Weinstein Noise Sensitivity Scale}\footnote[2]{Items marked with single asterisks (*) were reverse coded. In other words, ``Strongly disagree'' was coded as ``5'' and ``Strongly agree'' was coded as ``1'' for these items. The asterisks were not shown in the digital form presented to the participants.}: Below are a number of statements addressing individual reactions to noise. After reading each statement, please select the option that best represents your level of agreement with the statement.\\\\
	\begin{enumerate*}[label={[\arabic*]}, start=1, itemjoin=; \hfill\quad]
		\item Strongly disagree
		\item Disagree
		\item Neither agree nor disagree
		\item Agree
		\item Strongly agree
	\end{enumerate*}\\
	\begin{enumerate}
		\item I complain once I have run out of patience with a noise.
		\item I am annoyed even by low noise levels.
		\item I would not want to live on a noisy street, even if the house was nice.
		\item I get mad at people who make noise that keeps me from falling asleep or getting work done.
		\item I cannot fall asleep easily when there is noise.
		\item I am sensitive to noise.
		\item I am easily awakened by noise.
		\item I get used to most noises without much difficulty.*
		\item I find it hard to relax in a place that's noisy.
		\item I'm good at concentrating no matter what is going on around me.*
	\end{enumerate}
\end{enumerate}
\end{mdframed}
\begin{mdframed}
\begin{enumerate}[start=6]
	\item \textbf{Shortened Perceived Stress Scale}: The questions in this scale ask you about your feelings and thoughts during the last month. In each case, you will be asked to indicate how often you felt or thought a certain way. Although some of the questions are similar, there are differences between them and you should treat each one as a separate question. The best approach is to answer each question fairly quickly. That is, don't try to count up the number of times you felt a particular way, but rather indicate the alternative that seems like a reasonable estimate.\\\\
	\begin{enumerate*}[label={[\arabic*]}, start=0, itemjoin=; \hfill\qquad]
		\item Never 
		\item Almost never 
		\item Sometimes 
		\item Fairly often 
		\item Very often
	\end{enumerate*}\\\\
	In the last month, how often have you...
	\begin{enumerate}
	    \item been upset because of something that happened unexpectedly?
		\item felt that you were unable to control the important things in your life?
		\item felt nervous and ``stressed''?
		\item felt confident about your ability to handle your personal problems?
		\item felt that things were going your way?
		\item found that you could not cope with all the things that you had to do?
		\item been able to control irritations in your life?
		\item felt that you were on top of things?
		\item been angered because of things that were outside of your control?
		\item felt difficulties were piling up so high that you could not overcome them?
	\end{enumerate}
\end{enumerate}
\end{mdframed}

\clearpage
\begin{mdframed}
\begin{enumerate}[start=7]
	\item \textbf{WHO-5 Well Being Index}: For each of the statements below, which is the closest to how you have been feeling over the last two weeks?\\\\
	\begin{enumerate*}[label={[\arabic*]}, start=0, itemjoin=; \qquad]
		\item At no time
		\item Some of the time
		\item Less than half of the time
		\item More~than~half~of~the~time
		\item Most of the time
		\item All of the time
	\end{enumerate*}\\
	\begin{enumerate}
		\item I have felt cheerful and in good spirits.
		\item I have felt calm and relaxed.	
		\item I have felt active and vigorous.
		\item I woke up feeling fresh and rested.
		\item My daily life has been filled with things that interest me.
	\end{enumerate}
\end{enumerate}
\end{mdframed}
\begin{mdframed}
\begin{enumerate}[start=8]	
	\item \textbf{Positive and Negative Affect Schedule}\footnote[3]{Items corresponding to Positive Affect (marked with the oplus symbol $^\oplus$) and Negative Affect (marked with the ominus symbol $^\ominus$) were tallied to give separate Positive Affect and Negative Affect scores. The oplus and ominus symbols were not shown in the digital form presented to the participants.}: In the last two weeks, to what extent have you felt this way?\\\\
	\begin{enumerate*}[label={[\arabic*]}, start=1, itemjoin=; \qquad]
		\item Very slightly or not at all
		\item A little 
		\item Moderately 
		\item Quite a bit
		\item Extremely
	\end{enumerate*}\\\\
    \begin{enumerate*}[itemjoin=\qquad, label={\parbox{1.2em}{\alph*)}}]
		\item \parbox{7em}{Interested$^\oplus$  }
		\item \parbox{7em}{Distressed$^\ominus$ }
		\item \parbox{7em}{Excited$^\oplus$     }
		\item \parbox{7em}{Upset$^\ominus$      }
		\item \parbox{7em}{Strong$^\oplus$      }
		\item \parbox{7em}{Guilty$^\ominus$     }
		\item \parbox{7em}{Scared$^\ominus$     }
		\item \parbox{7em}{Hostile$^\ominus$    }
		\item \parbox{7em}{Enthusiastic$^\oplus$}
		\item \parbox{7em}{Proud$^\oplus$       }
		\item \parbox{7em}{Irritable$^\ominus$  }
		\item \parbox{7em}{Alert$^\oplus$       }
		\item \parbox{7em}{Ashamed$^\ominus$    }
		\item \parbox{7em}{Inspired$^\oplus$    }
		\item \parbox{7em}{Nervous$^\ominus$    }
		\item \parbox{7em}{Determined$^\oplus$  }
		\item \parbox{7em}{Attentive$^\oplus$   }
		\item \parbox{7em}{Jittery$^\ominus$    }
		\item \parbox{7em}{Active$^\oplus$      }
		\item \parbox{7em}{Afraid$^\ominus$     }
	\end{enumerate*}
\end{enumerate}
\end{mdframed}

\setcounter{figure}{0}
\setcounter{table}{0}

\FloatBarrier
\clearpage
\section{Distributions and Statistical Test Results for PIQ and ARQ}
\label{sec:Appendix/Distributions and Statistical Test Results for PIQ and ARQ}

\begin{figure*}[h]
	\centering
	\includegraphics[width=0.32\linewidth]{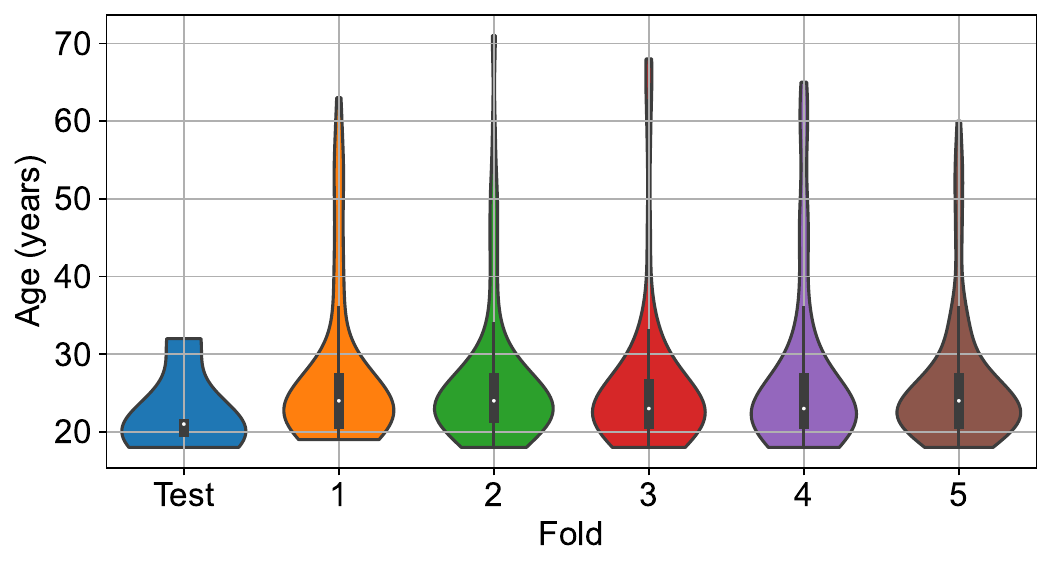}
	\includegraphics[width=0.32\linewidth]{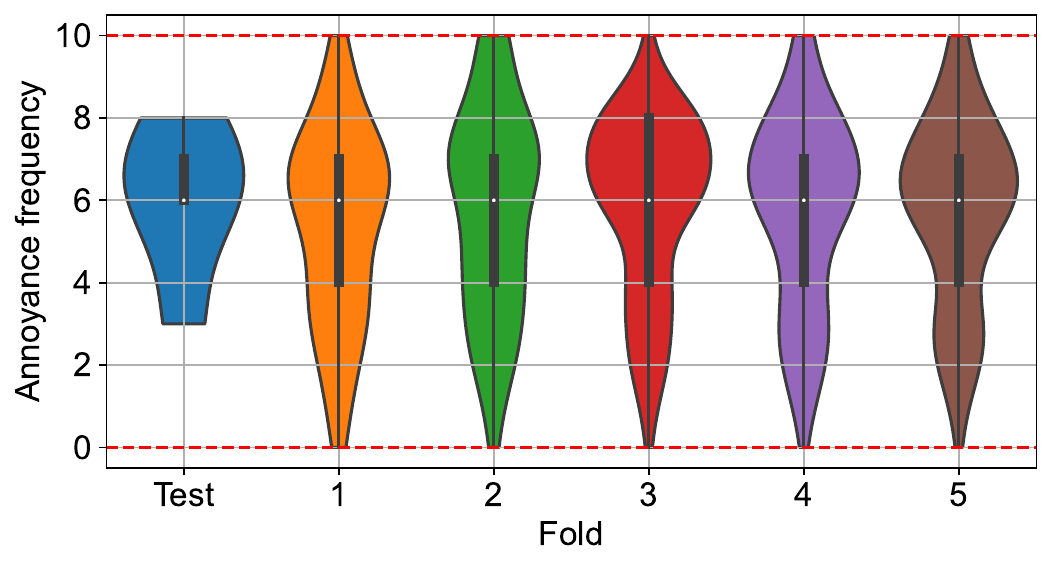}
	\includegraphics[width=0.32\linewidth]{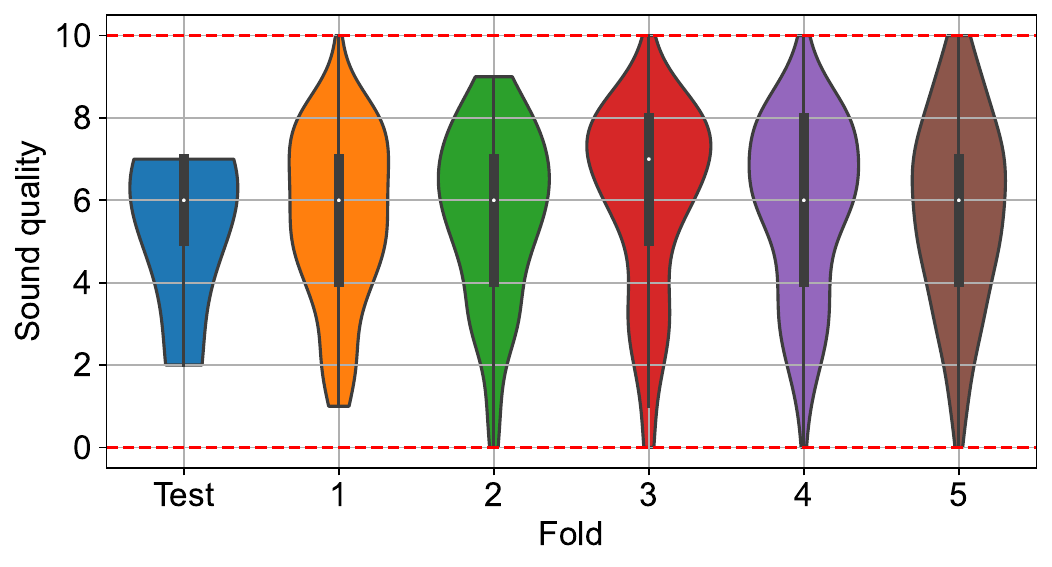}
	\includegraphics[width=0.32\linewidth]{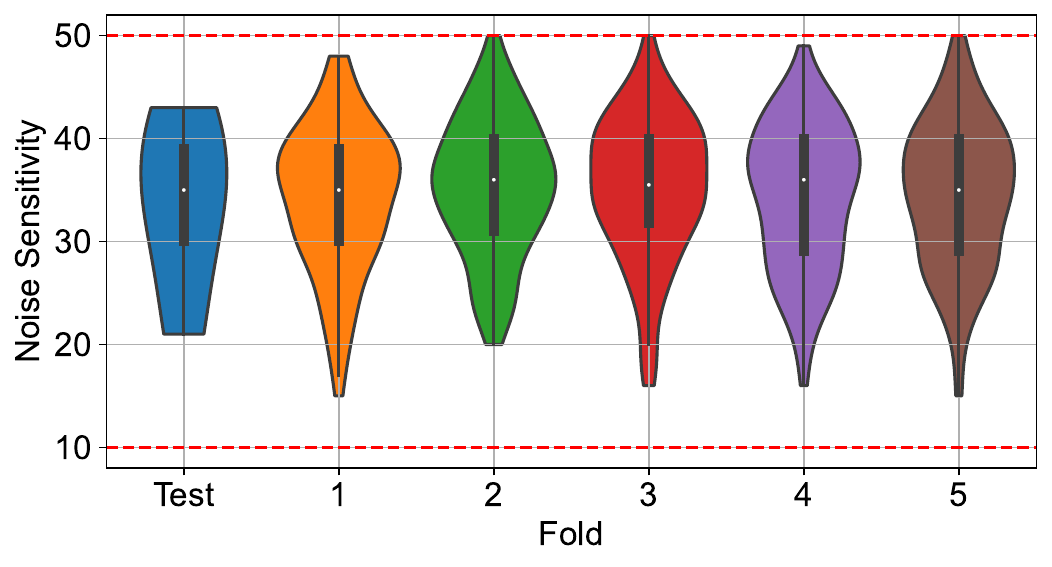}
	\includegraphics[width=0.32\linewidth]{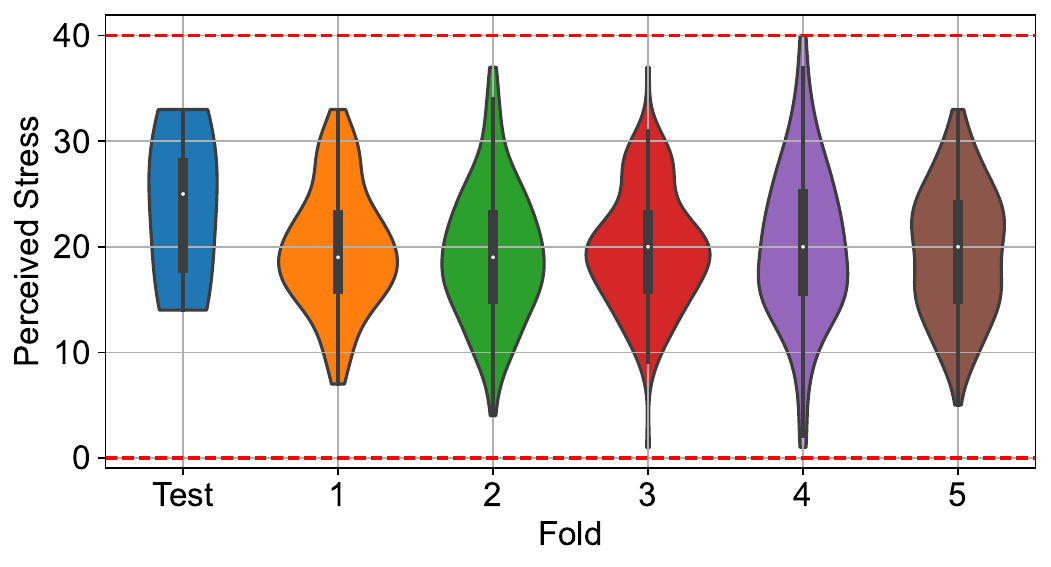}
	\includegraphics[width=0.32\linewidth]{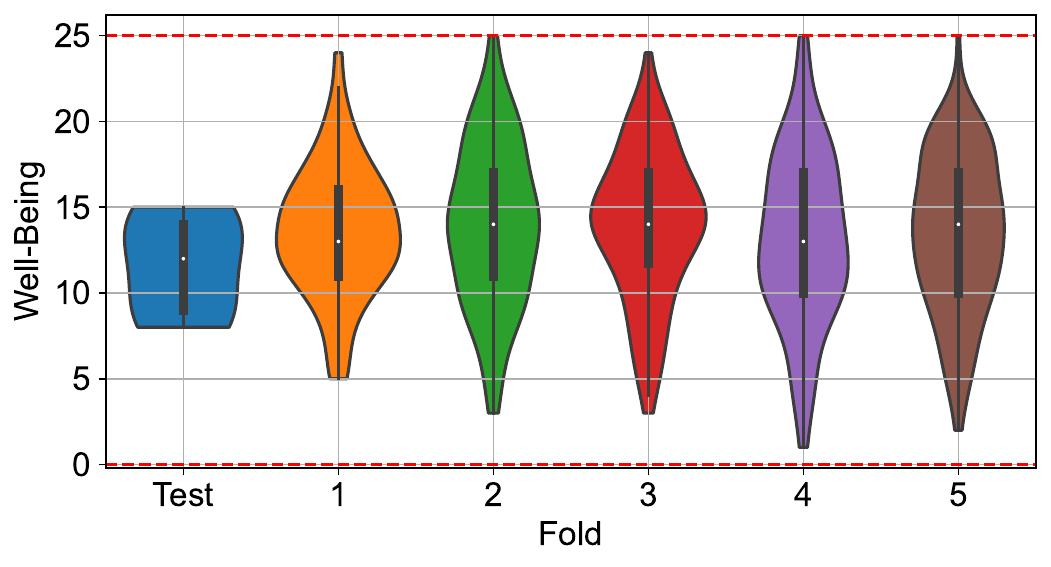}
	\includegraphics[width=0.32\linewidth]{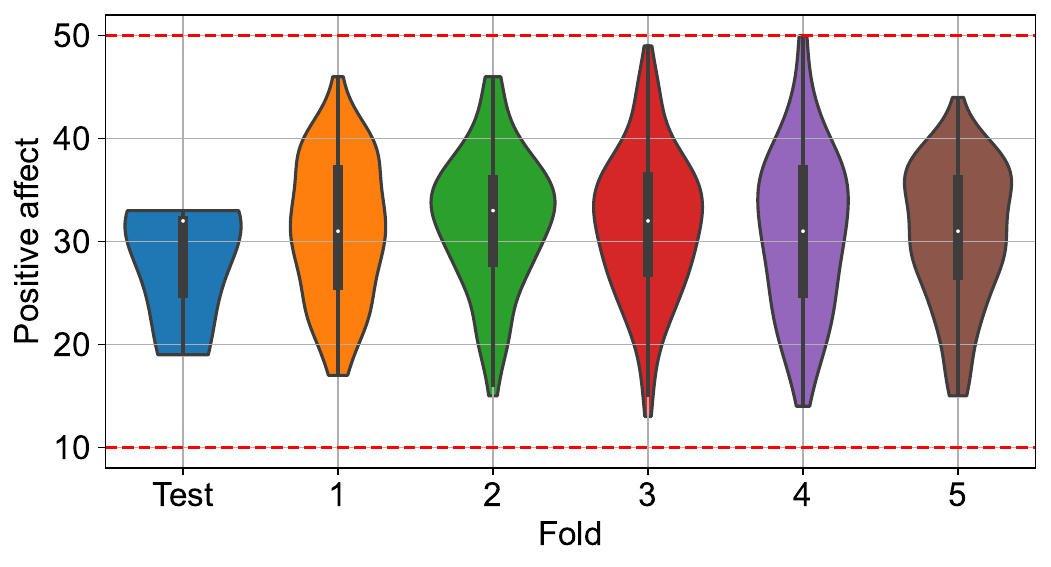}
	\includegraphics[width=0.32\linewidth]{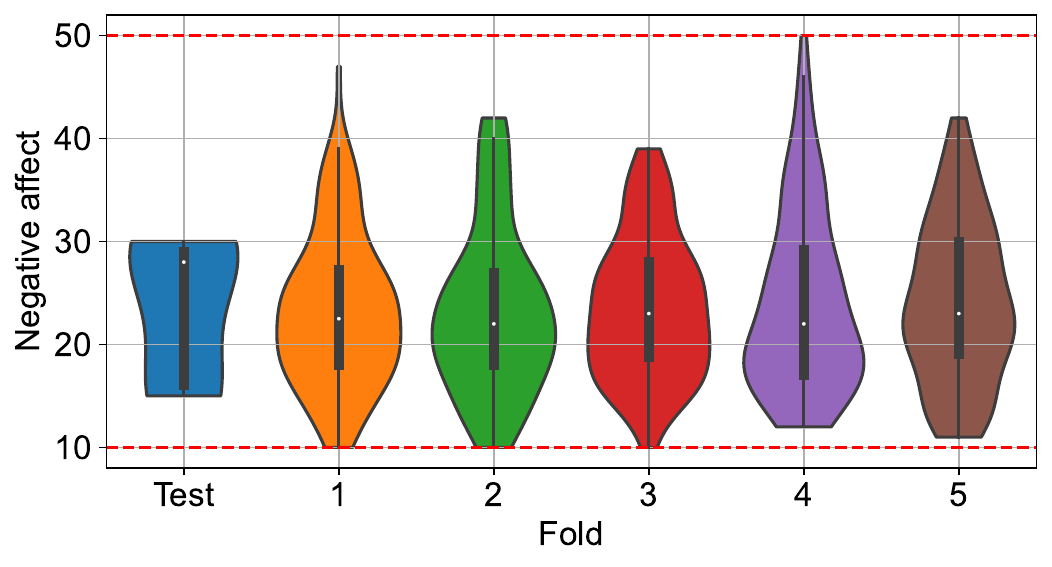}
	\caption{Violin plots of distributions of PIQ responses coded as continuous variables, by fold: (a) Age, (b) Extent annoyed by noise over past 12 months, (c) Satisfaction of overall acoustic environment in Singapore, (d) Score on WNSS-10, (e) Score on PSS-10, (f) Score on WHO-5, (g) Score on PANAS for Positive Affect, (h) Score on PANAS for Negative Affect. Red horizontal dotted lines denote the maximum and minimum values theoretically attainable by the variables. Kruskal-Wallis tests found no significant differences ($p>0.05$) between distributions of all variables by fold.}
	\label{fig:continuous_responses}
\end{figure*}

\begin{landscape}
\begin{figure*}[h]
	\centering
	\newcommand{\thisfigurescale}{0.28} 
	\includegraphics[width=\thisfigurescale\linewidth]{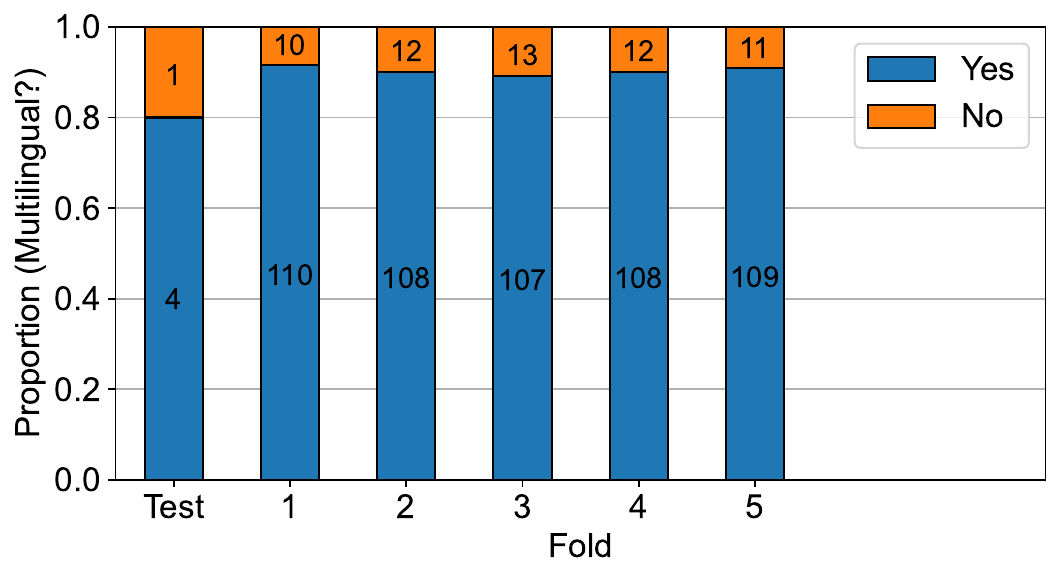}
	\includegraphics[width=\thisfigurescale\linewidth]{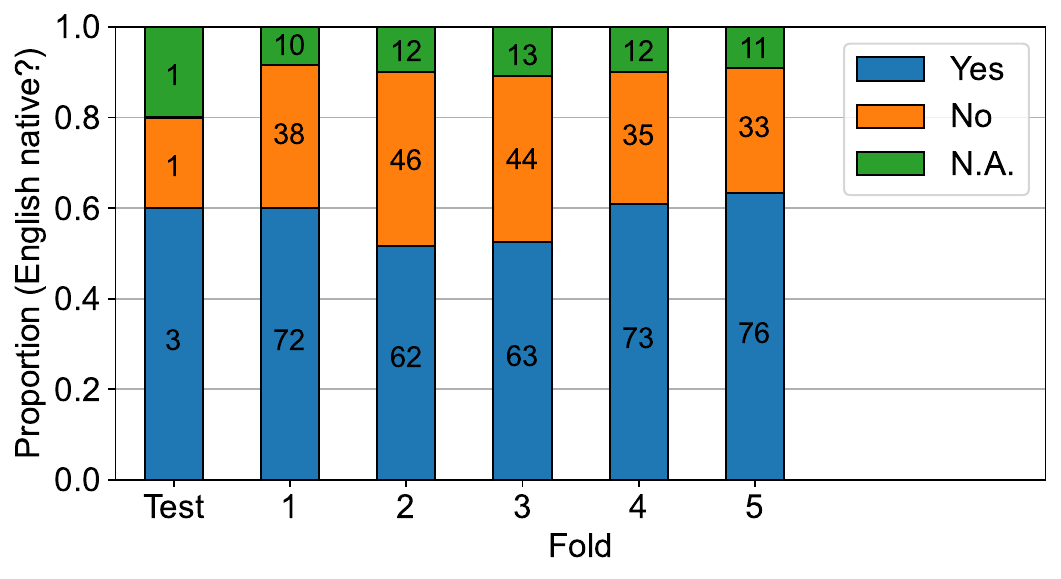}
	\includegraphics[width=\thisfigurescale\linewidth]{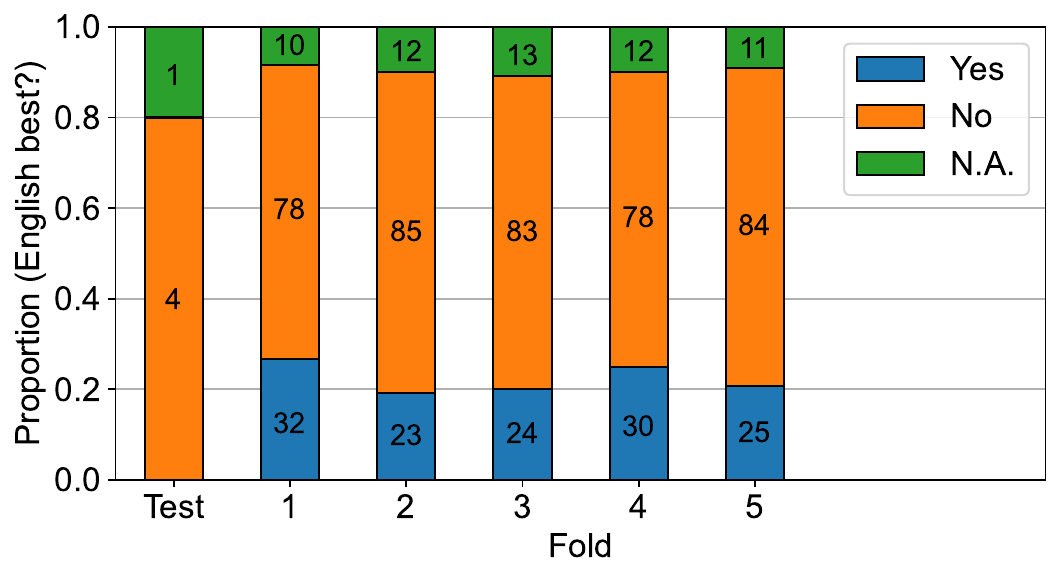}
	\includegraphics[width=\thisfigurescale\linewidth]{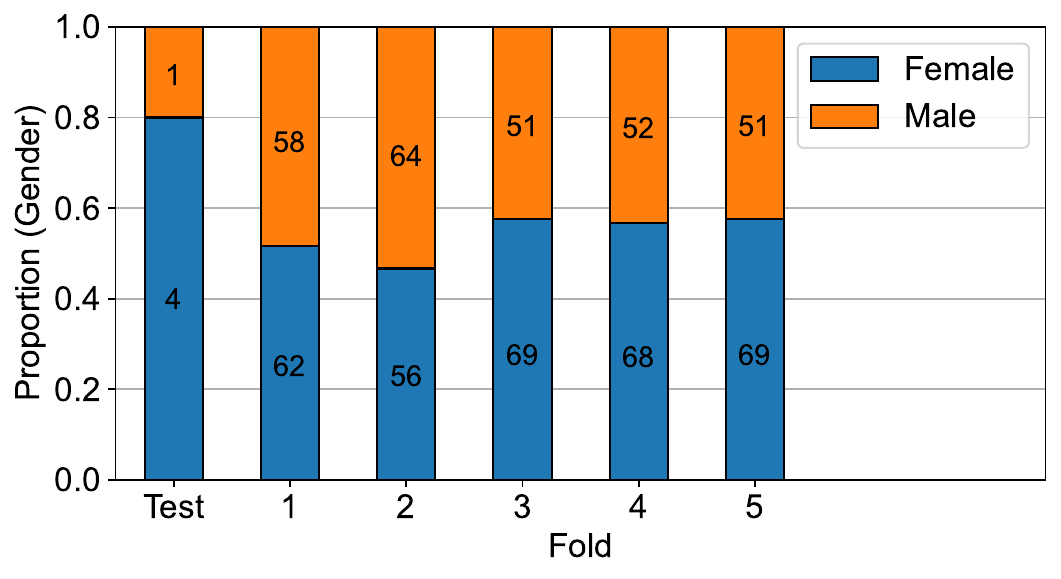}
	\includegraphics[width=\thisfigurescale\linewidth]{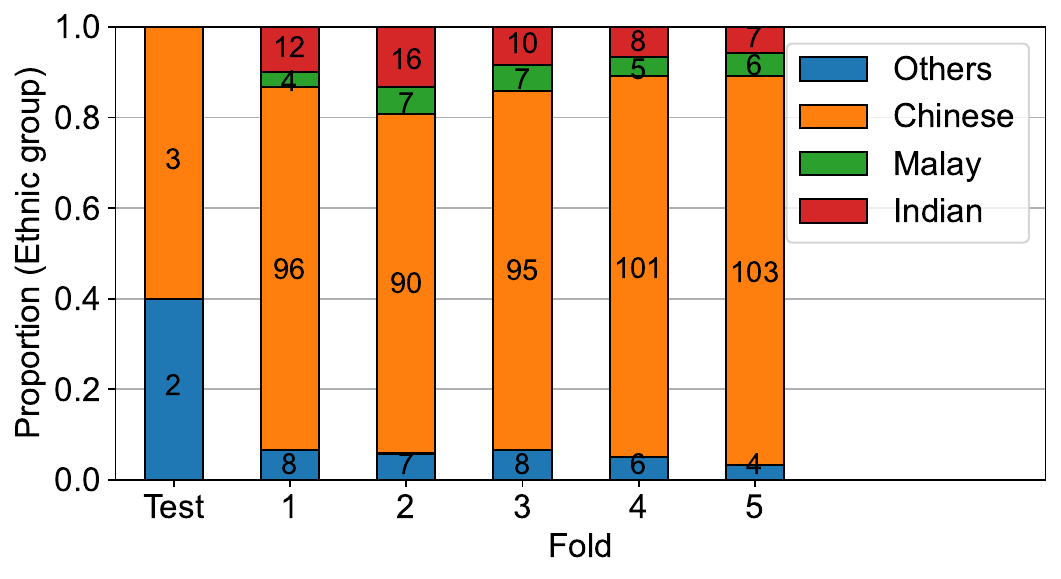}
	\includegraphics[width=\thisfigurescale\linewidth]{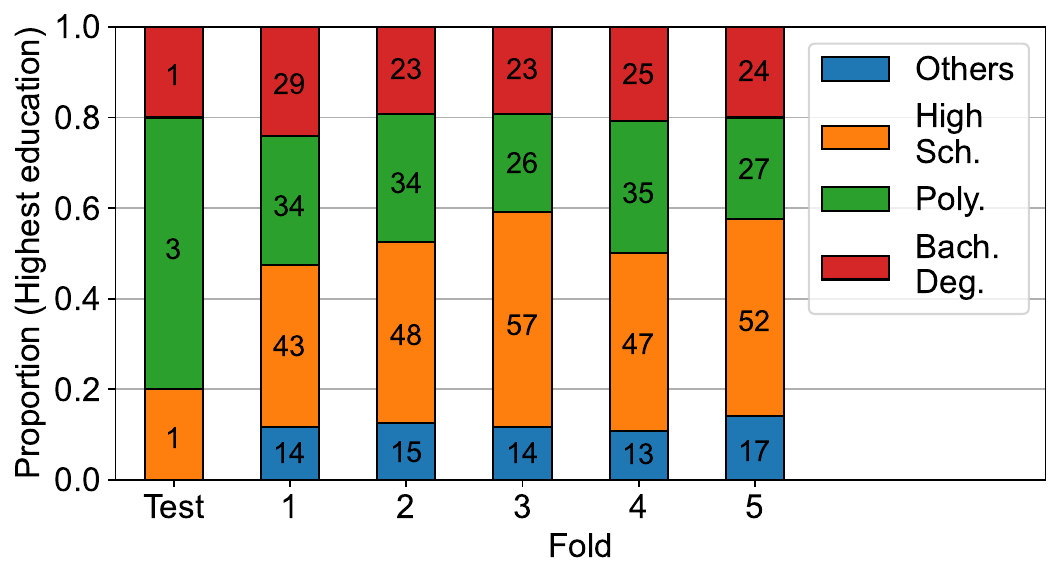}
	\includegraphics[width=\thisfigurescale\linewidth]{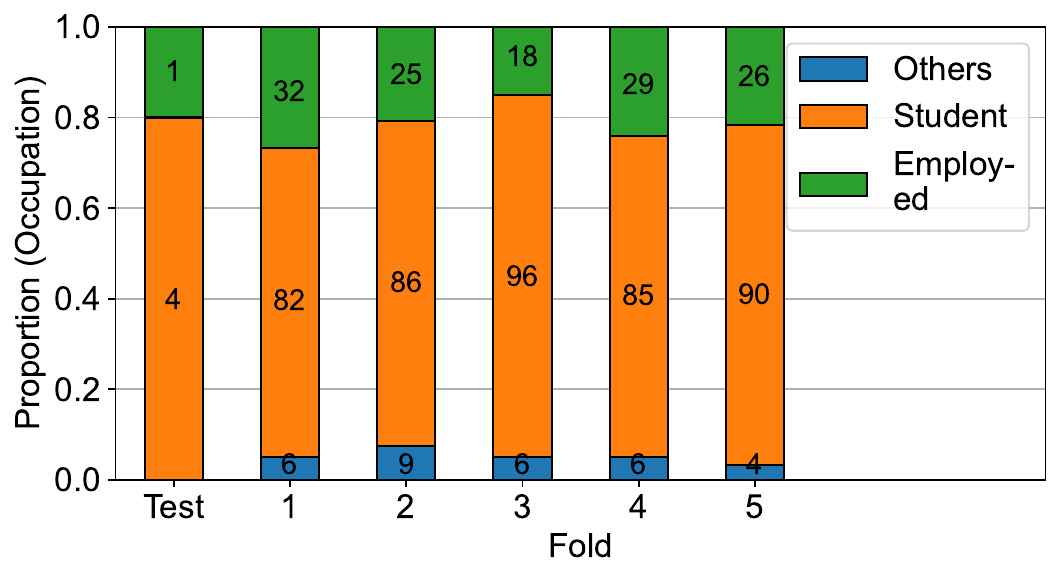}
	\includegraphics[width=\thisfigurescale\linewidth]{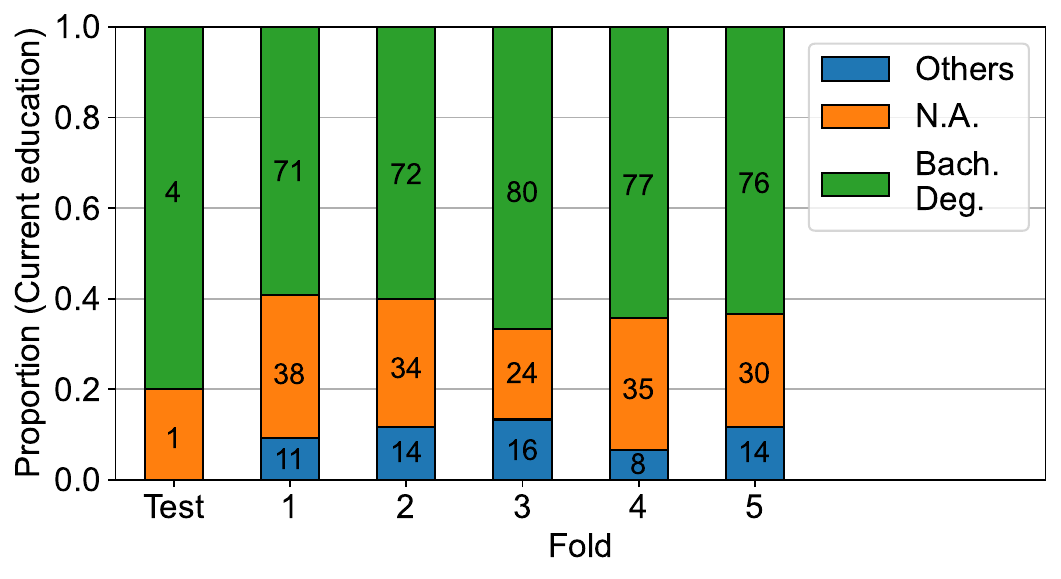}
	\includegraphics[width=\thisfigurescale\linewidth]{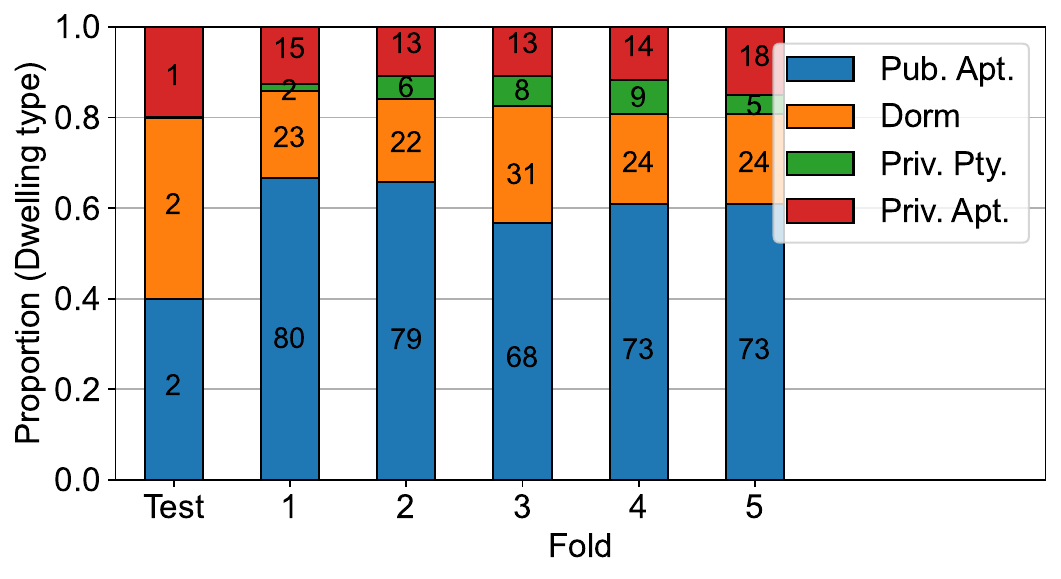}
	\includegraphics[width=\thisfigurescale\linewidth]{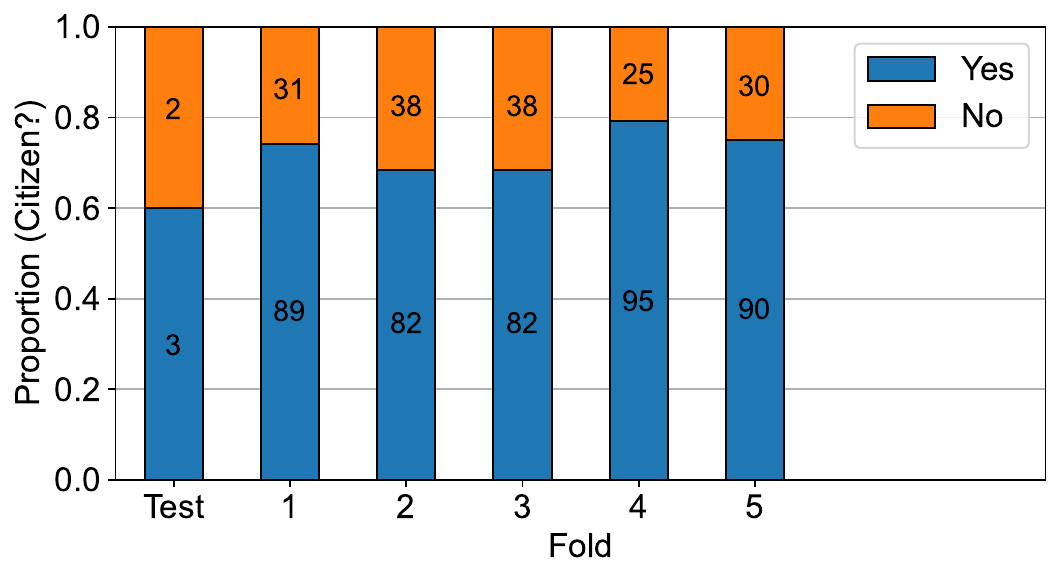}
	\includegraphics[width=\thisfigurescale\linewidth]{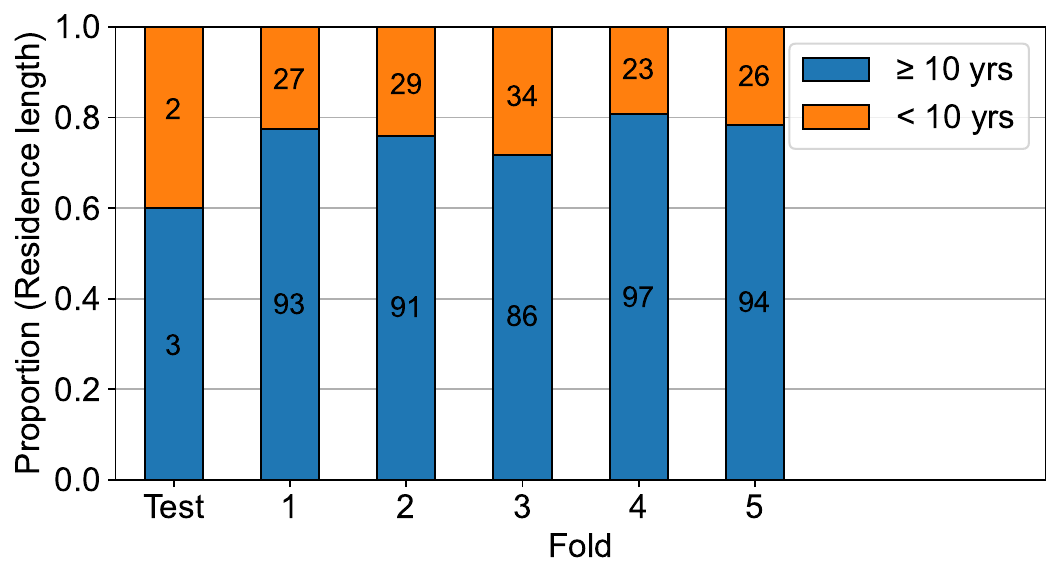}
	\caption{Stacked bar plots of distributions of PIQ responses coded as categorical variables, by fold: (a) If participant was multilingual, (b) If participant was a native English speaker (N.A. if monolingual), (c) If participant considered themselves most fluent in English (N.A. if monolingual), (d) Gender, (e) Ethnicity, (f) Highest education level obtained, (g) Occupational status, (h) Current education level (N.A. if not student), (i) Dwelling type, (j) If participant was a Singapore citizen, (k) Length of residence in Singapore. Chi-squared tests found no significant differences ($p>0.05$) between distributions of all variables by fold, except for ethnicity in the test set. Abbreviations in legend entries: Poly. = Polytechnic, Bach. Deg. = Bachelor's degree, Pub. Apt. = Public apartment, Priv. Pty. = Private property, Priv. Apt = Private apartment.}
	\label{fig:categorical_responses}
\end{figure*}
\end{landscape}

\begin{figure*}[h]
	\centering
	\includegraphics[width=0.32\linewidth]{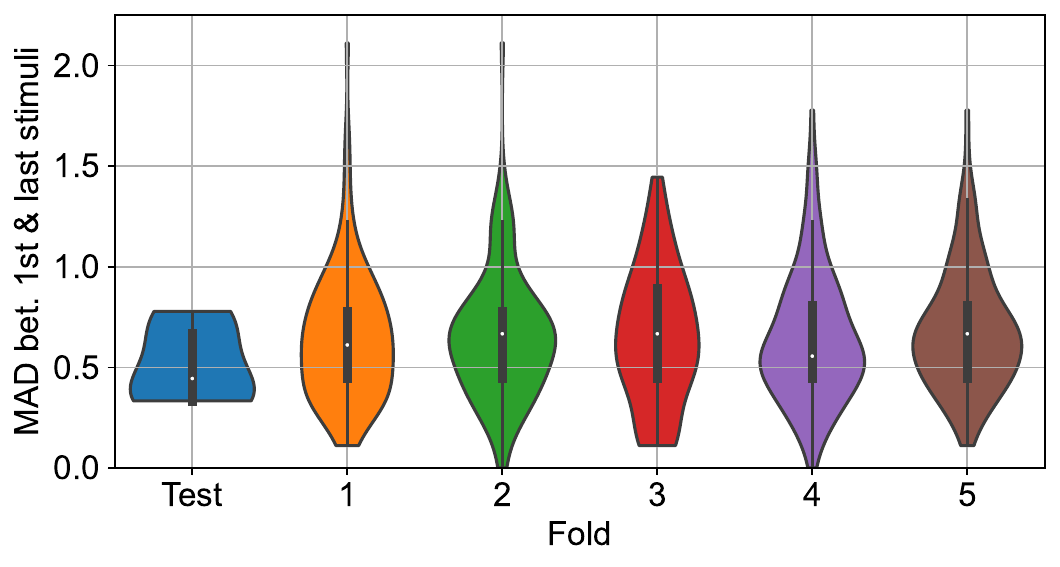}
	\includegraphics[width=0.32\linewidth]{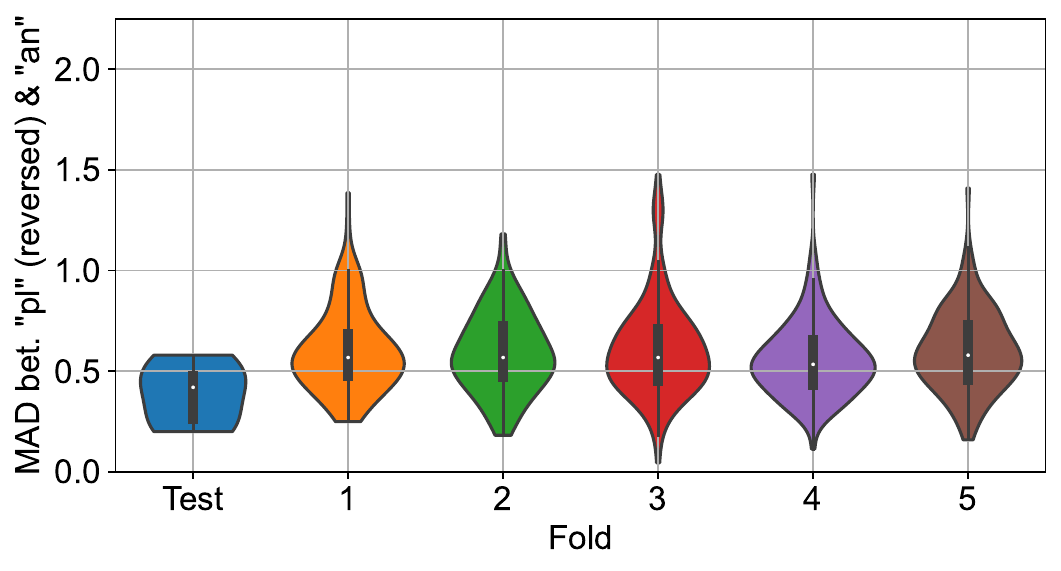}
	\includegraphics[width=0.32\linewidth]{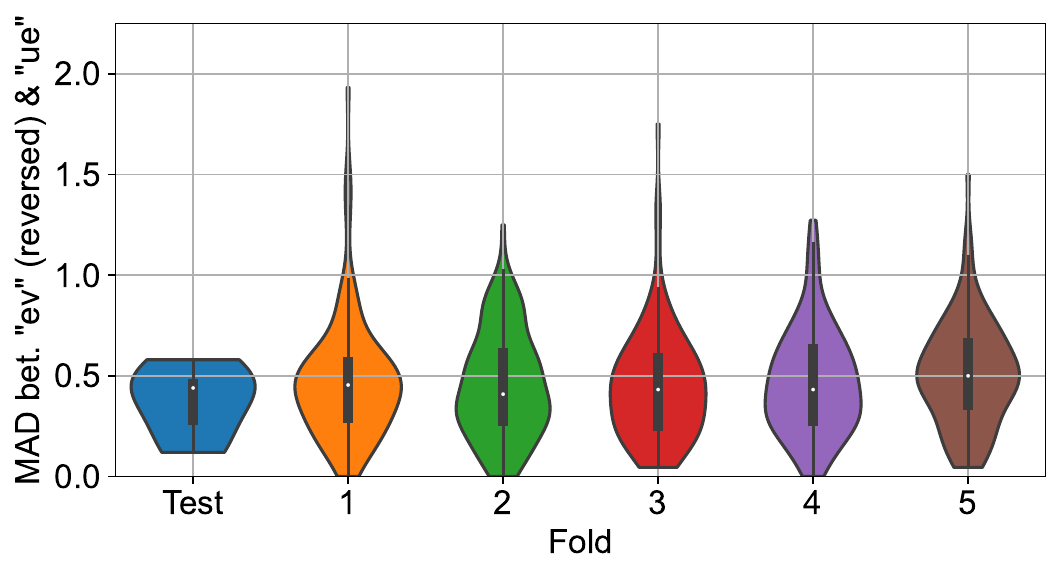}
	\includegraphics[width=0.32\linewidth]{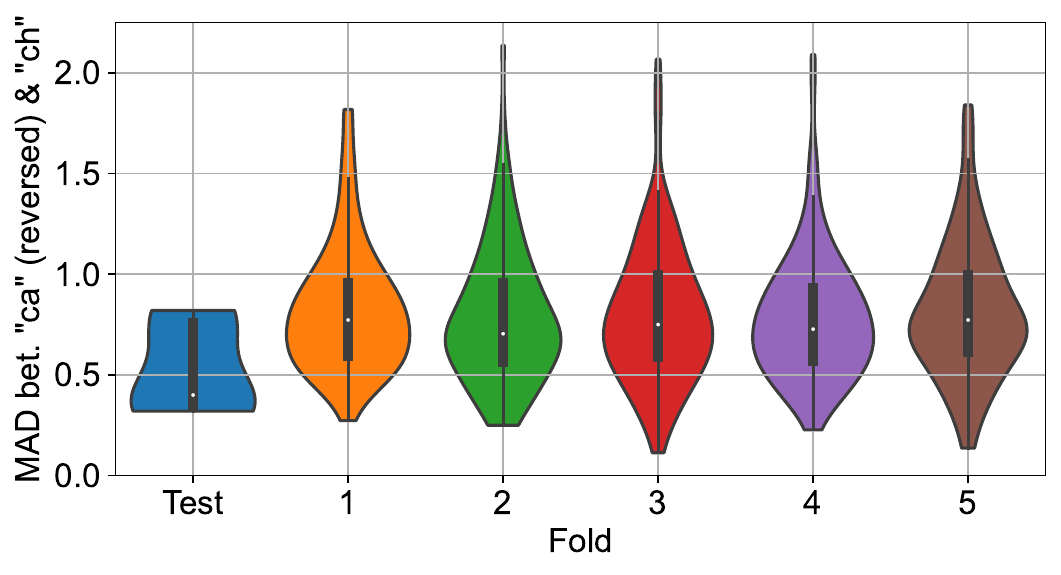}
	\includegraphics[width=0.32\linewidth]{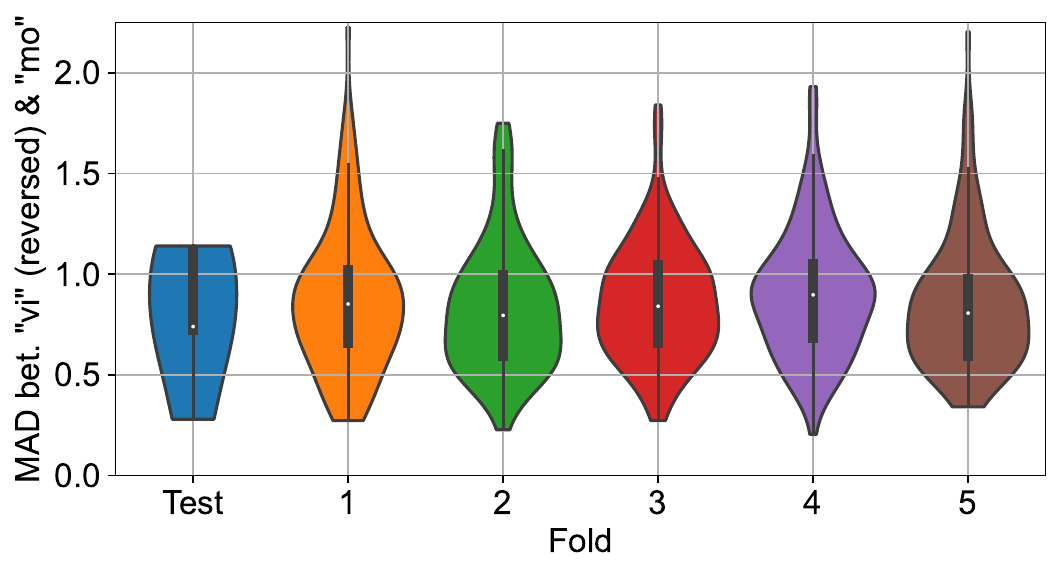}
	\includegraphics[width=0.32\linewidth]{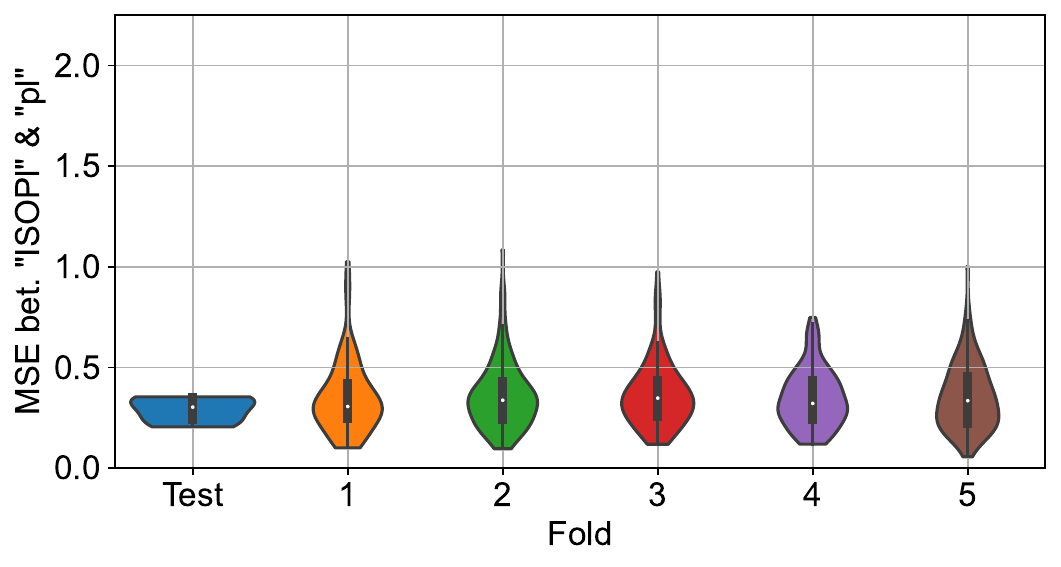}
	\includegraphics[width=0.32\linewidth]{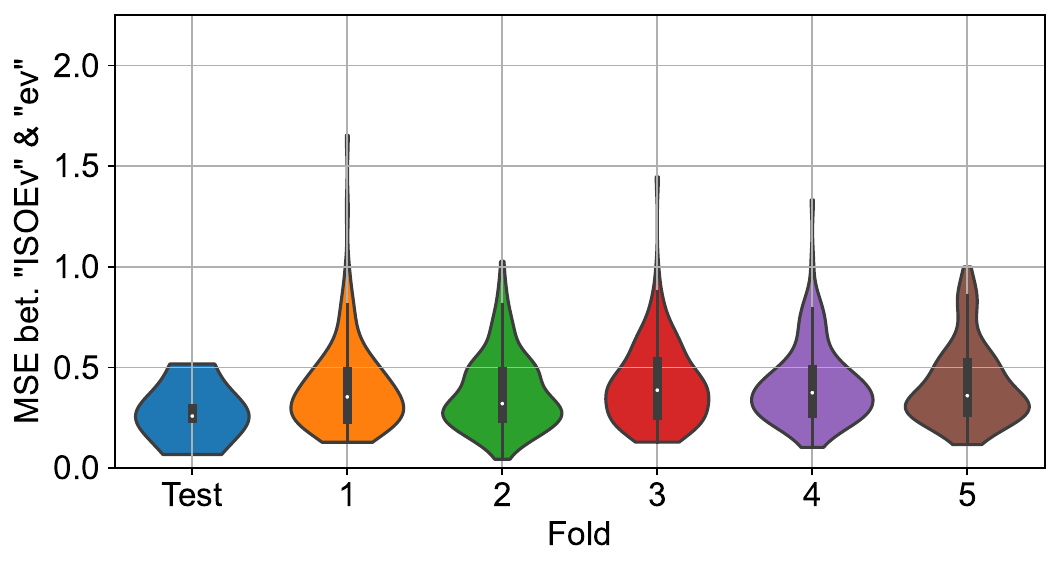}
	\caption{Violin plots of distributions of consistency metrics by fold:
	(a) Mean absolute difference (MAD) between ARQ responses on first (``practice'') and last (``consistency check'') stimuli,
	(b) MAD between reversed ``pleasant'' ($6-r_{\text{pl}}$) and ``annoying'' ($r_{\text{an}}$) responses across all stimuli,
	(c) MAD between reversed ``eventful'' ($6-r_{\text{ev}}$) and ``uneventful'' ($r_{\text{ue}}$) responses across all stimuli,
	(d) MAD between reversed ``calm'' ($6-r_{\text{ca}}$) and ``chaotic'' ($r_{\text{ch}}$) responses across all stimuli,
	(e) MAD between reversed ``vibrant'' ($6-r_{\text{vi}}$) and ``monotonous'' ($r_{\text{mo}}$) responses across all stimuli,
	(f) Mean squared error (MSE) between ``pleasant'' ($r_{\text{pl}}$) and rescaled ``ISO Pleasantness'' ($3+2P$) responses across stimuli,
	(g) MSE between ``eventful'' ($r_{\text{ev}}$) and rescaled ``ISO Eventfulness'' ($3+2E$) responses across stimuli.}
	\label{fig:consistency_checks}
\end{figure*}

This appendix collates the detailed distributions and statistical test results for the PIQ and ARQ described in Section 4. \Cref{fig:continuous_responses} and \Cref{fig:categorical_responses} respectively consolidate the distributions of the PIQ items coded as continuous and categorical variables, as violin plots by fold in the cross-validation set. \Cref{fig:consistency_checks} presents the distributions of the single-value metrics described in Section 3.9, as violin plots by fold in the cross-validation set and test set. Tables \ref{tab:continuous_responses} and \ref{tab:categorical_responses} respectively consolidate the $\chi^2$ and Kruskal-Wallis test results for the PIQ items coded as continuous and categorical variables, and Table \ref{tab:consistency_checks} presents the Kruskal-Wallis test results for the single-value metrics described in Section 3.9.

\begin{table}[h]
	\centering
	\caption{Summary of results of Kruskal-Wallis tests for PIQ responses coded as continuous variables, by fold. The test statistic for the Kruskal-Wallis test is denoted by $H$ and the corresponding $p$-values are given as $p$. No significant differences ($p>0.05$) were observed between distributions of all variables by fold, regardless of whether the test set was included as an additional fold.}
	\label{tab:continuous_responses}
	\def\arraystretch{1.0}
	\begin{tabular}{c@{ }lcccc}
		\toprule
		&& \multicolumn{2}{c}{Cross-validation set} & \multicolumn{2}{c}{Cross-validation set + test set} \\
		\cmidrule(lr){3-6}
		&Variable & $H$ & $p$ & $H$ & $p$ \\
		\midrule
		(a) & Age                                                       & \hspace{3mm} 1.6519 \hspace{3mm} & \hspace{3mm} 0.7994 \hspace{3mm} & \hspace{4mm} 4.4477 \hspace{4mm} & \hspace{4mm} 0.4869 \hspace{4mm} \\
		(b) & Extent annoyed by noise over past 12 months               & \hspace{3mm} 3.2587 \hspace{3mm} & \hspace{3mm} 0.5155 \hspace{3mm} & \hspace{4mm} 3.3215 \hspace{4mm} & \hspace{4mm} 0.6506 \hspace{4mm} \\
		(c) & Satisfaction of overall acoustic environment in Singapore & \hspace{3mm} 4.7949 \hspace{3mm} & \hspace{3mm} 0.3090 \hspace{3mm} & \hspace{4mm} 5.0618 \hspace{4mm} & \hspace{4mm} 0.4084 \hspace{4mm} \\
		(d) & Score on WNSS-10                                          & \hspace{3mm} 2.3861 \hspace{3mm} & \hspace{3mm} 0.6651 \hspace{3mm} & \hspace{4mm} 2.4252 \hspace{4mm} & \hspace{4mm} 0.7877 \hspace{4mm} \\
		(e) & Score on PSS-10                                           & \hspace{3mm} 1.1454 \hspace{3mm} & \hspace{3mm} 0.8870 \hspace{3mm} & \hspace{4mm} 2.5283 \hspace{4mm} & \hspace{4mm} 0.7722 \hspace{4mm} \\
		(f) & Score on WHO-5                                            & \hspace{3mm} 2.7427 \hspace{3mm} & \hspace{3mm} 0.6018 \hspace{3mm} & \hspace{4mm} 4.0850 \hspace{4mm} & \hspace{4mm} 0.5372 \hspace{4mm} \\
		(g) & Score on PANAS for Positive Affect                        & \hspace{3mm} 2.0961 \hspace{3mm} & \hspace{3mm} 0.7181 \hspace{3mm} & \hspace{4mm} 3.2124 \hspace{4mm} & \hspace{4mm} 0.6673 \hspace{4mm} \\
		(h) & Score on PANAS for Negative Affect                        & \hspace{3mm} 1.2815 \hspace{3mm} & \hspace{3mm} 0.8645 \hspace{3mm} & \hspace{4mm} 1.3094 \hspace{4mm} & \hspace{4mm} 0.9340 \hspace{4mm} \\
		\bottomrule
	\end{tabular}
\end{table}
\vfill
\FloatBarrier

\begin{landscape}
\begin{table}[h]
	\centering
	\caption{Summary of results of $\chi^2$-tests for PIQ responses coded as categorical variables, by fold. The test statistic for the $\chi^2$-test is denoted by $\chi^2$ and the corresponding $p$-values are given as $p$. Significant differences ($p<0.05$) were observed only for the test set in terms of the ethnicity of the participants.}
	\label{tab:categorical_responses}
	\def\arraystretch{1.0} 
	\begin{tabularx}{0.925\linewidth}{c@{ }Xcccccccccccc}
		\toprule
		&& \multicolumn{12}{c}{Fold} \\
		\cmidrule(lr){3-14}
		&& \multicolumn{2}{c}{Test} & \multicolumn{2}{c}{1} & \multicolumn{2}{c}{2} & \multicolumn{2}{c}{3} & \multicolumn{2}{c}{4} & \multicolumn{2}{c}{5} \\
		\cmidrule(lr){3-14}
		& Variable & $\chi^2$ & $p$ & $\chi^2$ & $p$ & $\chi^2$ & $p$ & $\chi^2$ & $p$ &$\chi^2$ & $p$ &$\chi^2$ & $p$ \\
		\midrule
		(a) & If participant was multilingual                             &  0.6114 & 0.4343\hphantom{$^*$} & 0.2443 & 0.6211 & 0.0153 & 0.9017 & 0.1870 & 0.6654 & 0.0153 & 0.9017 & 0.0344 & 0.8530 \\ 
		(b) & If participant was a native English speaker                 &  0.8026 & 0.6694\hphantom{$^*$} & 0.3707 & 0.8308 & 1.9425 & 0.3786 & 1.3122 & 0.5189 & 0.6725 & 0.7145 & 1.6799 & 0.4317 \\ 
		(c) & If participant considered themselves most fluent in English &  1.7748 & 0.4117\hphantom{$^*$} & 1.3885 & 0.4995 & 0.6943 & 0.7067 & 0.4855 & 0.7845 & 0.5547 & 0.7578 & 0.2225 & 0.8947 \\ 
		(d) & Gender                                                      &  1.3607 & 0.2434\hphantom{$^*$} & 0.2630 & 0.6081 & 2.5980 & 0.1070 & 0.5918 & 0.4417 & 0.3435 & 0.5578 & 0.5918 & 0.4417 \\ 
		(e) & Ethnicity                                                   & 11.7723 & 0.0082$^*$            & 1.0508 & 0.7890 & 3.5286 & 0.3171 & 0.6204 & 0.8917 & 0.9676 & 0.8091 & 2.6249 & 0.4531 \\ 
		(f) & Highest education level obtained                            &  3.3766 & 0.7603\hphantom{$^*$} & 3.8880 & 0.6918 & 1.5046 & 0.9592 & 3.7658 & 0.7083 & 1.7228 & 0.9433 & 2.9474 & 0.8154 \\ 
		(g) & Occupational status                                         &  0.2967 & 0.9900\hphantom{$^*$} & 3.0213 & 0.5543 & 3.3845 & 0.4957 & 5.3639 & 0.2520 & 2.8398 & 0.5850 & 0.8643 & 0.9296 \\ 
		(h) & Current education level                                     &  0.8517 & 0.9736\hphantom{$^*$} & 2.1174 & 0.8327 & 0.6463 & 0.9858 & 5.0025 & 0.4156 & 2.5056 & 0.7756 & 4.2588 & 0.5128 \\ 
		(i) & Dwelling type                                               &  1.8017 & 0.6146\hphantom{$^*$} & 3.1992 & 0.3619 & 0.7510 & 0.8612 & 2.9759 & 0.3954 & 1.5848 & 0.6628 & 1.0186 & 0.7968 \\ 
		(j) & If participant was a Singapore citizen                      &  0.4287 & 0.5126\hphantom{$^*$} & 0.0829 & 0.7734 & 1.3259 & 0.2495 & 1.3259 & 0.2495 & 2.3152 & 0.1281 & 0.2435 & 0.6217 \\ 
		(k) & Length of residence in Singapore                            &  0.7960 & 0.3723\hphantom{$^*$} & 0.0300 & 0.8626 & 0.0674 & 0.7951 & 1.7997 & 0.1798 & 1.0787 & 0.2990 & 0.1517 & 0.6969 \\
		\bottomrule
	\end{tabularx}
\end{table}
\begin{table}[h] 
	\centering
	\caption{Summary of results of Kruskal-Wallis tests for consistency checks on ARQ responses, by fold. The test statistic for the Kruskal-Wallis test is denoted by $H$ and the corresponding $p$-values are given as $p$. No significant differences ($p>0.05$) were observed between distributions of all variables by fold, regardless of whether the test set was included as an additional fold. MAD stands for mean absolute deviation. MSE stands for mean squared error. Except (a), the MAD and MSE were computed by taking the mean across all stimuli presented to single participant.}
	\label{tab:consistency_checks}
	\def\arraystretch{1.0}
	\makebox[\textwidth][c]{ %
		\begin{tabular}{c@{ }l@{ between }l@{ and }lcccc}
			\toprule
			\multicolumn{4}{l}{ } & \multicolumn{2}{c}{Cross-validation set} & \multicolumn{2}{c}{Cross-validation set + test set} \\
			\cmidrule(lr){5-8}
			& \multicolumn{3}{l}{\hspace{-2mm}Variable} & $H$ & $p$ & $H$ & $p$ \\
			\midrule
			(a) & MAD & ARQ responses on first (``pre-experiment'') & last (``post-experiment'') stimuli     & \hspace{2mm} 1.7164 \hspace{2mm} & \hspace{2mm} 0.7877 \hspace{2mm} & \hspace{4mm} 2.9306 \hspace{4mm} & \hspace{4mm} 0.7107 \hspace{4mm} \\
			(b) & MAD & reversed ``pleasant'' ($6-r_{\text{pl}}$)   & ``annoying'' ($r_{\text{an}}$)         & \hspace{2mm} 5.0281 \hspace{2mm} & \hspace{2mm} 0.2844 \hspace{2mm} & \hspace{4mm} 9.7446 \hspace{4mm} & \hspace{4mm} 0.0828 \hspace{4mm} \\
			(c) & MAD & reversed ``eventful'' ($6-r_{\text{ev}}$)   & ``uneventful'' ($r_{\text{ue}}$)       & \hspace{2mm} 5.1821 \hspace{2mm} & \hspace{2mm} 0.2691 \hspace{2mm} & \hspace{4mm} 5.6611 \hspace{4mm} & \hspace{4mm} 0.3406 \hspace{4mm} \\
			(d) & MAD & reversed ``calm'' ($6-r_{\text{ca}}$)       & ``chaotic'' ($r_{\text{ch}}$)          & \hspace{2mm} 1.8261 \hspace{2mm} & \hspace{2mm} 0.7677 \hspace{2mm} & \hspace{4mm} 5.5218 \hspace{4mm} & \hspace{4mm} 0.3556 \hspace{4mm} \\
			(e) & MAD & reversed ``vibrant'' ($6-r_{\text{vi}}$)    & ``monotonous'' ($r_{\text{mo}}$)       & \hspace{2mm} 5.3364 \hspace{2mm} & \hspace{2mm} 0.2545 \hspace{2mm} & \hspace{4mm} 5.3441 \hspace{4mm} & \hspace{4mm} 0.3753 \hspace{4mm} \\
			(f) & MSE & ``pleasant'' ($r_{\text{pl}}$)              & rescaled ``ISO Pleasantness'' ($3+2P$) & \hspace{2mm} 1.3749 \hspace{2mm} & \hspace{2mm} 0.8485 \hspace{2mm} & \hspace{4mm} 2.0420 \hspace{4mm} & \hspace{4mm} 0.8433 \hspace{4mm} \\
			(g) & MSE & ``eventful'' ($r_{\text{ev}}$)              & rescaled ``ISO Eventfulness'' ($3+2E$) & \hspace{2mm} 4.8070 \hspace{2mm} & \hspace{2mm} 0.3077 \hspace{2mm} & \hspace{4mm} 6.9141 \hspace{4mm} & \hspace{4mm} 0.2271 \hspace{4mm} \\
			\bottomrule
		\end{tabular}
	}
\end{table}
\end{landscape}
\vfill
\FloatBarrier

\vfill
\setcounter{figure}{0}
\setcounter{table}{0}

\FloatBarrier
\clearpage
\section{Details of Principal Component Analysis for Fold Allocation}
\label{sec:Appendix/Details of Principal Component Analysis for Fold Allocation}

This appendix presents heat maps of the weights given to each input feature when the principal component analysis (PCA) in the fold allocation process described in Section 3.3 was conducted for the set of base urban soundscapes, as well as the sets of maskers in each class. Only the first $k$ principal components for each set are shown in the heat maps, where $k$ is the number of principal components that together explained at least 90\% of the observed variance.

Since the PCA was conducted independently for the set of base urban soundscapes, as well as the sets of maskers in each class, the assumption here was that each set had its own independent PCA that best summarized the input features for that set. We can observe from the heat maps that this is indeed the case, with the weights being significantly different between the sets. In addition, we can also see that weights with relatively higher absolute values are clustered together in blocks representing each class of psychoacoustic indicators (e.g., principal component \#2 for the set of soundscape tracks has weights with comparatively high absolute values for the sharpness-related indicators), which indicates that the principal component dimensions appear to be summarizing related features into the individual principal components. This is as expected since the purpose of applying PCA in Section 3.3 was purely for dimensionality reduction, in order to reduce the number of (redundant) parameters for the subsequent clustering via self-organizing maps.

\begin{figure*}[!ht]
	\centering
	\includegraphics[width=\linewidth]{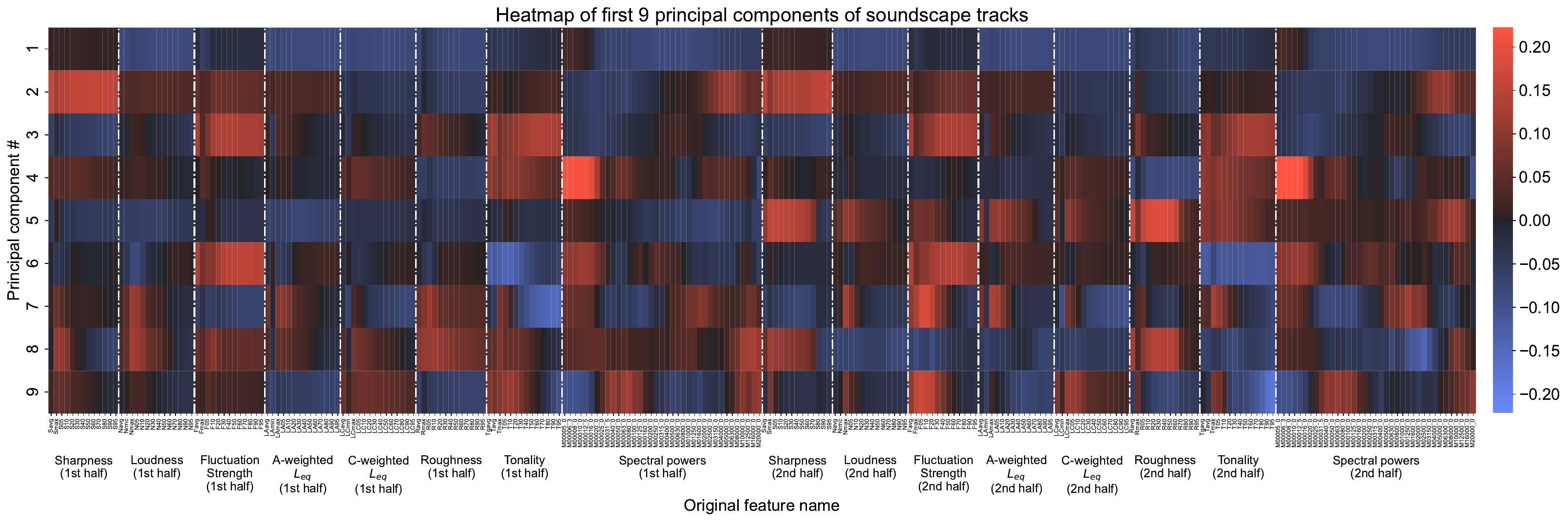}
	\includegraphics[width=\linewidth]{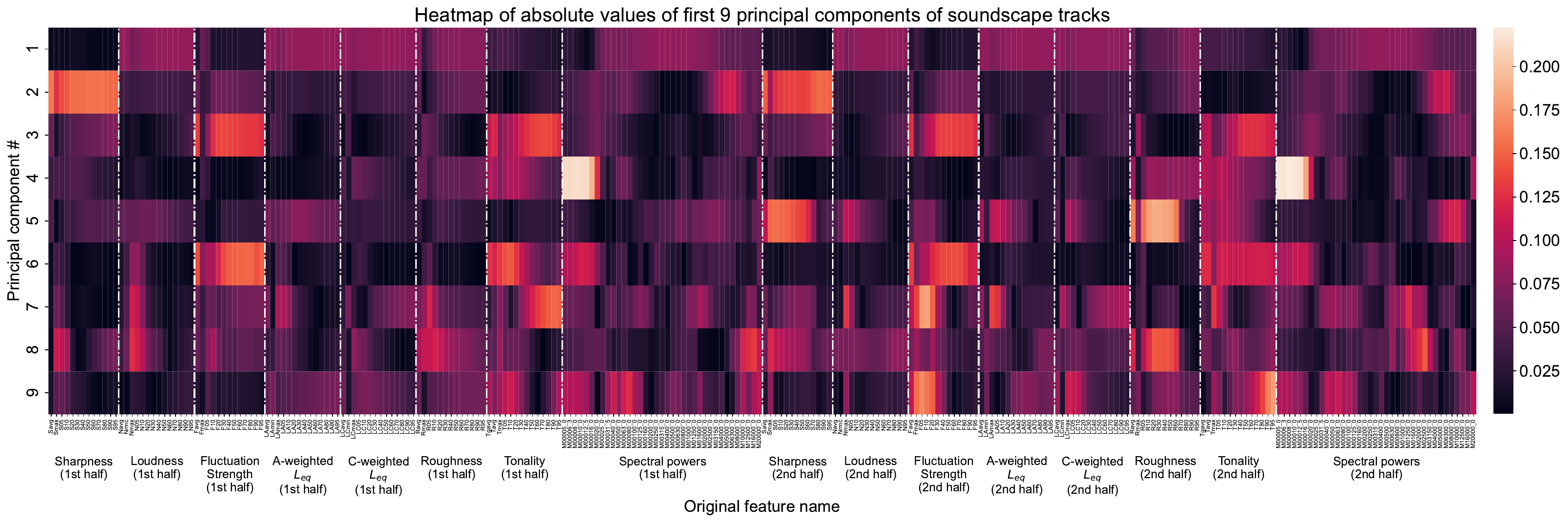}
	\caption{Heat maps of (top) actual values, (bottom) absolute values of the first 9 principal components of the base urban soundscape recordings.}
	\label{fig:pca_soundscape}
\end{figure*}

\begin{figure*}[!ht]
	\centering
	\includegraphics[width=\linewidth]{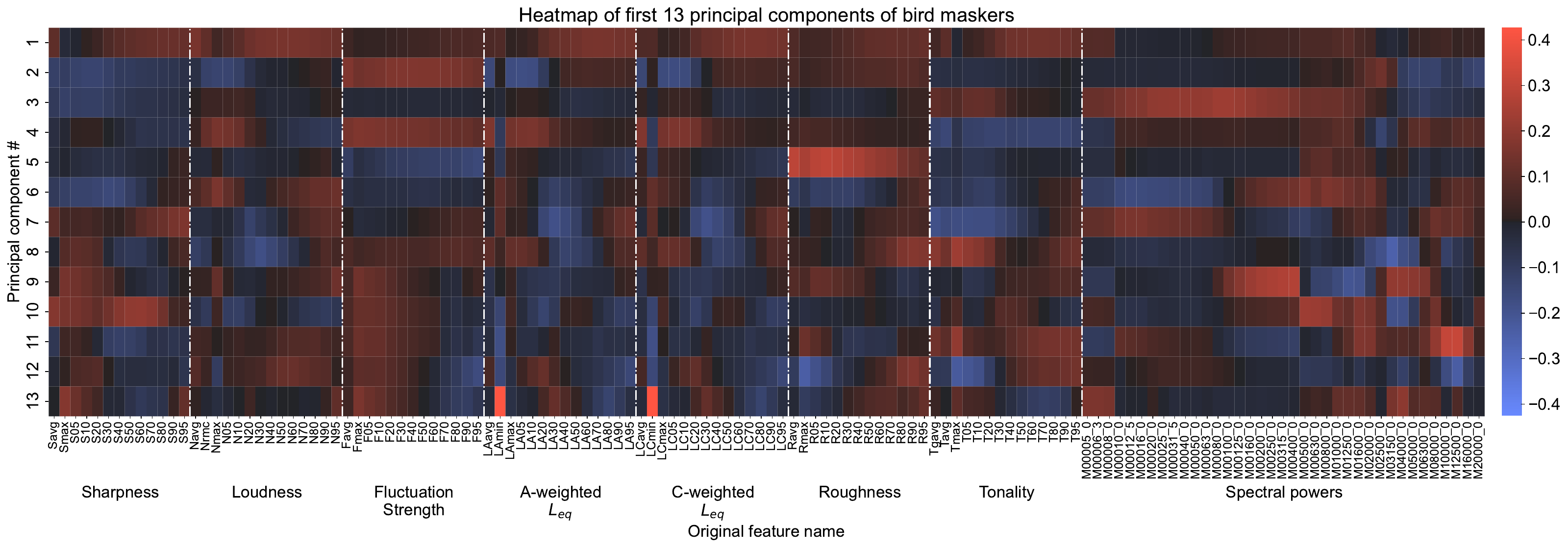}
	\includegraphics[width=\linewidth]{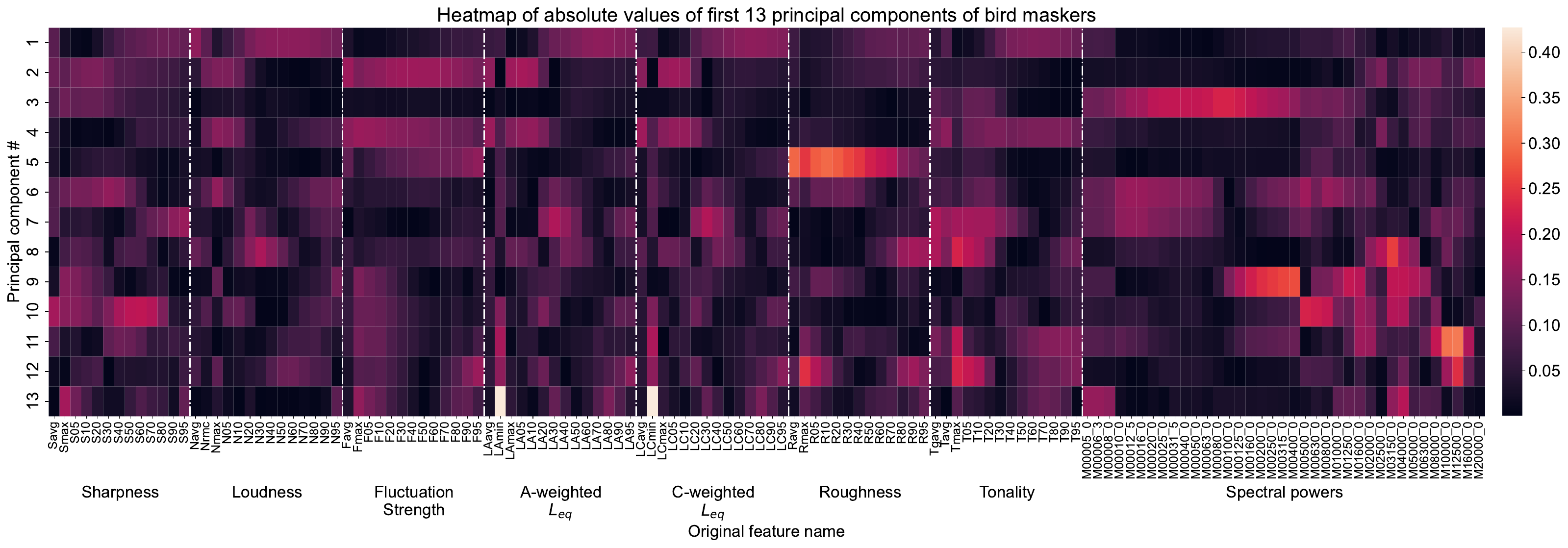}
	\caption{Heat maps of (top) actual values, (bottom) absolute values of the first 13 principal components of the maskers in the bird class.}
	\label{fig:pca_bird}
\end{figure*}

\begin{figure*}[!ht]
	\centering
	\includegraphics[width=\linewidth]{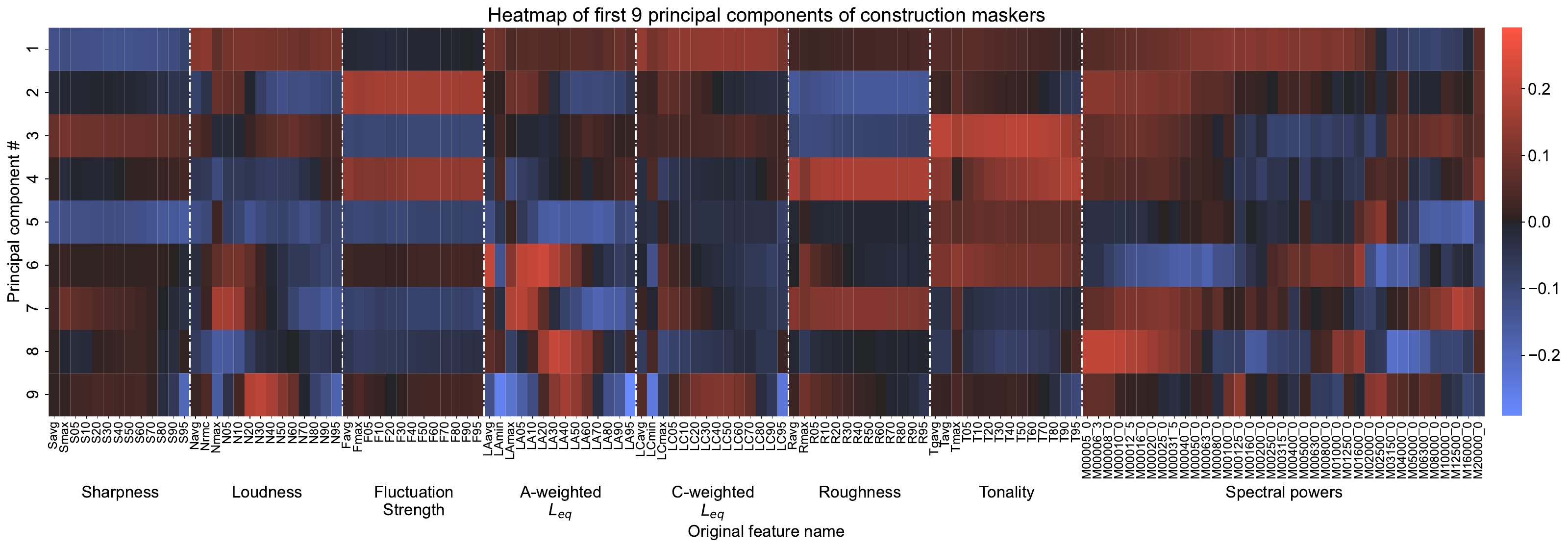}
	\includegraphics[width=\linewidth]{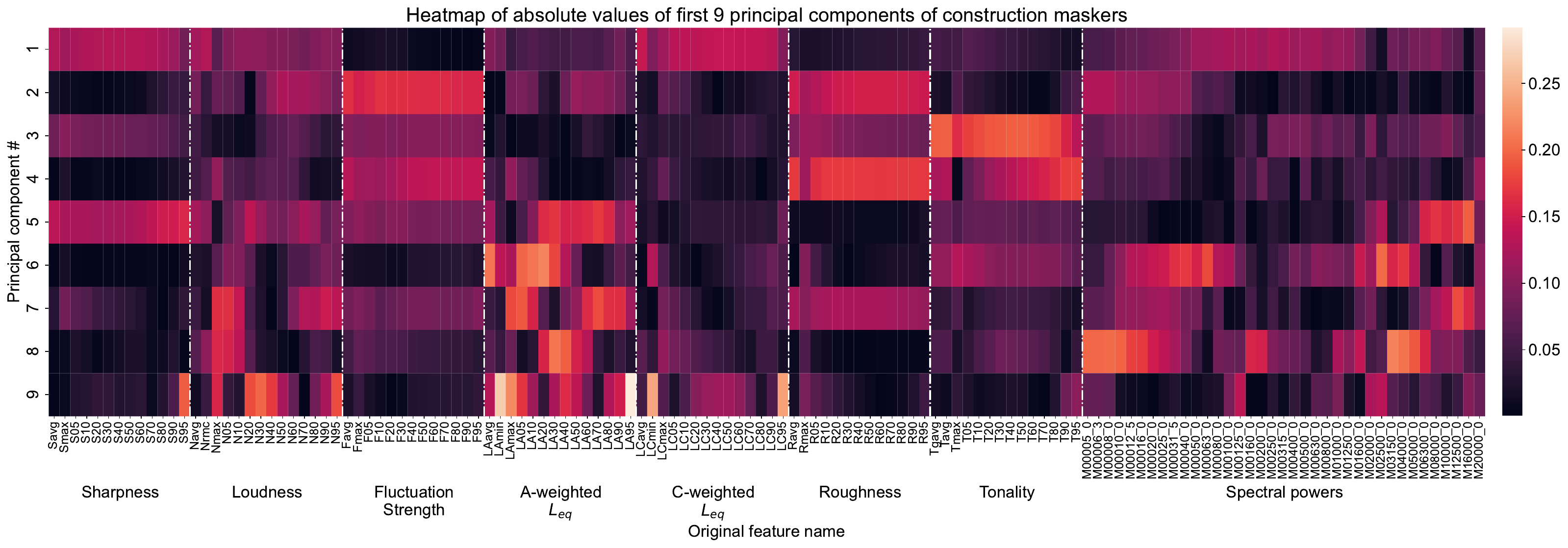}
	\caption{Heat maps of (top) actual values, (bottom) absolute values of the first 9 principal components of the maskers in the construction class.}
	\label{fig:pca_construction}
\end{figure*}

\begin{figure*}[!ht]
	\centering
	\includegraphics[width=\linewidth]{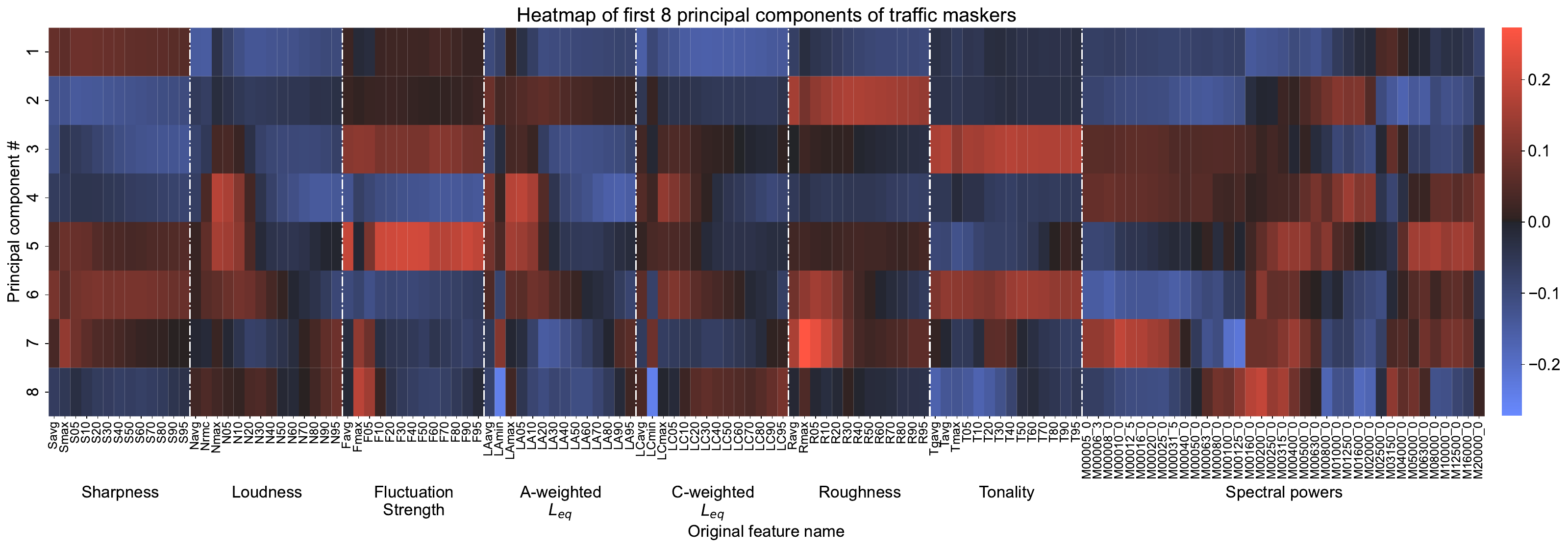}
	\includegraphics[width=\linewidth]{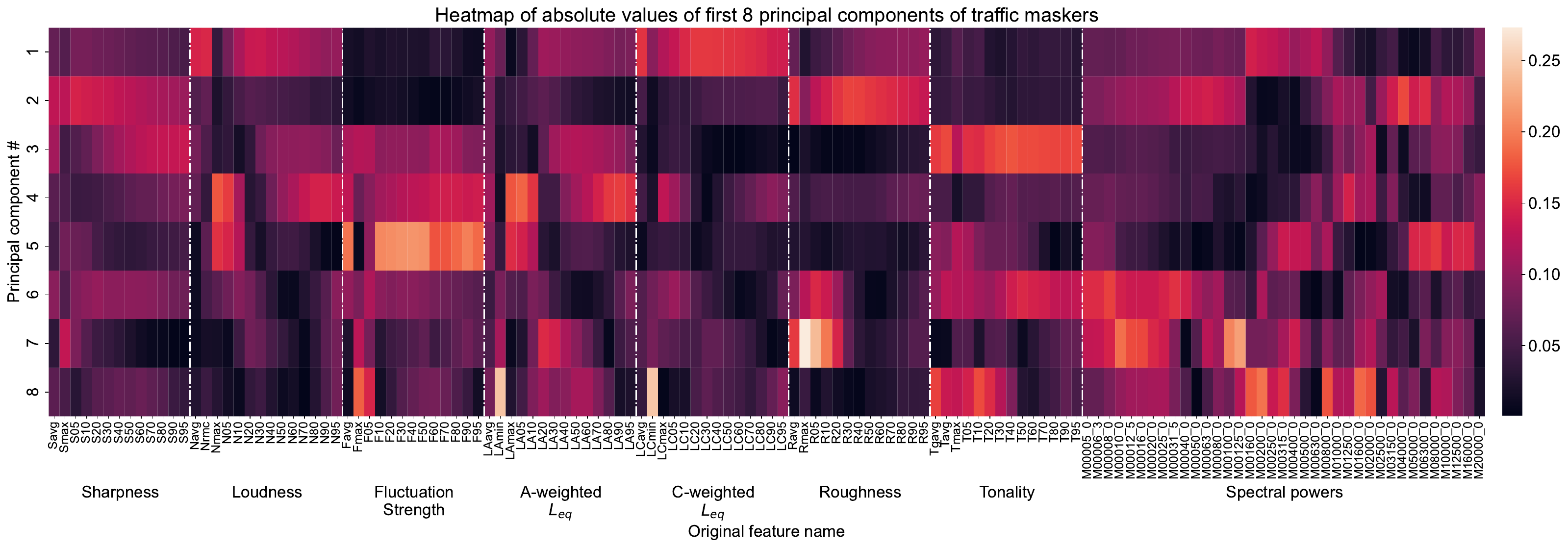}
	\caption{Heat maps of (top) actual values, (bottom) absolute values of the first 8 principal components of the maskers in the traffic class.}
	\label{fig:pca_traffic}
\end{figure*}

\begin{figure*}[!ht]
	\centering
	\includegraphics[width=\linewidth]{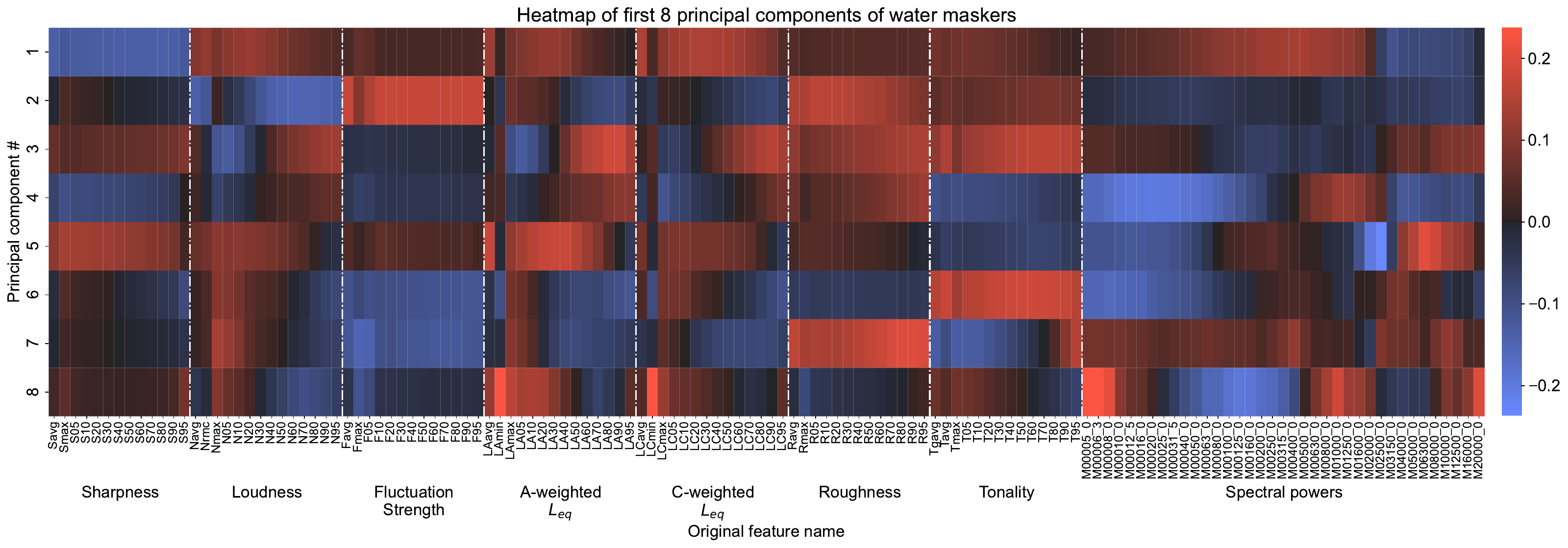}
	\includegraphics[width=\linewidth]{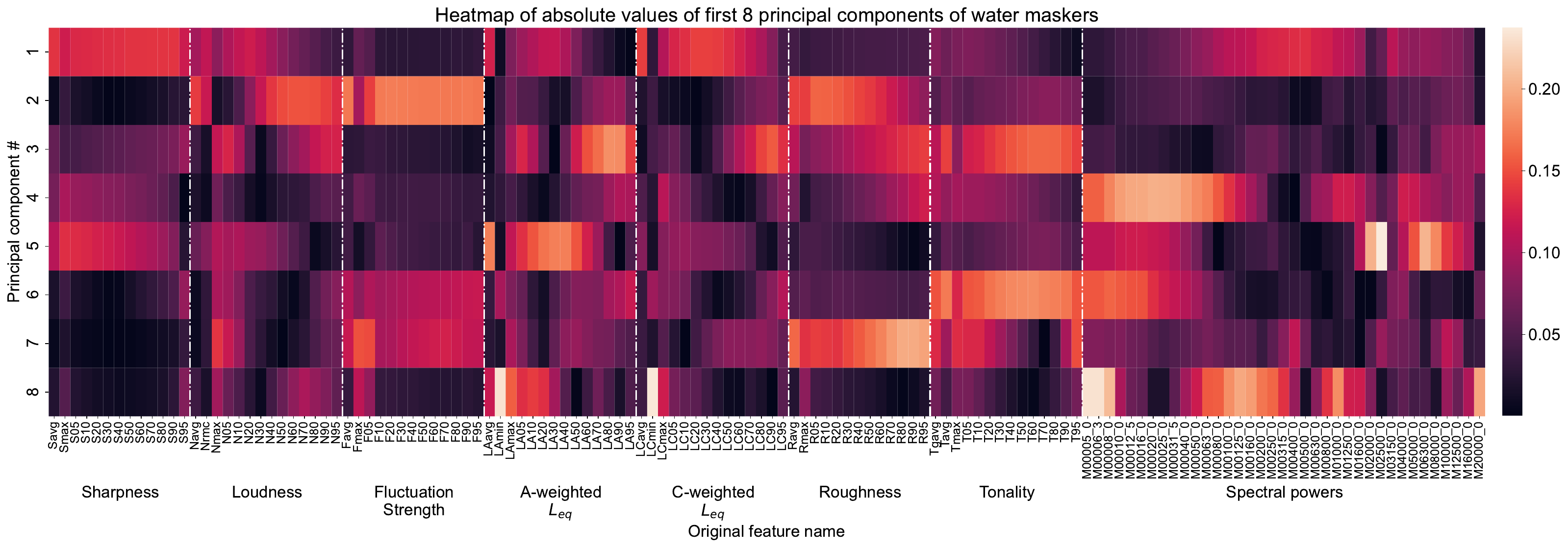}
	\caption{Heat maps of (top) actual values, (bottom) absolute values of the first 8 principal components of the maskers in the water class.}
	\label{fig:pca_water}
\end{figure*}

\begin{figure*}[!ht]
	\centering
	\includegraphics[width=\linewidth]{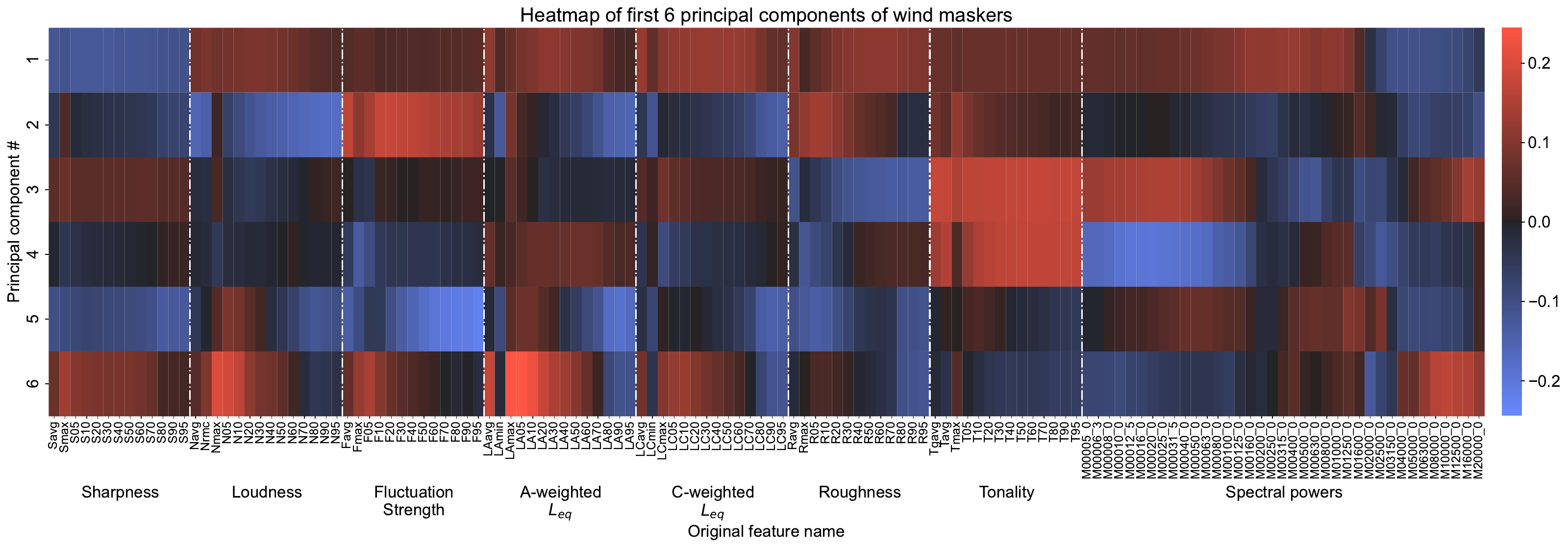}
	\includegraphics[width=\linewidth]{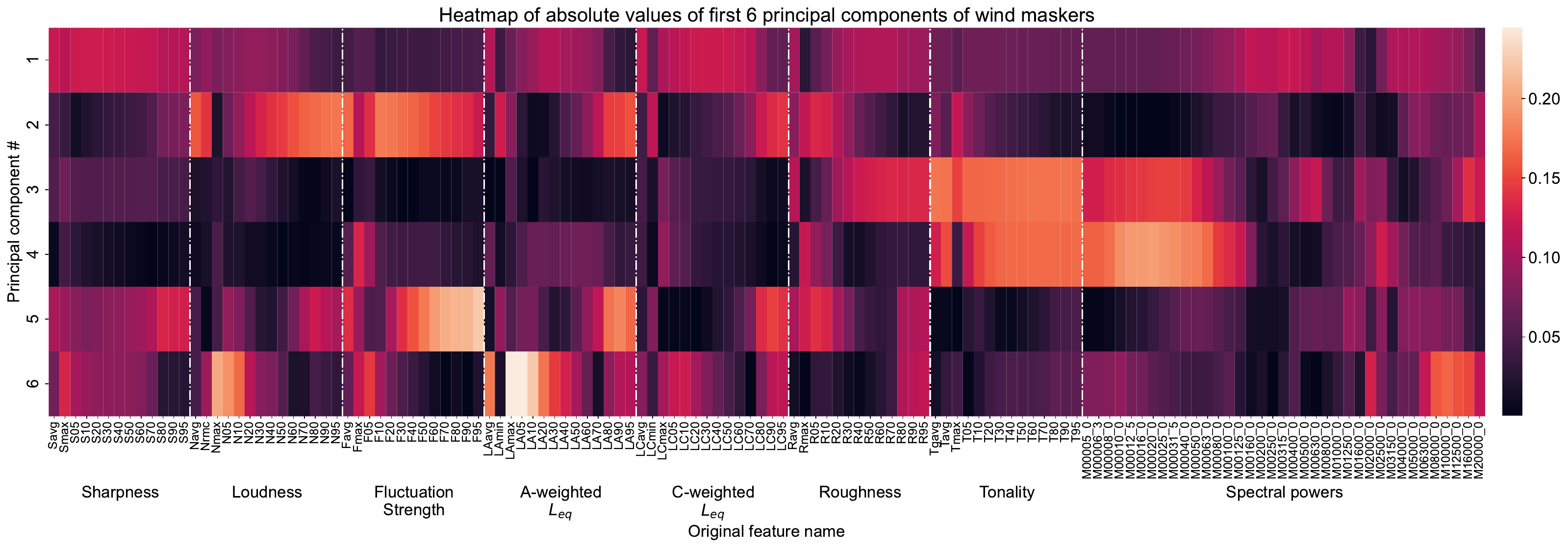}
	\caption{Heat maps of (top) actual values, (bottom) absolute values of the first 6 principal components of the maskers in the wind class.}
	\label{fig:pca_wind}
\end{figure*}
\setcounter{figure}{0}
\setcounter{table}{0}

\end{document}